\documentclass[
reprint,
superscriptaddress,
 amsmath,amssymb,
 aps,
 prc,
 showkeys,
]{revtex4-2}

\usepackage{graphicx}
\usepackage{dcolumn}
\usepackage{bm}
\usepackage{hyperref}
\usepackage[mathlines]{lineno}
\usepackage{textgreek}
\usepackage{enumerate}
\usepackage{wasysym}
\hypersetup{colorlinks=true, urlcolor=blue, linkcolor=blue, citecolor=blue}

\usepackage{subfiles}
\usepackage{soul}

\begin{document}

\newcommand{\cevns}{CE\textnu{}NS} 
\newcommand\isotope[2]{\textsuperscript{#2}#1}

\title{A measurement of the sodium and iodine scintillation quenching factors across multiple NaI(Tl) detectors to identify systematics}

\newcommand{\chicago}{Enrico Fermi Institute and Kavli Institute for Cosmological Physics, University of Chicago, Chicago, IL 60637, USA}
\newcommand{\duke}{Department of Physics, Duke University, Durham, NC 27708, USA}
\newcommand{\tunl}{Triangle Universities Nuclear Laboratory, Duke University, Durham, NC 27708, USA}
\newcommand{\yale}{Department of Physics and Wright Laboratory, Yale University, New Haven, CT 06520, USA}
\newcommand{\zaragoza}{Centro de Astropart\'{\i}culas y F\'{\i}sica de Altas Energ\'{\i}as (CAPA), Universidad de Zaragoza, 50009 Zaragoza, SPAIN}

\author{D.~Cintas}
\affiliation{\zaragoza}

\author{S.~Hedges}
\affiliation{\duke}
\affiliation{\tunl}

\author{W.~G.~Thompson}
\affiliation{\yale}

\author{P.~An}
\affiliation{\duke}
 \affiliation{\tunl}

\author{C.~Awe}
\affiliation{\duke}
 \affiliation{\tunl}

\author{P.~S.~Barbeau}
\affiliation{\duke}
\affiliation{\tunl}

\author{E.~Barbosa de Souza}
\affiliation{\yale}

\author{J.~H.~Jo}
\affiliation{\yale}

\author{L.~Li}
\affiliation{\duke}
\affiliation{\tunl}

\author{M.~Mart\'inez}
\affiliation{\zaragoza}

\author{R.~H.~Maruyama}
\affiliation{\yale}

\author{G.~C.~Rich}
\affiliation{\chicago}

\author{J.~Runge}
\affiliation{\duke}
\affiliation{\tunl}

\author{M.~L.~Sarsa}
\affiliation{\zaragoza}
\email{mlsarsa@unizar.es}
\thanks{Corresponding author}

\date{\today}

\begin{abstract}

The amount of light produced by nuclear recoils in scintillating targets is strongly quenched compared to that produced by electrons. A precise understanding of the quenching factor is particularly interesting for WIMP searches and \cevns~measurements since both rely on nuclear recoils, whereas energy calibrations are more readily accessible from electron recoils. There is a wide variation among the current measurements of the quenching factor in sodium iodide (NaI) crystals, especially below 10\,keV, the energy region of interest for dark matter and \cevns~studies. A better understanding of the quenching factor in NaI(Tl) is of particular interest for resolving the decades-old puzzle in the field of dark matter between the null results of most WIMP searches and the claim for dark matter detection by the DAMA/LIBRA collaboration. In this work, we measured sodium and iodine quenching factors for five small NaI(Tl) crystals grown with similar thallium concentrations and growth procedures. Unlike previous experiments, multiple crystals were tested, with measurements made in the same experimental setup to control systematic effects. The quenching factors agree in all crystals we investigated, and both sodium and iodine quenching factors are smaller than those reported by DAMA/LIBRA. The dominant systematic effect was due to the electron equivalent energy calibration originating from the non-proportional behavior of the NaI(Tl) light yield at lower energies, potentially the cause for the discrepancies among the previous measurements.

\end{abstract}

\maketitle


\section{Introduction}
\label{sec:intro}

Thallium-doped sodium iodide (NaI(Tl)) crystals have been used as particle detectors since the middle of the past century and are used widely  in nuclear and particle physics~\cite{knoll-ch8}. These detectors are commonly used in rare event searches, such as the direct detection of the dark matter composing the galactic dark halo in form of weakly interacting massive particles (WIMPs)~\cite{naiad-final-result,cosine100-detector,anais-detector,dama-libra2020, sabre, COSINUS:2023kqd} and the measurement of coherent elastic neutrino-nucleus scattering (\cevns)~\cite{COHERENT:2023ffx,Choi2023}. 
\par
Despite its long history, there are several properties that are yet to be precisely measured, correctly modelled, or understood. In particular, measurements of the response of NaI(Tl) detectors to nuclear scattering events show discrepancies among past measurements~\cite{spooner-quench,dama-quench-original,gerbier-quench,simon-quench,chagani-quench,collar-quench,xu-quench,stiegler-quench,joo-quench,bignell-quench}. Several factors may contribute to the discrepancies: crystal properties such as doping, density of defects, or the growing method; or systematic effects in the measurement and analysis procedures that are yet unaccounted for.  Clarifying this issue is essential for WIMPs and \cevns~searches, because both WIMPs and neutrinos deposit energy in NaI(Tl) detectors mostly via nuclear recoils. 
\par
Only a fraction of the energy deposited by a particle in NaI(Tl) leads to scintillation. This fraction is different if the interacting particle is an electron, an alpha particle, or a heavier nucleus. This results in a very different conversion between light collected and energy deposited, i.e.~calibration of the energy scale, depending on the type of interacting particle. Usually, gamma rays are used for calibration, and because they deposit the energy via electrons, the energy scale derived from such calibration procedure is named {\it electron-equivalent energy} ($E_{ee}$)~\footnote{Throughout this article, we will use keV\textsubscript{ee} for {\it electron-equivalent energy}, while keV\textsubscript{nr} will be used for nuclear recoil energy.}.
Electrons are much more effective than alpha particles or nuclei in the production of light in most scintillating materials, and the reduced light yield for nuclear recoils with respect to electronic recoils is parameterized via the so-called {\it quenching factor} (QF): the ratio of light yields between nuclear and electronic recoil depositions of the same energy~\cite{knoll-ch8}. 
\par
Knowledge of both sodium (QF$\mathrm{_{Na}}$) and iodine (QF$\mathrm{_{I}}$) quenching factors is essential to the physical interpretation of results obtained from NaI(Tl)-based \cevns\ and WIMP searches. To compare results between experiments using the same target, it is usually assumed the intrinsic character of these factors, implying they take same values from crystal-to-crystal. However, if it is the case that the values of the quenching factors are variable across different detectors,
a dedicated nuclear recoil calibration for each detector would be mandatory for any intended application. 
\par
For example, the nuclear recoil energy spectrum induced by \cevns\ can be used as a probe of physics beyond the standard model, as demonstrated by the COHERENT collaboration using CsI(Na) detectors~\cite{coherent-first-csi}. The precision to which the quenching factors are measured directly affects the sensitivity of these searches~\cite{collar-csi-quench}, motivating the need for an accurate low-energy measurement of the quenching factors for any interesting target, such as NaI(Tl).
\par
In WIMP searches, much of the interest in the measurement of NaI(Tl) quenching factors stems from the long-standing DAMA collaboration's claim of dark matter detection with NaI(Tl) detectors, observing an annual modulation in the detection rate for more than two decades \cite{dama-libra2020}, which has not been confirmed by any other experimental search. However, comparisons with results from experiments using different target nuclei depend on the dark matter particle and dark halo models and the DAMA/LIBRA puzzle has not been settled, although many relevant dark matter scenarios have been ruled out. 
Only recently have other experiments using NaI(Tl) reached threshold and background conditions to test the DAMA/LIBRA result with high statistical significance: because these experiments use the same target as DAMA/LIBRA, they can be compared directly without depending on the potential differences in dark matter interaction models. The ANAIS-112 \cite{anais-first-results,anais-two-years,anais-three-years,anais-reanalysis} and COSINE-100 \cite{cosine100-nature,cosine100-first-mod} experiments have carried out model-independent annual modulation searches and model-dependent dark matter searches. ANAIS-112 has approached 3$\sigma$ sensitivity with the analysed exposure \cite{anais-ml,anais-reanalysis}, while the accumulated exposure guarantees a much higher sensitivity approaching 5$\sigma$ by 2025. COSINE-100 has set stringent bounds on different compatibility scenarios and has demonstrated the effect of assuming different quenching factors for COSINE-100 and DAMA/LIBRA detectors in the testing of DAMA's discovery claim~\cite{cosine-quench} in a model-dependent approach. Summarizing the present situation, the accurate determination of the scintillation quenching factors for the different detectors is the most relevant systematic affecting the testing of the DAMA/LIBRA claim. 
\par
An early determination of the sodium and iodine quenching factors was performed by the DAMA collaboration, and relied on measuring the response of a NaI(Tl) detector to neutrons emitted by \isotope{Cf}{252} and comparing the observed response to that from a Monte Carlo simulation, assuming the quenching factor is independent of the energy \cite{dama-quench-original}. With this approach, DAMA determined a sodium recoil quenching factor of QF$\mathrm{_{Na}}=0.30\pm0.01$ from 6.5 to 97 keV\textsubscript{nr} and an iodine recoil quenching factor of QF$\mathrm{_{I}}=0.09\pm0.01$ from 22 to 330 keV\textsubscript{nr}. 
\par
In contrast, most recent quenching factor measurements utilize a monoenergetic neutron beam scattering off the detector \cite{collar-quench,xu-quench,joo-quench}. Through the selection of beam energy and neutron scattering angle (by detecting the scattered neutron), different nuclear recoil energies can be studied and thus the energy-dependence of the quenching factor can be investigated. Over the past decade, experiments utilizing this approach have consistently measured values of QF$\mathrm{_{Na}}$ and QF$\mathrm{_{I}}$ smaller than those reported by DAMA. They also observe QF$\mathrm{_{Na}}$ decreasing with decreasing recoil energy. While these experiments consistently observe this decreasing trend, the resulting QF$\mathrm{_{Na}}$ values are in tension with each other for recoil energies below $\approx$20 keV\textsubscript{nr}, which is particularly relevant as it occurs in the energy range where the signal from WIMPs and \cevns~is expected. This discrepancy could be due to intrinsic differences between sodium iodide crystals in the production of light for the different particles or unaccounted for systematic measurement errors. Resolution of this discrepancy is thus essential to the physical interpretation of results from both WIMP searches and \cevns\ measurements, highlighting the necessity of a quenching factor evaluation across multiple detectors in a single experiment with a consistent approach, designed to reduce the systematic uncertainties.
\par
In this paper, we present measurements of the QF$\mathrm{_{Na}}$ and QF$\mathrm{_{I}}$ in multiple NaI(Tl) detectors performed in the same experimental setup. Particular emphasis was placed on identifying and removing potential sources of systematic error during the design of the measurement protocol. The specific features of the experiment carried out are:

\begin{enumerate}[(i)]
    \item A triggering scheme that does not rely on internal triggers from the NaI(Tl) light signal. As discussed by Collar in Refs.~\cite{collar-quench,collar-csi-quench}, trigger inefficiencies at low energies can result in artificially increased quenching factor values. We utilize a triggering scheme based only on triggers generated by the detection of the scattered neutron in an array of backing detectors and a selection procedure based on the neutron time-of-flight. The latter helps to reduce the contribution from multiple-scattered neutrons, which are slower. Further details on this approach can be found in Section~\ref{sec:setup}. 
    \item Small NaI(Tl) detectors to minimize the rate of multiple-scatter events that could pose a relevant background. As is detailed in Section \ref{sec:setup}, cylindrical crystals of two sizes having the same diameter and length (1.5 and 2.5~cm, respectively) have been used in the measurements. 
    \item Geant4 \cite{geant4} and MCNP-PoliMi \cite{mcnp-polimi} simulations of the setup have been developed to have good estimates of the neutron beam energy and nuclear recoil energy deposition distributions in the NaI(Tl) detector resulting from neutron scattering. This approach allows the use of simulated recoil spectral shapes to fit the experimental data, taking into account the precise geometrical disposition of the beam, NaI(Tl) crystal, backing detectors, as well as the uncertainties in all of the previous properties. The result of these simulations translates into non-Gaussian recoil energy distributions, and moreover, enables the estimate of the associated uncertainties in the final derived quenching factors. This helps to remove potential systematic biases observed in other experiments that have modeled the recoil spectrum as a Gaussian distribution and those related with finite-size effects.
    \item A relatively low energy neutron beam of about 1~MeV. The use of a low energy neutron beam decreases the uncertainty because low recoil energies correspond to relatively large scattering angles. As described by Xu {\it et al.}~\cite{xu-quench}, probing low recoil energies with higher energy incident neutrons requires measuring recoils at smaller scattering angles, which contribute more to the systematic uncertainty.
    \item A direct measurement of both the neutron beam's spatial profile and its full energy distribution. Both of these measurements were fed directly into the simulation, which increased its accuracy compared with simplified neutron beam models. 
    \item Different approaches for calibrating in electron equivalent energy the NaI(Tl) light signal have been explored. NaI(Tl) light yield is known to be non-proportional with energy. Nevertheless, proportionality is often assumed, and the peak resulting from neutron inelastic scattering on \isotope{I}{127} used as reference. In this work we introduced alternative calibration approaches, using an external gamma source of \isotope{Ba}{133}. 
    \item Testing multiple detectors in the same experimental setup. This allows us to test possible intrinsic differences between the five crystals measured, which had different properties, in terms of powder quality, as described in Section~\ref{sec:setup}. All crystals had similar thallium content and were grown using the same growth method at Alpha Spectra, Inc.~(AS), in Grand Junction, Colorado, US. To confirm that the tension between previously reported values of the sodium and iodine quenching factors is due to intrinsic differences between crystals, similar strategies should be considered for crystals grown by different techniques and having different thallium content, as proposed by Bharadwaj {\it et al.}~\cite{Bharadwaj:2023aoz}.
\end{enumerate}

The article is structured as follows. Section~\ref{sec:setupgen} describes the experimental setup, including the beam and detectors configuration, data acquisition system, and measurement protocols . Section~\ref{sec:simulations} describes the simulations of the full experimental setup, which are an essential input for the scintillation quenching factors estimates. In Section~\ref{sec:analisis}, the data analysis is described for the beam energy determination, electron equivalent energy calibration, selection of events compatible with neutrons scattering both in the NaI(Tl) and backing detectors, and fitting procedure followed for the quenching factor determination. This data analysis strongly relies on the work of D.~Cintas~\cite{cintas-thesis}, S.~Hedges~\cite{hedges-thesis}, and W.~Thompson~\cite{thompson-thesis}.
\par Finally, results for the QF$\mathrm{_{Na}}$(E) and QF$\mathrm{_{I}}$ are discussed in Section~\ref{sec:results}, while conclusions are drawn in Section~\ref{sec:conclusions}.

\section{Experimental Setup}
\label{sec:setupgen}

\subsection{Overview}
\begin{figure}[htbp]
\centering 
\includegraphics[width=.48\textwidth]{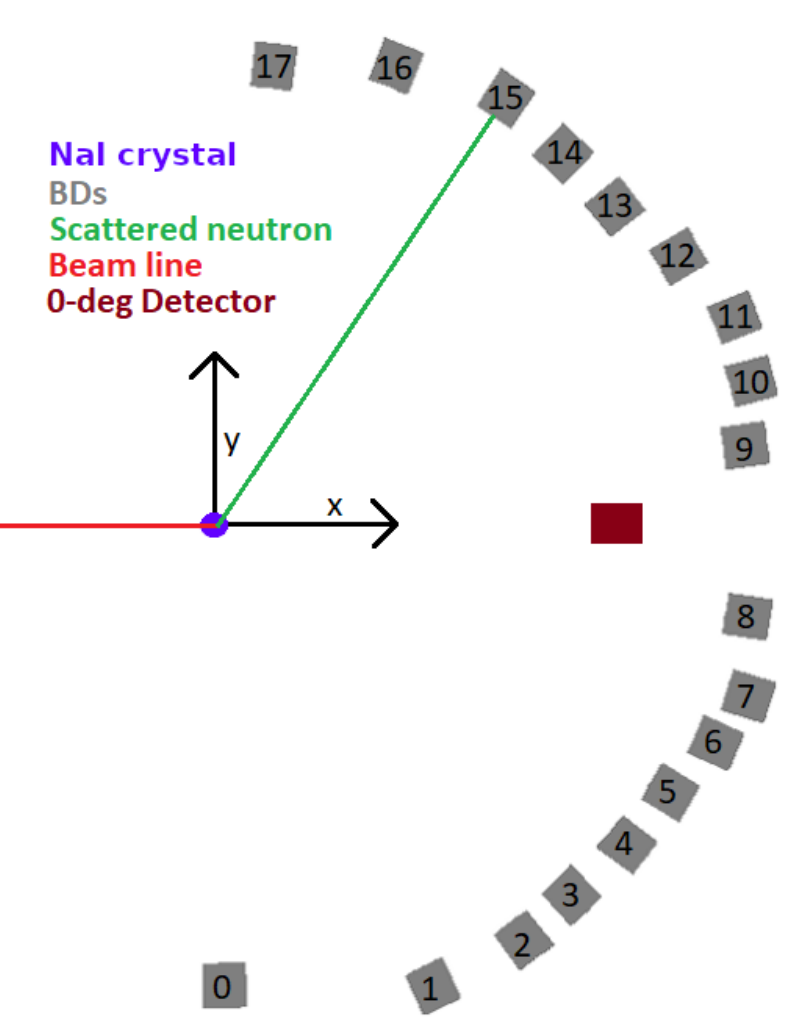}
\vspace{0.6cm}
\includegraphics[width=.48\textwidth]{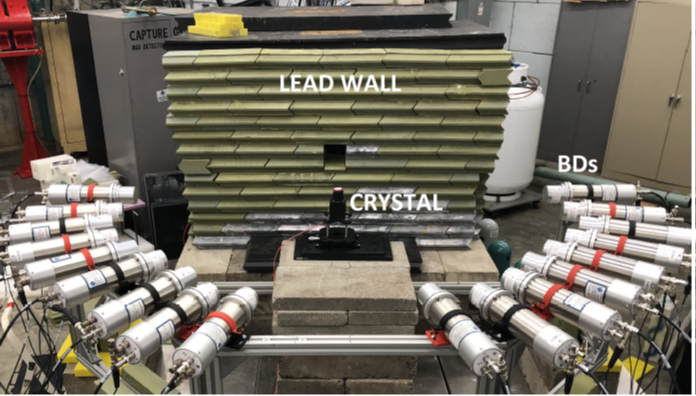}
\caption{Upper panel: Backing detector positions around the NaI(Tl) crystal corresponding to the October run. The on-axis backing detector (labelled as 0-deg detector) is also shown in one of the positions used for the beam energy measurement. Lower panel: picture of the experimental setup.  
} 
\label{fig:BDarc} 
\end{figure}

The measurements reported in this paper took place at the Advanced Neutron Calibration Facility at the Triangle Universities Nuclear Laboratory (TUNL), North Carolina (US), in 2018. The NaI(Tl) crystals to be tested were coupled to the same photomultiplier tube (PMT) in a very similar configuration, and placed in the beamline of a quasi-monoenergetic neutron beam. 
The neutron scattering angle relates directly to the energy deposited in the NaI(Tl) via kinematics and the known monochromatic beam energy. This angle is measured with 18 liquid scintillator-based backing detectors (BDs) placed along a semi-circle. 
A schematic view of this setup is shown in Figure~\ref{fig:BDarc}. Five different NaI(Tl) crystals were tested throughout the course of the experiment. Due to beamtime constraints at TUNL, the measurements took place over two separate runs, referred to as the August and October runs.
\par

\subsection{Neutron Beam Generation}
The quasi-monoenergetic neutron beam used in the measurements is produced via the \isotope{Li}{7}(p,n)\isotope{Be}{7} reaction. A pulsed $\approx$2.7~MeV proton beam generated using the 10~MV tandem van de Graff accelerator at TUNL is driven towards a lithium fluoride target (LiF), resulting in a pulsed neutron beam with energies of the order of 1~MeV. In the August run, the time between pulses was 800~ns, while in the October run was 400~ns, allowing to increase the number of nuclear scatters observed in the NaI(Tl). In both cases, the typical pulse width was about 2~ns. The pulsed nature of the beam aided in the removal of background events through the application of time-of-flight (TOF) techniques, discussed in Section~\ref{sec:analisisTOF}. 
\par
Before reaching the LiF target, the proton beam passed through an induction coil, known as the beam pickoff monitor (BPM). The corresponding BPM signal provided timing information for use in the neutron beam energy measurement (Section~\ref{sec:analisisEn}) and background rejection via TOF cuts (Section~\ref{sec:analisisTOF}).
\par
The LiF target consisted of a thin 500~nm (August) or 750~nm (October) thick layer deposited on a 0.1~mm-thick tantalum backing. Tantalum was chosen to effectively stop protons that have passed through the lithium fluoride layer, as it does not produce a large gamma background. 
\par
Off-axis neutrons were removed by a neutron collimator made of layers of borated polyethylene and high density polyethylene. The side of the collimator facing the NaI(Tl) detector was shielded by at least 4 inches of lead to absorb the gammas produced from neutron capture in the collimator. A dedicated measurement of the neutron beam profile was carried out at two different distances from the LiF target, 47.4 and 98.6~cm. Both measurements provided compatible results for the neutron beam divergence, with the far distance providing the most accurate measurement, corresponding to a half-angle of 2.0\textdegree$\pm$0.3\textdegree, modeling the source as a point-like and located at the mean LiF target position.
\par
The energy of the neutron beam was directly measured by an on-axis backing detector using the neutron TOF in a dedicated run, as explained in Section~\ref{sec:analisisEn}. In addition, TOF information from the on-axis detector was recorded throughout the full data run, with the on-axis backing detector positioned downstream of the NaI(Tl) detector. This permitted us to measure and correct for possible instabilities of the neutron beam over the course of the full run. Measurements of the neutron beam profile and energy were fed directly into the simulations, as detailed in Section~\ref{sec:simulations}.

\subsection{Detector Configuration and Data Acquisition}
\label{sec:setup}
Over the course of the experiment, five NaI(Tl) crystals were tested that varied in characteristics such as size and type of NaI powder used in the growth procedure. These detectors were produced by Alpha Spectra Inc.~(AS), all sharing the thallium content and the growth mechanism, but produced in different ingots. Crystals~No.~1, 2, and 3 will be referred to in the following as COSINE's crystals, whereas crystals No.~4 and No.~5, as ANAIS's crystals. Four of these NaI(Tl) crystals were grown using the AS-WIMPScint class of NaI powder. WIMPScint-III is the same powder quality used to grow most of the crystals used in the COSINE-100~\cite{cosine100-detector} and ANAIS-112~\cite{anais-detector} experiments. In particular, crystal~No.~5, was cut from the same ingot as several of the ANAIS-112 crystals. Crystal~No.~4, on the other hand, was grown with standard Alpha Spectra powder. In three of the five detectors, the crystallographic orientation of the NaI was known, allowing the search for effects of channeling, which are not presented in this article. Details of the five crystals measured can be found in Table~\ref{tab:1}.
\par
\begin{table*}[htbp]
\centering 
\begin{tabular}{|c|c|c|c|c|}
\hline
  Crystal  & Measurement  & Proprietary & Powder  & Dimensions \\
  number & period & & quality & (mm) \\
  \hline
1 & August & COSINE & AS-WIMPScint-I & 25 \\
2 & August & COSINE & AS-WIMPScint-II & 25 \\
3 & October & COSINE & AS-WIMPScint-III & 25 \\
4 & October & ANAIS & AS-Standard & 15 \\
5 & October & ANAIS & AS-WIMPScint-III & 15 \\
\hline
\end{tabular}
\caption{\label{tab:1} Characteristics of the five NaI(Tl) crystals measured. All of them were grown by Alpha Spectra Inc.~with different quality starting powder but a similar thallium content and growth mechanism. The last column provides the length and diameter of the crystals, which were cylindrical in shape with length equal to the diameter in all the cases. 
} 
\end{table*}

In all cases, the particular sodium iodide detector under investigation was optically coupled to a square 1$\times$1~inch Hamamatsu ultra-bialkali H11934-200-10 PMT using EJ-550 optical grade silicone grease from Eljen Technology. This particular PMT was chosen for its high peak quantum efficiency of 43\%, well suited for detecting the emission peak of NaI(Tl). The linearity of the response of the specific PMT used has been tested in detail in Ref.~\cite{pmt-linearity-tests}. The main difference between COSINE's and ANAIS's crystals was that the latter were designed with two optical windows to allow the coupling of two PMTs. As in these measurements only one PMT was used, one of the windows was covered with Teflon and copper, which resulted in a poorer light collection than that of COSINE's crystals. 
\par
The NaI(Tl) crystal was placed on the beamline with its center 112~cm downstream of the LiF target in the August run and 66~cm downstream in the October run. The BDs surrounding the NaI(Tl) in order to tag the neutron scattering angle were Eljen Technology model 510 and featured 2$\times$2\diameter~inch EJ-309 liquid scintillator cells. EJ-309 features remarkable pulse shape discrimination (PSD) capabilities between gamma and neutron interactions, which allowed for the removal of gamma backgrounds. Details on the precise positions of these backing detectors can be observed in Figure~\ref{fig:BDarc} for the October run. Backing detector positions were chosen in order to probe sodium nuclear recoil energies between 10 and 80~keV\textsubscript{nr}, approximately. 
\par
\begin{figure}[htbp]
\centering 
\includegraphics[width=.45\textwidth]{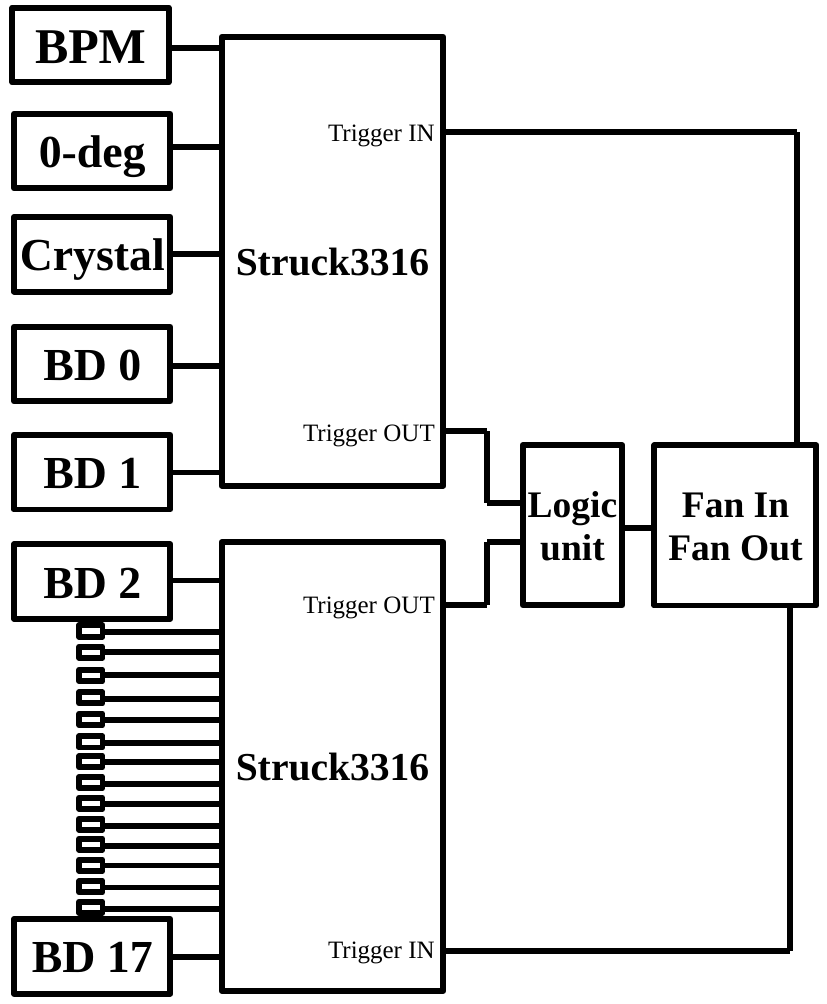}
\caption{\label{fig:DAQscheme} Scheme of the DAQ system for the measurements. The Struck 3316 digitizers sample the signals of all the detectors and the BPM and act also as discriminator, with a trigger configuration that depends on the measurement (see text for more detail).  
} 
\end{figure}

The data acquisition system (DAQ) is shown in Figure~\ref{fig:DAQscheme}. Full signal waveforms from the NaI(Tl) crystal, 18 backing detectors, on-axis detector (0-deg) and BPM were recorded by two Struck 3316 digitizers with 14 bits resolution operating at 250 MHz. The NaI(Tl) and backing detector signals were fed directly into the digitizer; to counter signal attenuation of the BPM over the relatively longer cable length the signal was first fed into a LeCroy 133B dual linear amplifier. 
Each digitizer acted as a discriminator with a trigger configuration that depended on the measurement, as described below. 
\par
The DAQ was designed to avoid trigger bias in the selection of nuclear recoils in the NaI(Tl) detectors. In the beam-on measurements, the DAQ trigger was generated by the backing detectors, which utilized the internal finite-impulse response (FIR) trigger of the Struck digitizer. When this internal trigger condition was met, waveforms in the backing detectors, NaI(Tl) detector, and beam-pulse monitor were recorded. In the August run, only the backing detector issuing the trigger was recorded, along with the BPM and NaI(Tl) detector, while for the October run, all backing detectors were recorded whenever a single backing detector triggered. The output trigger from the digitizers was sent to a logical unit in OR mode which generated the global trigger for the DAQ, used as external trigger for both digitizers. 
\par
The trigger strategy was different in calibration, background runs, and beam-energy measurements. In calibration runs, the trigger was generated and used only by the detector being calibrated, either NaI(Tl) or BD. In background runs, any of the detectors except the BPM could generate triggers. In beam energy measurements, only the on-axis detector triggered and was recorded. 
\par
\begin{figure*}[htbp]
\centering 
\includegraphics[width=.85\textwidth]{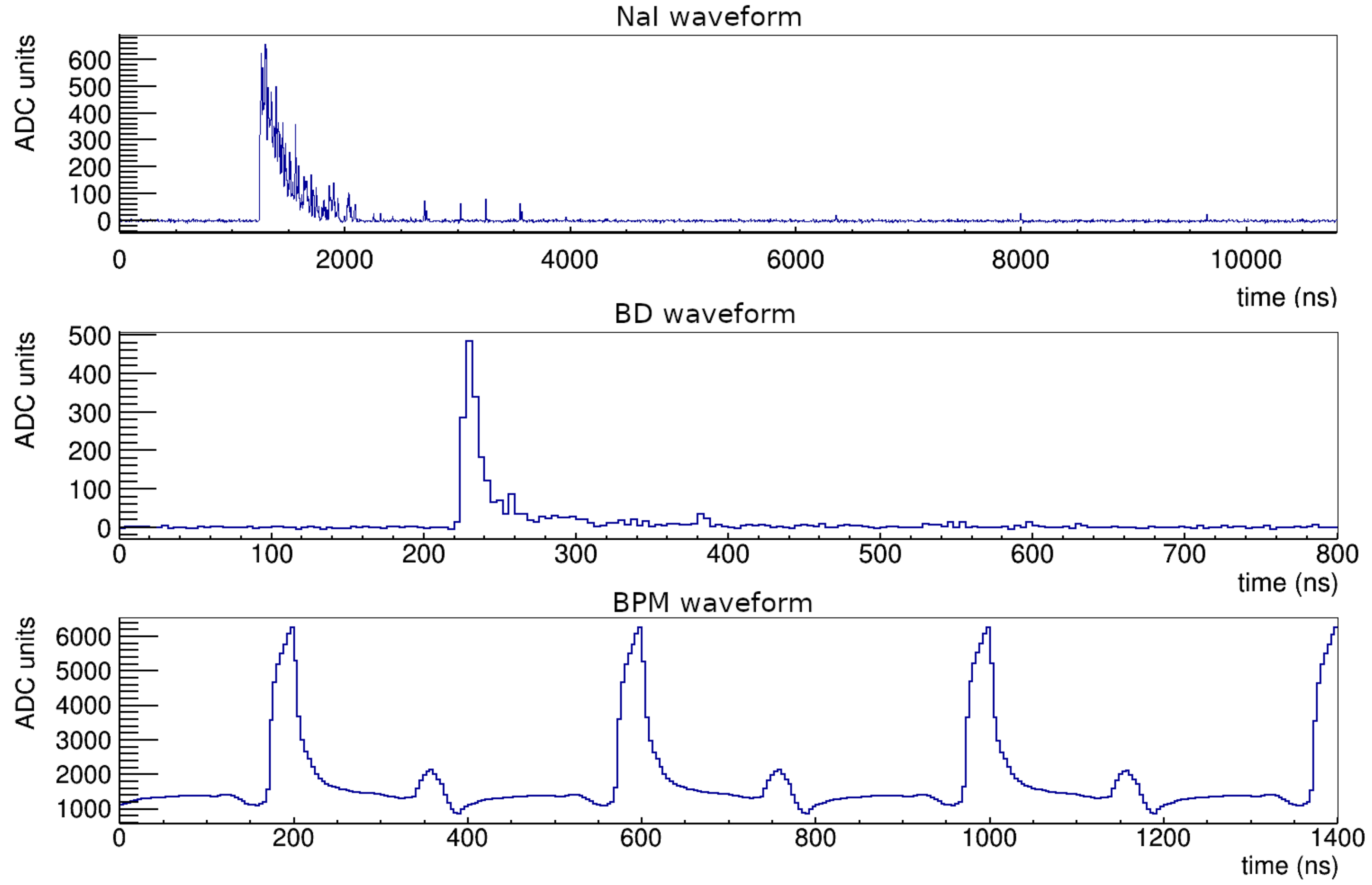}
\caption{\label{fig:event} Example waveforms from the NaI detector (NaI), backing detector (BD), and beam pickoff monitor (BPM) acquired by the DAQ system for an event triggered by a scattered neutron in the backing detectors (BD). The BD and NaI waveforms have been background-subtracted.
} 
\end{figure*}

Different digitization windows were used for each signal: 1,400~ns for the BPM, 10,800~ns for the NaI(Tl) crystal and 800~ns for the backing detectors. One example of waveform recorded for a beam-on event, triggered by one of the backing detectors, is shown in Figure~\ref{fig:event}.
\par

\subsection{Run Summary}
Neutron scattering data was collected with each NaI(Tl) detector for between 18 and 45~hours. The exposure time for each individual detector can be found in Table~\ref{tab:2}. Approximately every eight hours of beam-on measurement, NaI(Tl) crystals were rotated 30$^\circ$, trying to reduce the possible effect of channeling in the QF results. 
\par
A set of calibration measurements was made at the beginning of each detectors' data run and, subsequently, every eight hours. During these calibrations, the NaI(Tl) detector was exposed to a \isotope{Ba}{133} source, and the backing detectors, including the on-axis detector, were exposed to a \isotope{Cs}{137} source. Additionally, the beam-off background, both in NaI(Tl) and backing detectors, was measured. Finally, a dedicated beam energy measurement was conducted once during both the August and October runs. 

\vspace{0.7cm}

\begin{table}[htbp]
\centering 
\begin{tabular}{|c|c|c|c|c|c|c|}
\hline
  Crystal  & \multicolumn{6}{c|}{Beam-on measurement time (hours)}\\
  \cline{2-7}
  number & 0$^\circ$ & 30$^\circ$ & 60$^\circ$ & 90$^\circ$ & 120$^\circ$ & total \\
  \hline
1 & 10.02 & 7.33 & 7.58 & 9.32 & 0 & 34.25 \\
2 & 9.08 & 7.72 & 8.62 & 4.75 & 0 & 30.17 \\
3 & 8.17 & 7.90 & 7.53 & 8.30 & 0 & 31.90\\
4 & 8.35 & 8.18 & 8.33 & 9.62 & 8.70 & 43.18 \\
5 & 8.58 & 9.02 & 0 & 0 & 0 & 17.60 \\
\hline
\end{tabular}
\caption{\label{tab:2} Beam-on measurement time in hours for each orientation and crystal measured. Total measurement times for each crystal are shown in the last column. } 
\end{table}

\section{Simulations}
\label{sec:simulations}

\subsection{Overview}
To account for the complexities in the experimental setup geometry, full Monte Carlo (MC) simulations were developed. These simulations allowed us to take into account the uncertainties in the positions of the detectors, the angular size of the neutron beam, and the detectors' size, and at the same time evaluate the role of the possible backgrounds originating from multiple scattering of the neutrons in the different components of the setup. This simulation has been used to obtain the distribution of the recoil energies deposited in the NaI(Tl) crystal for scattered neutrons reaching each of the backing detectors, as described in section~\ref{sec:simb}. This information will be used in the calculation of the QF, as detailed in section~\ref{sec:fit}. 
\par
We also used the simulation to reproduce the energy spectrum corresponding to the interactions of the gamma and x-rays emissions from the external \isotope{Ba}{133} source used for the energy calibration of the NaI(Tl) crystal (see Section~\ref{sec:Ecalibration}), as described in section~\ref{sec:simc}. 
\par
Finally, simulations of the neutron generation in the LiF target were carried out in order to understand the energy profile of the neutron beam and the TOF measurements with the on-axis backing detector, as detailed in Section~\ref{sec:analisisEn}. 
\par

\subsection{Simulation of the nuclear recoil distributions in the NaI(Tl) for each BD channel}
\label{sec:simb}
The simulations of the experiment geometry for the presented analysis were performed within the GEANT4 simulation framework~\cite{geant4}. The simulation takes into account naturally the geometrical configuration of the BD array, angular acceptances of each channel and detector sizes, while the dispersion of the neutron beam (both in energy and in direction) can be easily introduced. This allows both a thorough understanding of the experimental measurements and the estimate of sources of systematic effects. 
\par
In these simulations, neutrons were generated at the LiF target location using the experimentally-measured beam energy and collimator divergence. Each run and crystal underwent dedicated simulations. Events generating depositions in both the NaI(Tl) and one of the liquid-scintillator backing detectors (referred as channel) were the main simulation output. The analysis of the simulation results was done in similar way as for the experimental data. 
\par
\begin{figure}[htbp]
\centering 
\includegraphics[width=.5\textwidth]{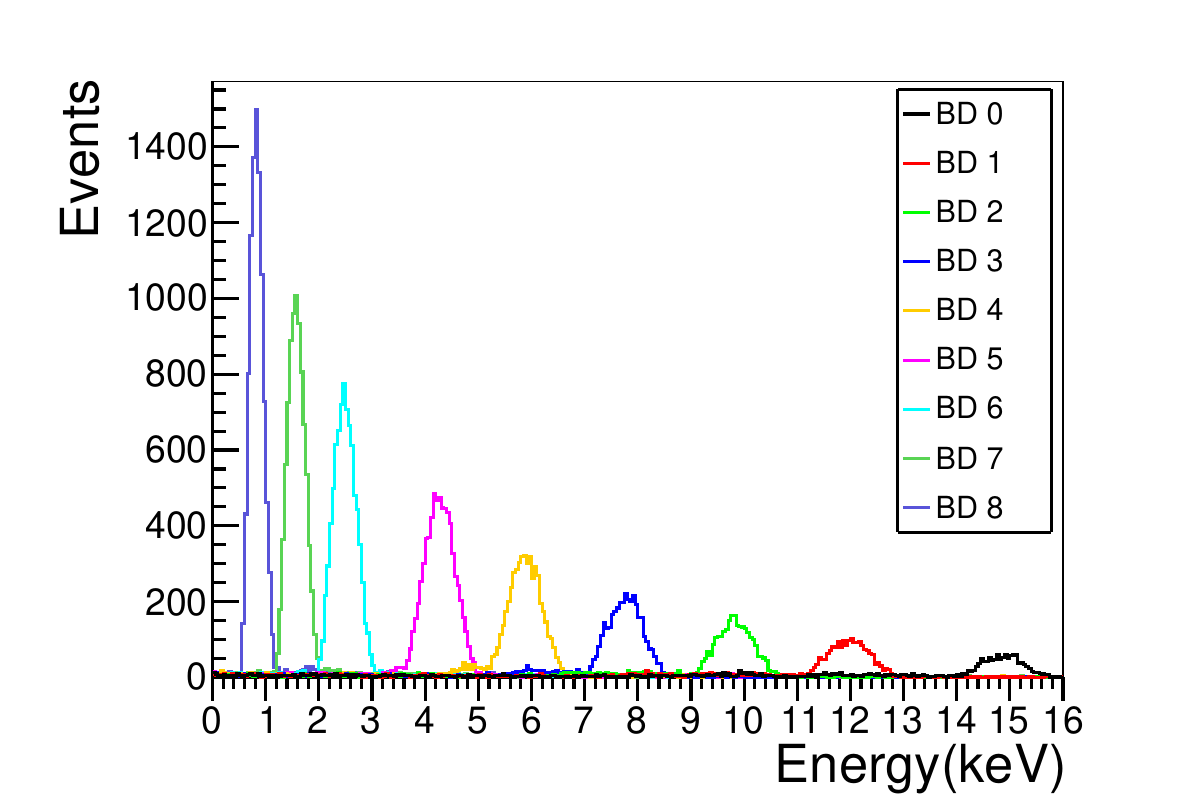}
\includegraphics[width=.5\textwidth]{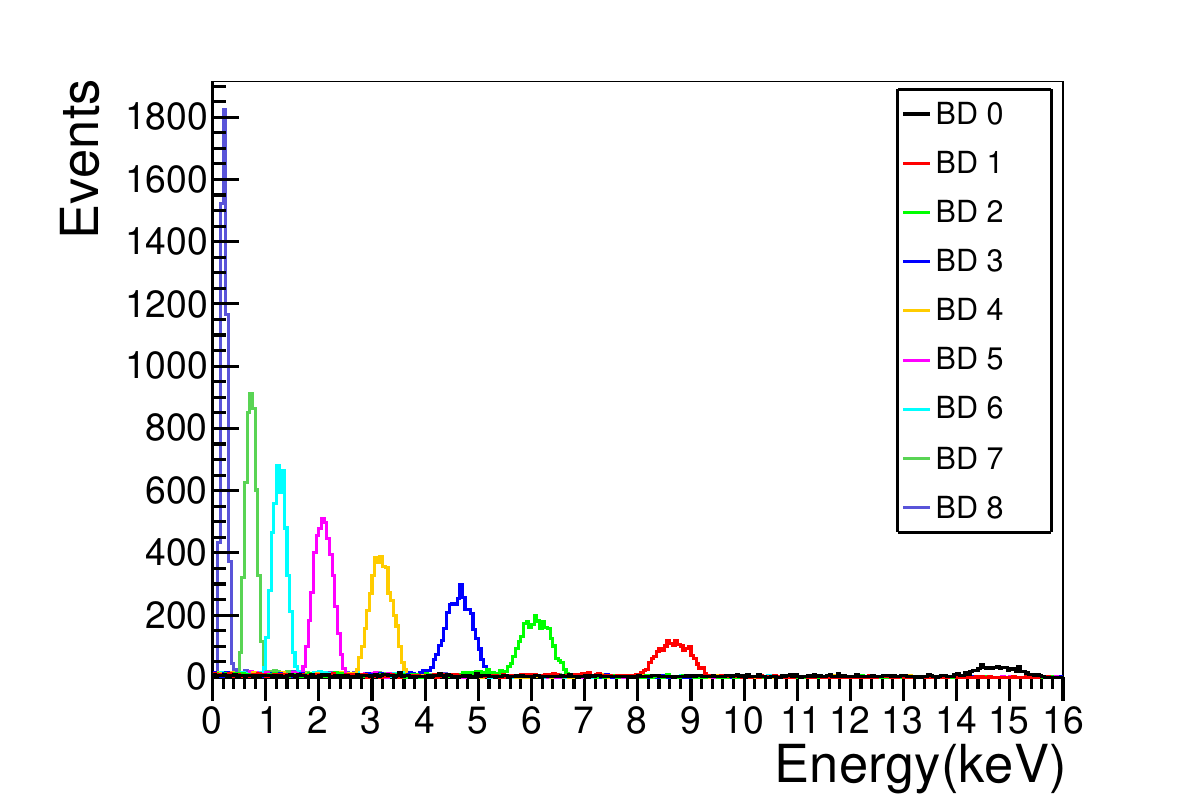}
\caption{\label{fig:NR_I} Simulated nuclear recoil energy distributions for the iodine nuclei in the NaI(Tl) crystal from 10$^9$~simulated neutrons. They are shown for each channel. BD positions correspond to the August (upper panel) and October (lower panel) runs. 
} 
\end{figure}
\begin{figure}[htbp]
\centering 
\includegraphics[width=.5\textwidth]{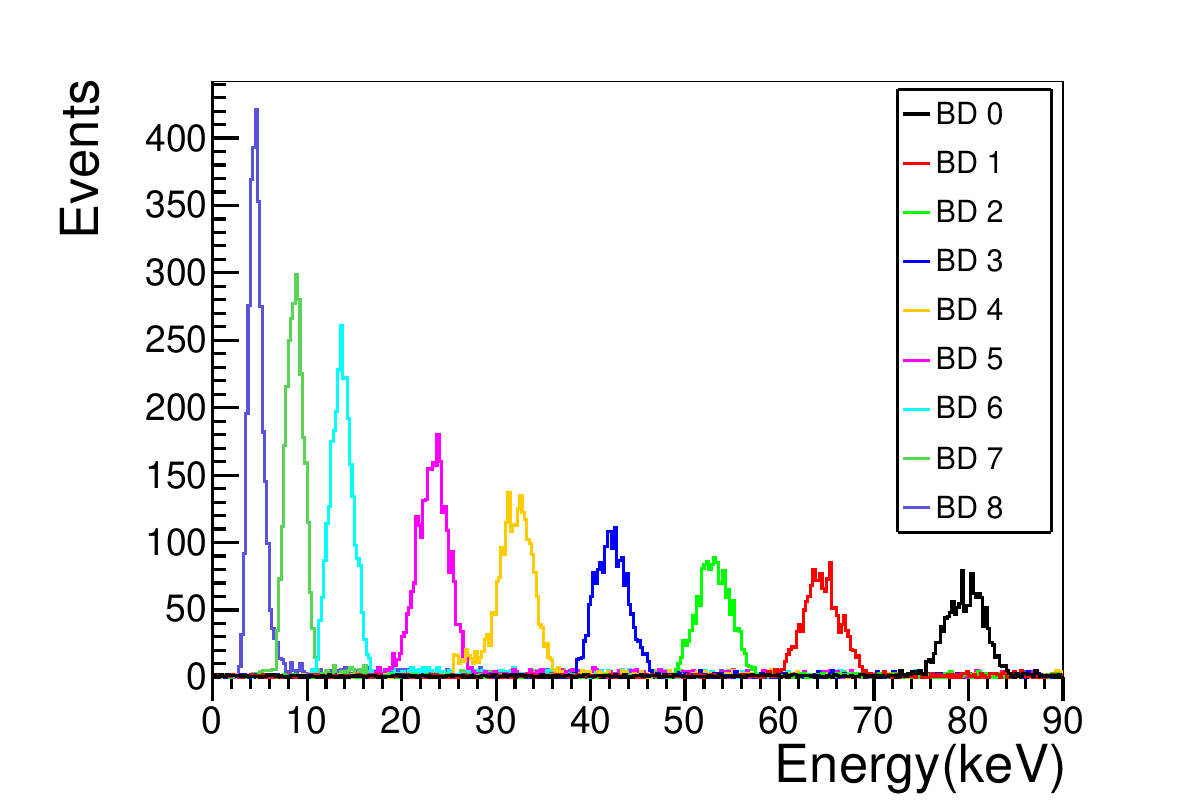}
\includegraphics[width=.5\textwidth]{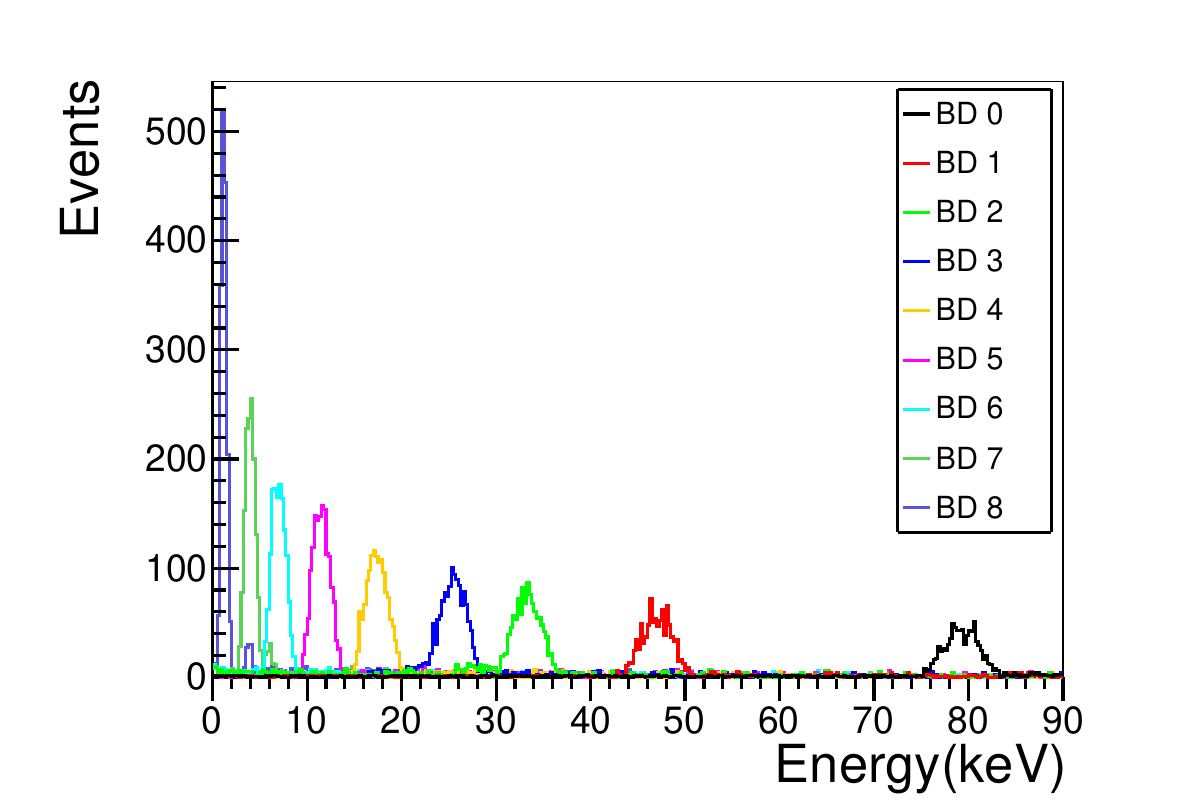}
\caption{\label{fig:NR_Na} Simulated nuclear recoil energy distributions for the sodium nuclei in the NaI(Tl) from 10$^9$~simulated neutrons. They are shown for each channel. BD positions correspond to the August (upper panel) and October (lower panel) runs. 
} 
\end{figure}

The simulated set-up consisted of the NaI crystal and housing (aluminium, with the inside covered by Teflon diffusor and the outside by insulating tape), the lead collimator, and the backing detectors, which were placed at different angles with respect to the beam, covering 180$^{\circ}$. 
The energy deposited in each experimental volume and the corresponding time is recorded for each simulated event, keeping track of electron and nuclear recoils' energy depositions separately. This allows us to build the TOF distributions for each BD and relate them to neutron energy distributions, distinguishing contributions from single and multiple scattering. The quenching factors of recoiling sodium and iodine nuclei are introduced in a subsequent step, to produce the electron equivalent energy spectra, which can be compared with the experimental data after being convolved with an energy resolution function. 
The nuclear recoil energy distributions in the crystal for each triggered BD derived
from the simulation are presented in Figures~\ref{fig:NR_I} and \ref{fig:NR_Na} for iodine and sodium nuclei recoils, respectively, for the BD positions corresponding to both August and October measurements. Energy resolution has still not been included in these distributions. 
\par
This approach allows us to consider quenching factors varying with the energy in the region of interest, as most of the recent estimates hint at. Additionally, the measured nuclear recoil distributions are skewed, and the Gaussian approximation does not provide good fitting in general, which is remedied by using the simulated recoil distribution. The simulation also allows us to properly convert recoil energy distributions into visible energy ones on an event by event basis. 
\par

Three different geometrical configurations were simulated: (i) the August configuration, (ii) the October configuration with the COSINE crystal (No.~3) and (iii) the October
configuration with the ANAIS crystals. To evaluate the systematic uncertainties in the nuclear recoil energies (and therefore, in the QF) due to the BD position uncertainties, two more simulations were run for each configuration, placing the BDs in the corresponding maximum and minimum scattering angles compatible with their position uncertainties.

\subsection{Simulations of the NaI(Tl) crystal calibration with \isotope{Ba}{133}}
\label{sec:simc}
One of the objectives of the simulation was to obtain the distribution of the energy depositions resulting from the NaI(Tl) crystal irradiation with the \isotope{Ba}{133} source, to better understand the experimental measurements and to improve the calibration in electron equivalent energy. 
\par
\begin{figure}[htbp]
\centering 
\includegraphics[width=.5\textwidth]{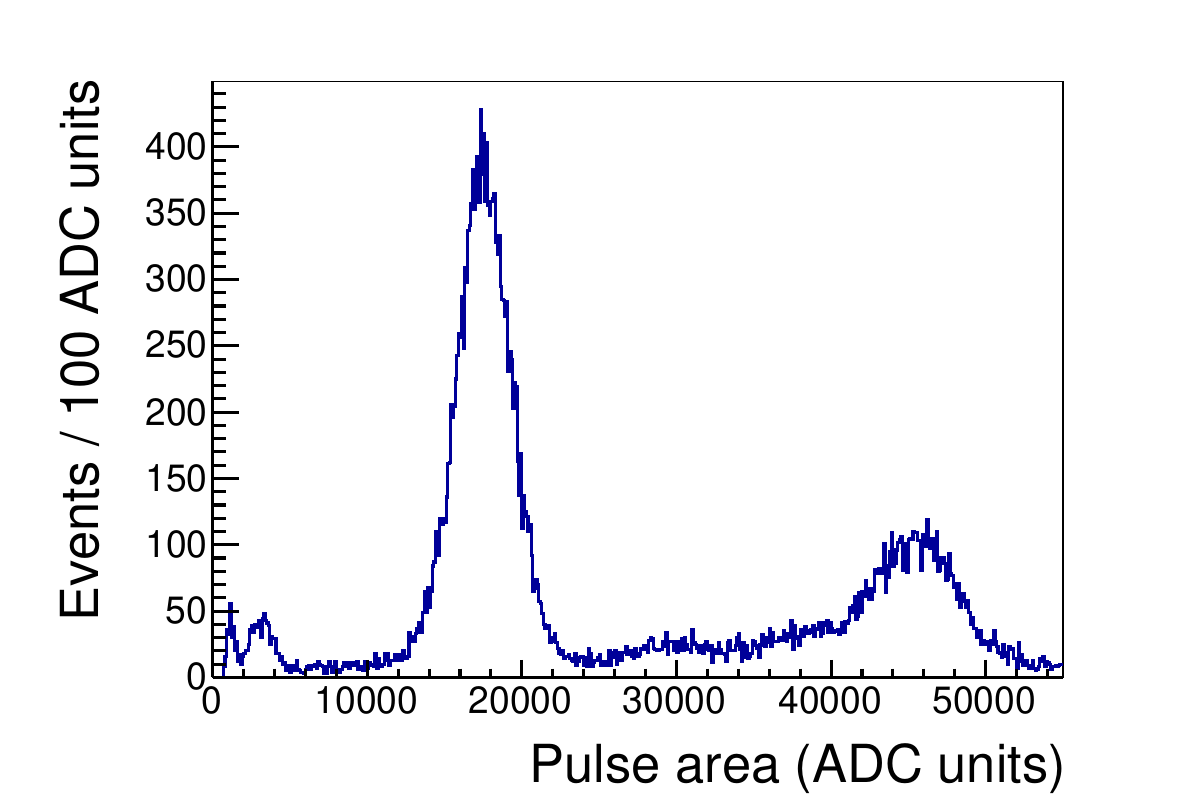}
\caption{\label{fig:Ba_esp_exp} Measured deposited energy distribution in the NaI(Tl) crystal No.~5 irradiated with an external \isotope{Ba}{133} source. 
} 
\end{figure}
\begin{figure}[htbp]
\centering 
\includegraphics[width=.5\textwidth]{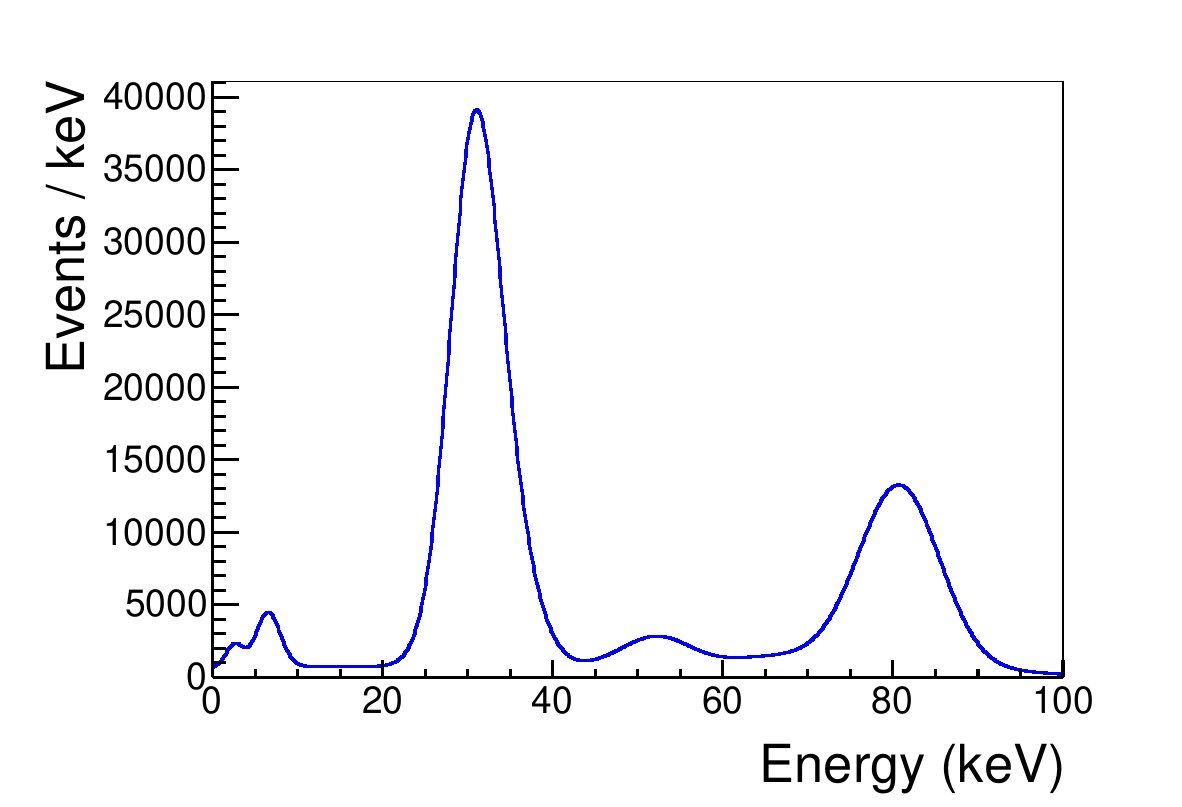}
\caption{\label{fig:Ba_esp_sim} Simulated deposited energy distribution in the NaI(Tl) irradiated with an external \isotope{Ba}{133} source. The energy resolution measured for crystal No.~5 has been used.
} 
\end{figure}

Due to the lack of a precise description of the encapsulation for the \isotope{Ba}{133} source, it was simulated as point-like and placed at the same position for all the experimental configurations. Consequently, it was not possible to reproduce precisely the measured calibration spectra, although simulation and measurements share key features. As an illustration, Figure~\ref{fig:Ba_esp_exp} shows the measured spectrum for crystal~No.~5, while in Figure~\ref{fig:Ba_esp_sim}, we present the result of the corresponding simulation. In the latter, the energy resolution was taken into account using that experimentally determined for crystal~No.~5.
\par
The simulation allowed us to determine the average energies of the peaks that will be used in the calibration process outlined in Section~\ref{sec:Ecalibration}: 6.6~keV, 30.9~keV, and 35.1~keV. The former is the result of the 35~keV x-rays escape, while the last two peaks are not being resolved in the measurement.

\section{Data Analysis}
\label{sec:analisis}

\subsection{Event analysis chain}
\label{sec:analysischain}
A similar analysis chain was applied for all collected data to identify pulse properties, such as pulse onset and pulse area, and to guarantee the quality of the data. 
\par
In the case of the NaI(Tl) waveforms, first the baseline level of each recorded waveform and the corresponding root mean square (RMS) were calculated. A quality cut was applied to select only those waveforms which were not affected either by baseline drift or by the presence of dark photoelectrons in the pretrigger region. This was done by comparing the baseline values obtained for the first and last 200~ns of the waveform and selecting only those pulses having a difference lower than three RMS. The baseline was then calculated by averaging the two values (from the beginning and the end of the waveform) to be used to derive the pulse area. Next, a pulse-finding algorithm was applied with a threshold at 5~RMS from the baseline value, after checking the stability of the baseline along the data taking. The waveform position where the pulse is above the threshold was stored as $t_{0,\mbox{NaI}}$, the pulse onset. However, to avoid threshold effects, fixed integration windows (2~$\mu$s width), independent from the $t_{0,\mbox{NaI}}$ value, were used for obtaining the area of the pulses that was used as energy estimator throughout this work.
For NaI(Tl) calibrations, the pulse area was calculated by integrating the pulse waveform from 1.5 to 3.5~$\mu$s, while in the beam-on measurements, because the trigger was done by the BDs, the TOF had to be taken into account. The latter, implied that the pulse area calculation was dependent on the type of interacting particle: pulses in NaI(Tl) correlated with gammas triggering the BDs should appear later in the waveform trace than those correlated with neutrons. For neutrons triggering the BDs, the integration of the NaI(Tl) pulse was done in a fixed window from 1.2 to 3.2~$\mu$s in the NaI waveform, as explained in Section~\ref{sec:correlationBDNaI}. This allowed us to include in the recoil spectra for each channel any energy deposition in a time window compatible with the neutron TOF. 
\par
\begin{figure}[htbp]
\centering 
\includegraphics[width=.5\textwidth]{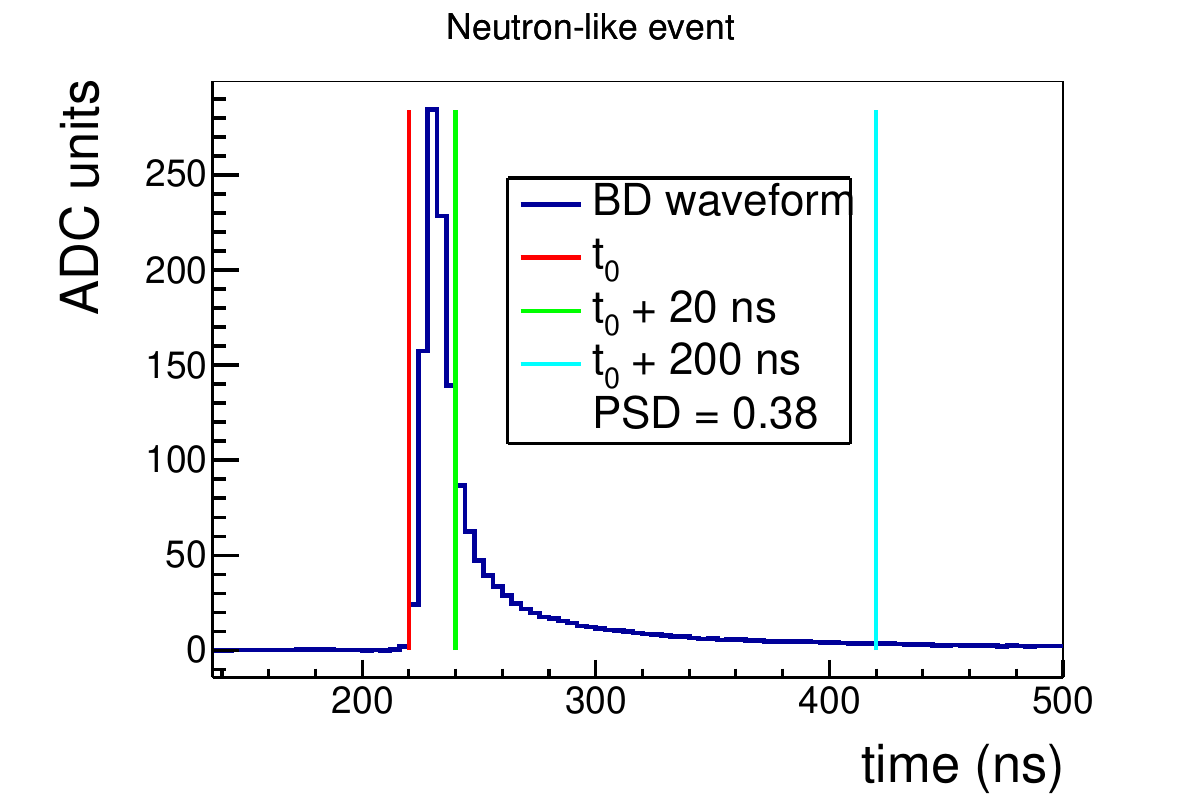}
\includegraphics[width=.5\textwidth]{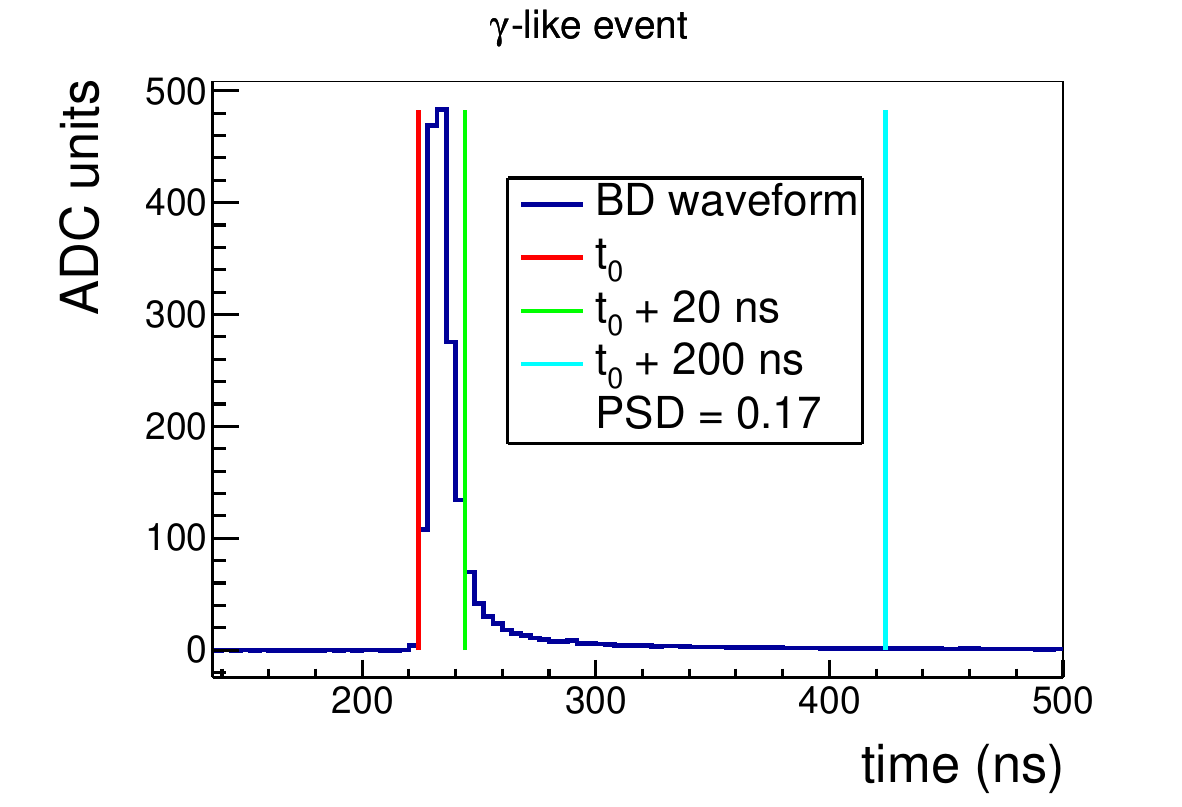}
\caption{\label{fig:PSD} Pulses in the BD corresponding to neutron (upper panel) and gamma (lower panel) events. PSD values are shown, as well as the integration ranges used for the tail and total pulse areas. 
} 
\end{figure}

For the backing detectors waveforms, the baseline, and corresponding RMS were also calculated in the first 160~ns of the pulse trace. The pulse-finding algorithm applied allowed to identify pulses in the waveforms as deviations above 5~RMS of the baseline in each of the BDs and to determine the corresponding pulse onset, t0. A variable called multiplicity was defined for each event as the number of BDs having a signal above that threshold. Events with multiplicity larger than 1, which accounted for around 1\% of the total number of events, were removed from the QF analysis dataset. In addition, saturated events in the BDs were also removed (about 0.005\% of the events). Next, a pulse shape discrimination parameter, PSD, was built for each BD waveform in order to profit from the ability of liquid scintillators for neutron-gamma discrimination, defined as the ratio between the area corresponding to the tail of the pulse (integral from $t0+20$~ns to $t0+200$~ns) and the total pulse area (integral from $t0$ to $t0+200$~ns). Examples of neutron and gamma events in the BDs are shown in Figure~\ref{fig:PSD} with the corresponding PSD values. 
\par

\begin{figure*}[htbp]
\centering 
\includegraphics[width=.65\textwidth]{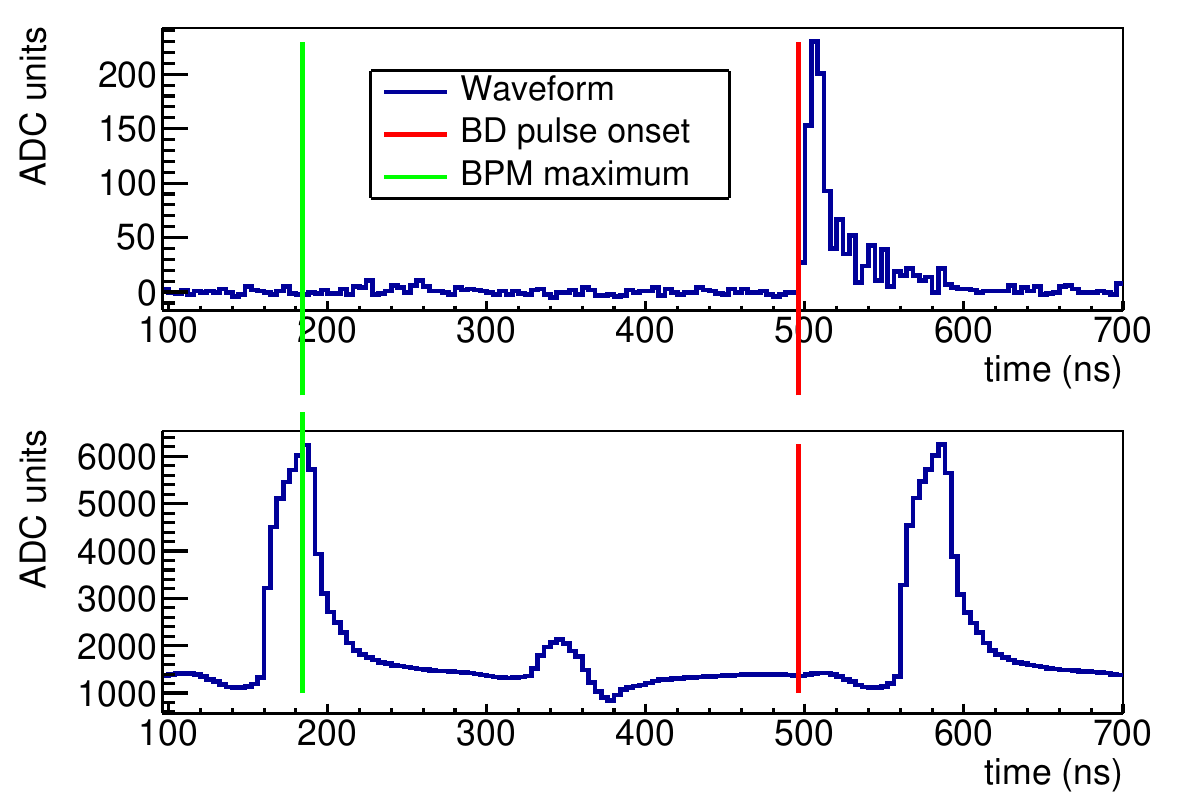}
\caption{\label{fig:timeSincePrevBPM} Pulses in the BD (top) and BPM (bottom) showing the calculation of the time after the last neutron beam pulse (timeSincePrevBPM) as the time difference between the BD pulse onset, t0 (red line) and the previous BPM waveform maximum (green line). 
} 
\end{figure*}

An important variable related with the TOF of the particle triggering the BDs is the time after the last neutron beam pulse (timeSincePrevBPM). It is calculated from the waveforms of the BDs and the BPM as the difference between t0 and the previous maximum of the BPM signal, as Figure~\ref{fig:timeSincePrevBPM} shows. It is not directly the TOF because this difference includes an offset related with the signal processing and the TOF of protons between the BPM and the LiF target. 
\par

\subsection{Beam Energy Measurement}
\label{sec:analisisEn}
The neutron beam energy distribution for both runs is one of the most relevant inputs in the simulations developed to obtain the expected nuclear recoil energy distributions and derive the QF estimates. 
These distributions can be obtained from the information on the TOF of the neutrons between the LiF target and the 0-deg detector in dedicated TOF measurements, done both in August and October runs. The 0-deg detector was placed at three different distances from the LiF target. These distances were 296.3~cm, 343.6~cm and 394.3~cm in the August run and 74.5~cm, 133.2~cm and 210.8~cm in the October run. 
\par
\begin{figure}[htbp]
\centering 
\includegraphics[width=.5\textwidth]{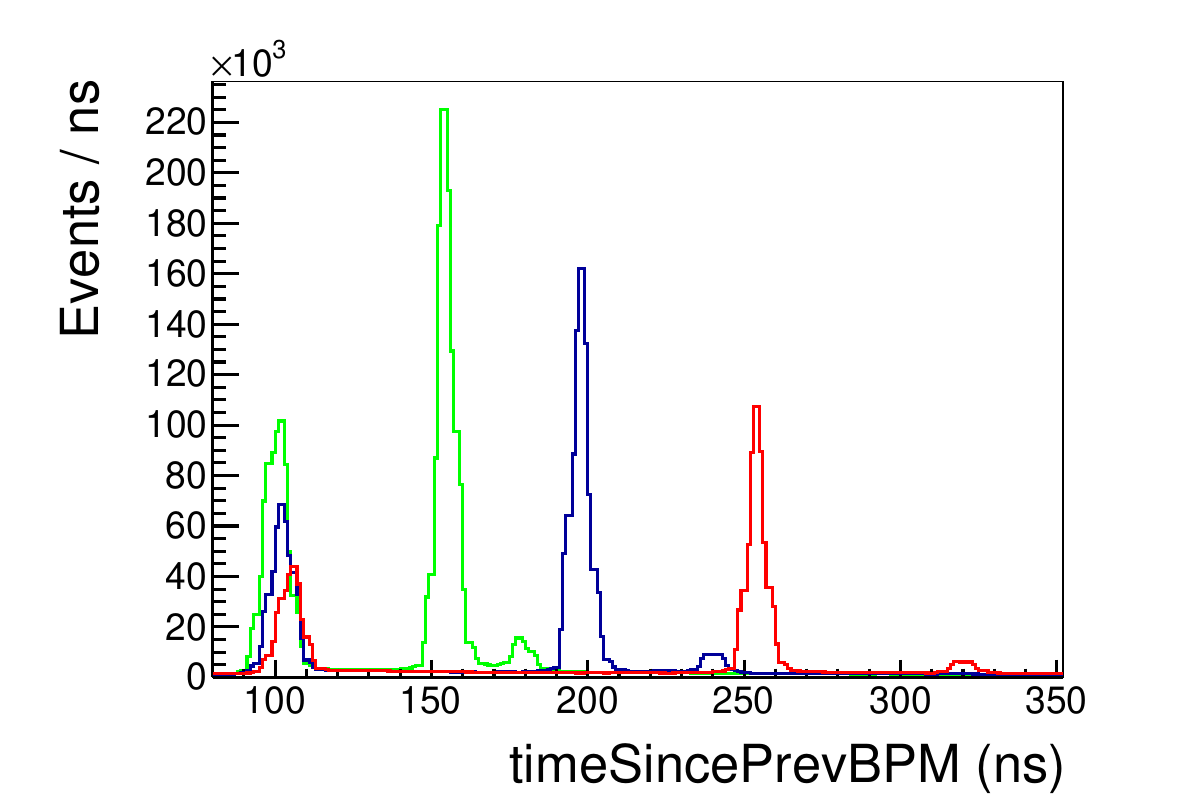}
\caption{\label{fig:timeSincePrevBPM_0deg} Distribution of the time after the BPM signal during dedicated TOF measurements for the October run: green for close, blue for middle, and red for far position of the 0-deg detector. The peaks at $\approx$100~ns correspond to photons produced in the LiF target.
} 
\end{figure}

The time after the BPM signal can be calculated for every event triggering the 0-deg detector, corresponding to the neutron TOF plus an offset. The distributions of this time for the three positions of the 0-deg detector for the October run are shown in Figure \ref{fig:timeSincePrevBPM_0deg}. Photons produced in the LiF target are identified easily, having TOF much shorter than neutrons. The corresponding distribution allows an estimate of the offset by taking into account the time required by photons to travel the distance between the LiF target and the 0-deg detector, D/c, where D is the distance from the 0-deg detector to the LiF target and c is the speed of light. Then, distributions of the TOF for neutrons can be built for every distance and run. The width and asymmetry of the TOF distributions can be understood as resulting from the finite size of the 0-deg detector (dispersion on the D parameter), time width of the pulsed proton beam, and detector time response. It can be observed in Figure \ref{fig:timeSincePrevBPM_0deg} that in addition to the $\approx$1~MeV neutrons there is an underabundant population with energies around 500~keV, corresponding to the $^7$Li(p,n)$^7$Be* process. 
\par

These TOF distributions can be directly converted into neutron energy distributions. However, energy loss of protons within the ${}^7$Li target and effects from the finite size of the backing detector affect the evaluation of the neutron energy. Instead, the strategy to derive the neutron beam energy is described below. Gamma rays of arbitrary energy, but sufficiently above backing detector thresholds (here 1~MeV) were simulated using MCNPX-PoliMi~\cite{mcnp-polimi}, originating at the LiF target. These were smeared with timing distributions (either Gaussian or Gumbel functions) with floating smearing parameter and offset to fit the gammas produced during the TOF calibrations. Next, protons close in energy to that expected by the beam settings were simulated with SRIM~\cite{srim}, and converted into neutron energy distributions using data from Ref.~\cite{LISKIEN197557}. These neutron distributions were simulated in MCNPX-PoliMi, and the best-fit offset and smearing parameters from fitting the gamma distributions were applied to the simulation output. A RooMomentMorph~\cite{verkerke2003roofit} was formed with proton energy as a floating parameter; the resulting PDFs were then fit to neutron TOF data from all three measured positions to obtain the incident proton energy. Using SRIM to model proton energy loss and Ref.~\cite{LISKIEN197557} to convert to neutron energy,  the neutron beam energy distribution was obtained. All the details of this calculation can be found in \cite{hedges-thesis}.
\par
Results are shown in Table \ref{tab:En}. These are the energies that will be considered throughout the rest of this paper as input for the simulations and to derive the QF estimates. 
\par

\begin{table*}[htbp]
\centering 
\begin{tabular}{|c|c|c|c|c|}
\hline
Run & Time response (ns) & $E_p$ (keV) & Mean $E_n$ (keV) & Std. Dev. $E_n$ (keV) \\
\hline
August & 3.4 $\pm$ 0.06  & 2670.9$^{+1.5}_{-3.1}$ & 958 $\pm$ 5 & 4 $\pm$ 3 \\
October & 1.2 $\pm$ 0.03  & 2696.8$^{+0.3}_{-0.8}$ & 982 $\pm$ 7 & 7 $\pm$ 5\\
\hline
\end{tabular}
\caption{\label{tab:En} FWHM of the time response, proton energy, mean and standard deviation of the neutron energy distributions derived for the two runs. } 
\end{table*}

\subsection{Calibrations}
\label{sec:Ecalibration}
Common approaches followed in previous measurements of the QF in NaI(Tl) for the conversion of the light collected into electron equivalent energy use either the 59.5~keV gamma from an \isotope{Am}{241} source or the 57.6~keV gamma from the $^{127}$I($n,n^{\prime}\gamma$) process. The latter line also allows for the continuous monitoring of the stability of the response of the NaI(Tl) crystal throughout the beam-on data collection. However, calibrating the energy with just one reference line that is far from the region of interest brings some relevant systematic effects into the analysis. 
\par
\begin{figure*}[htbp]
\centering 
\includegraphics[width=.3\textwidth]{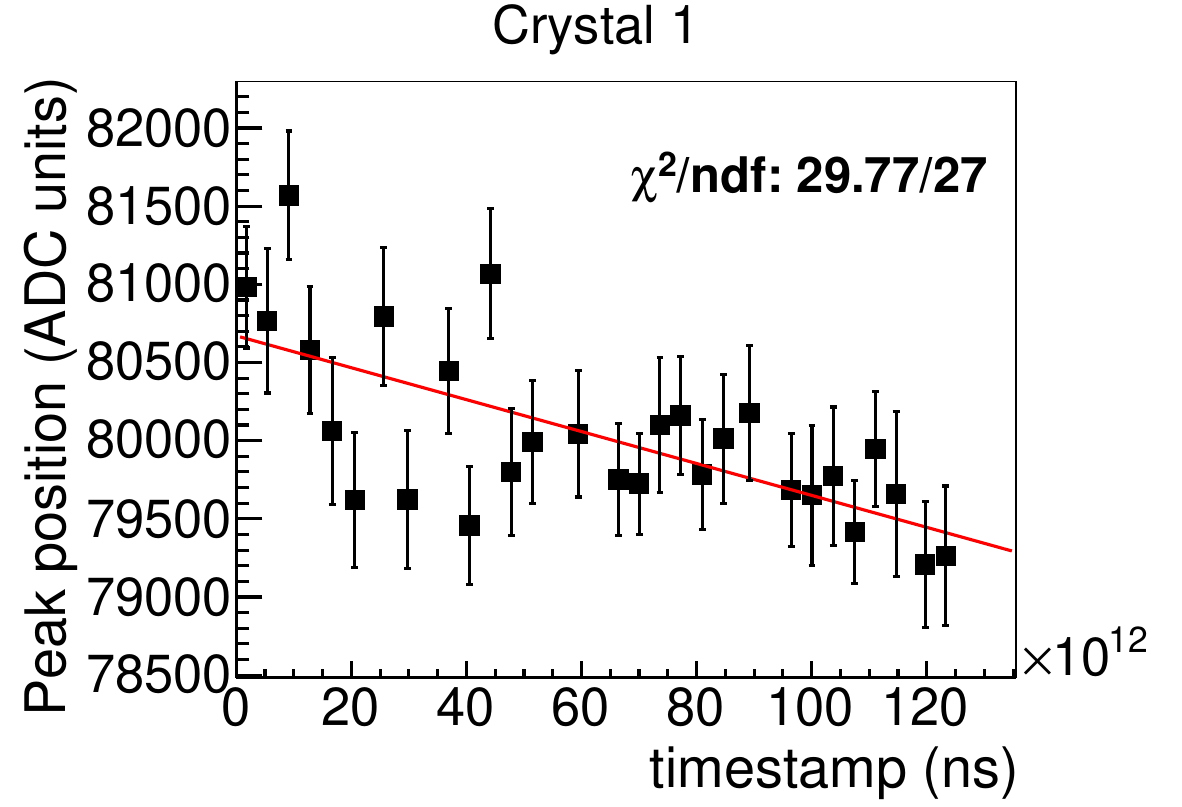}
\includegraphics[width=.3\textwidth]{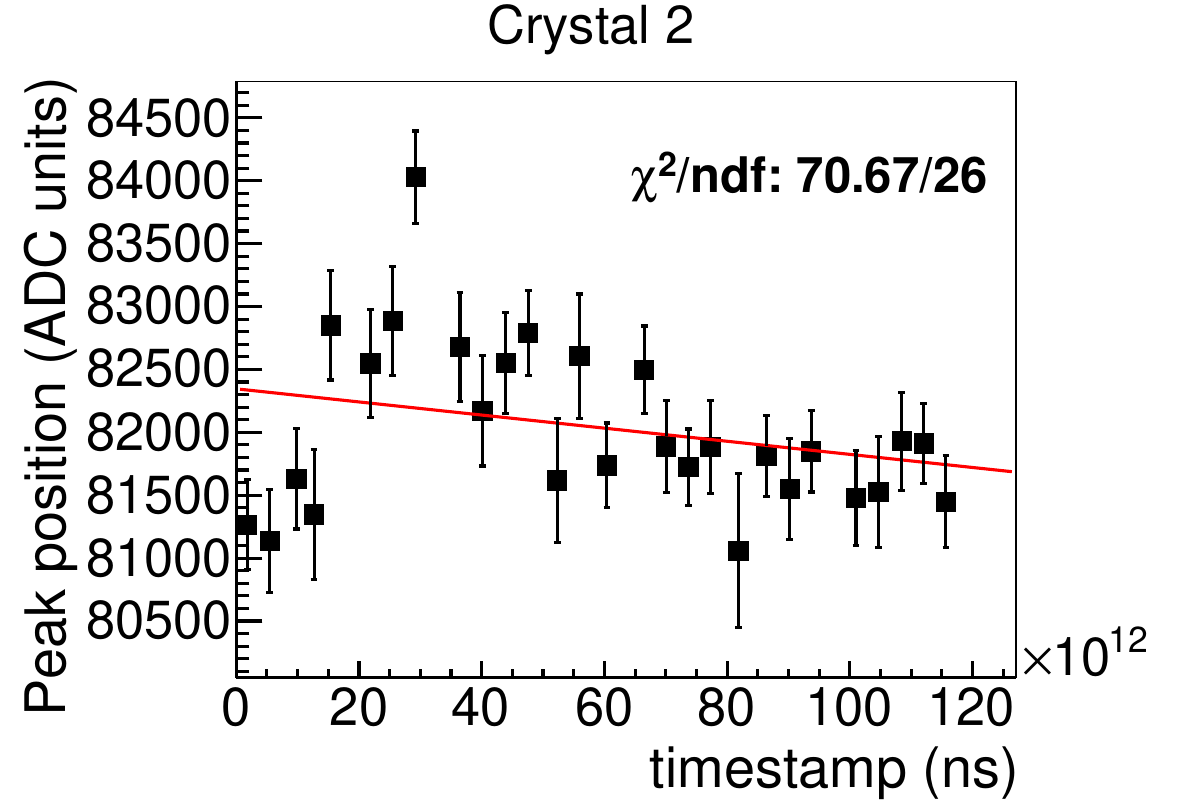} 
\includegraphics[width=.3\textwidth]{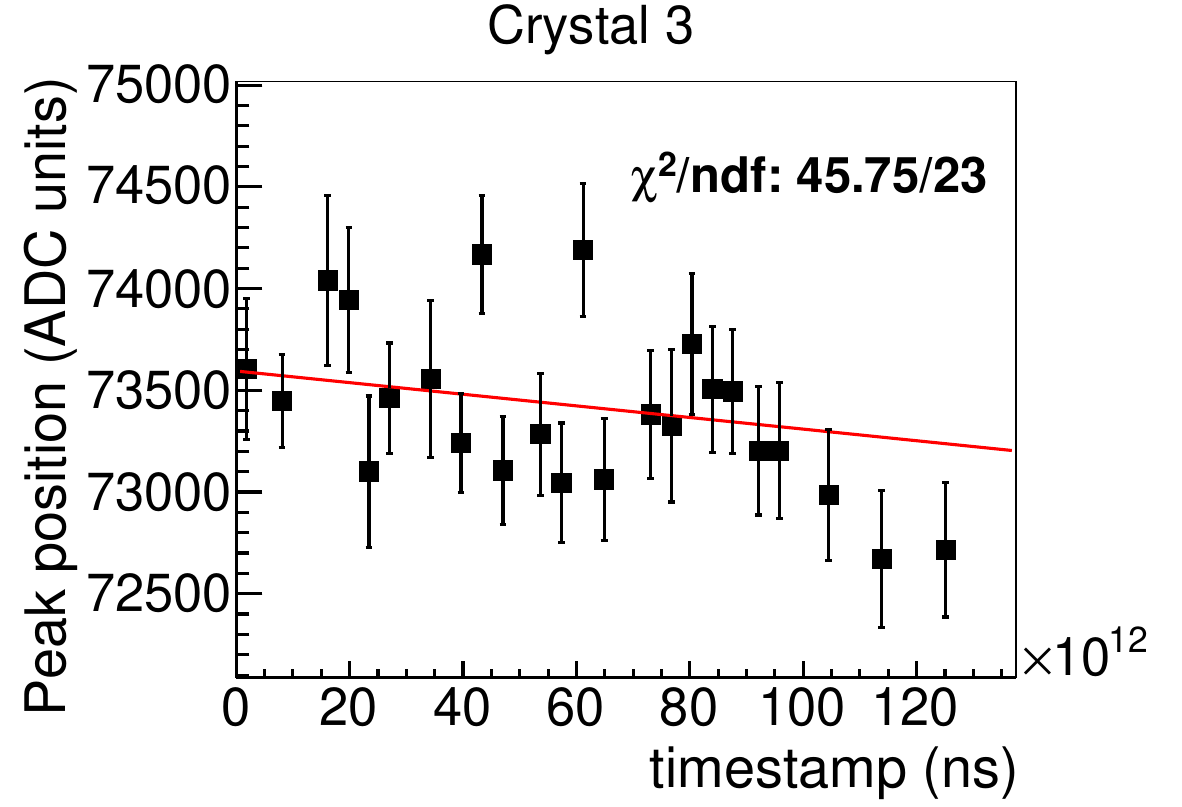}
\includegraphics[width=.3\textwidth]{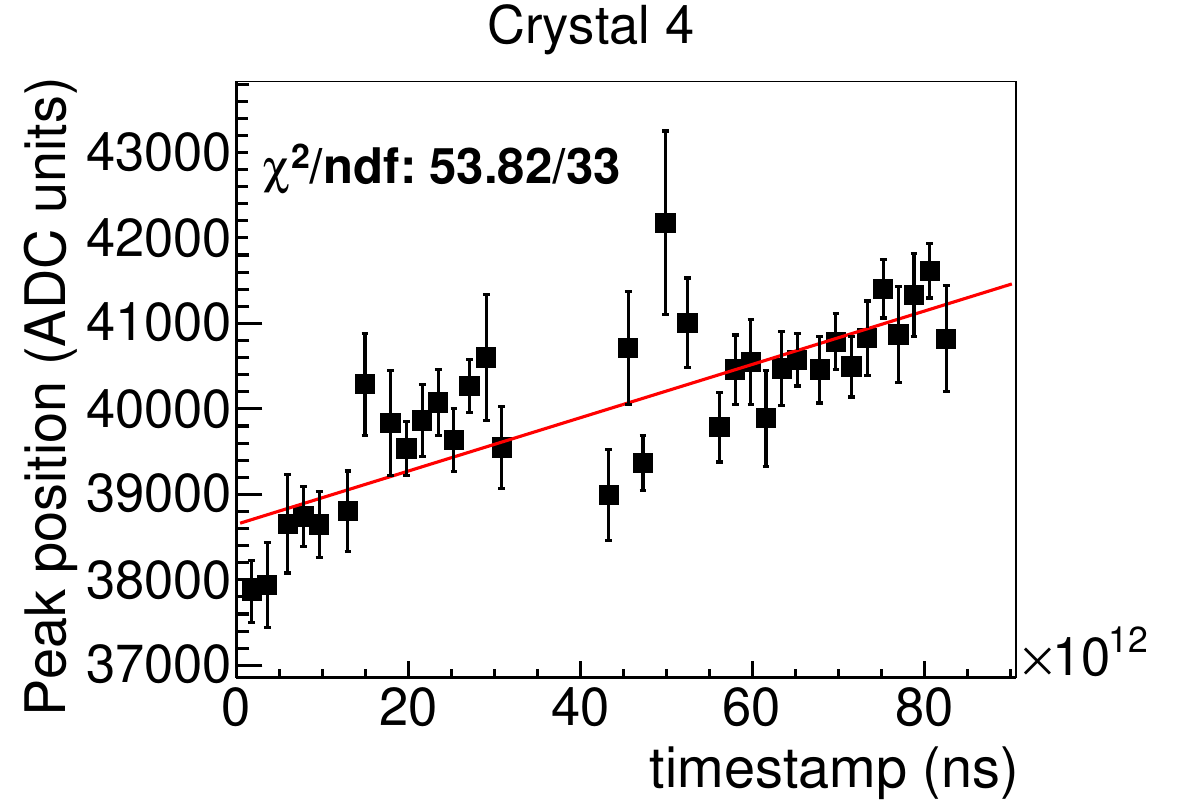} 
\includegraphics[width=.3\textwidth]{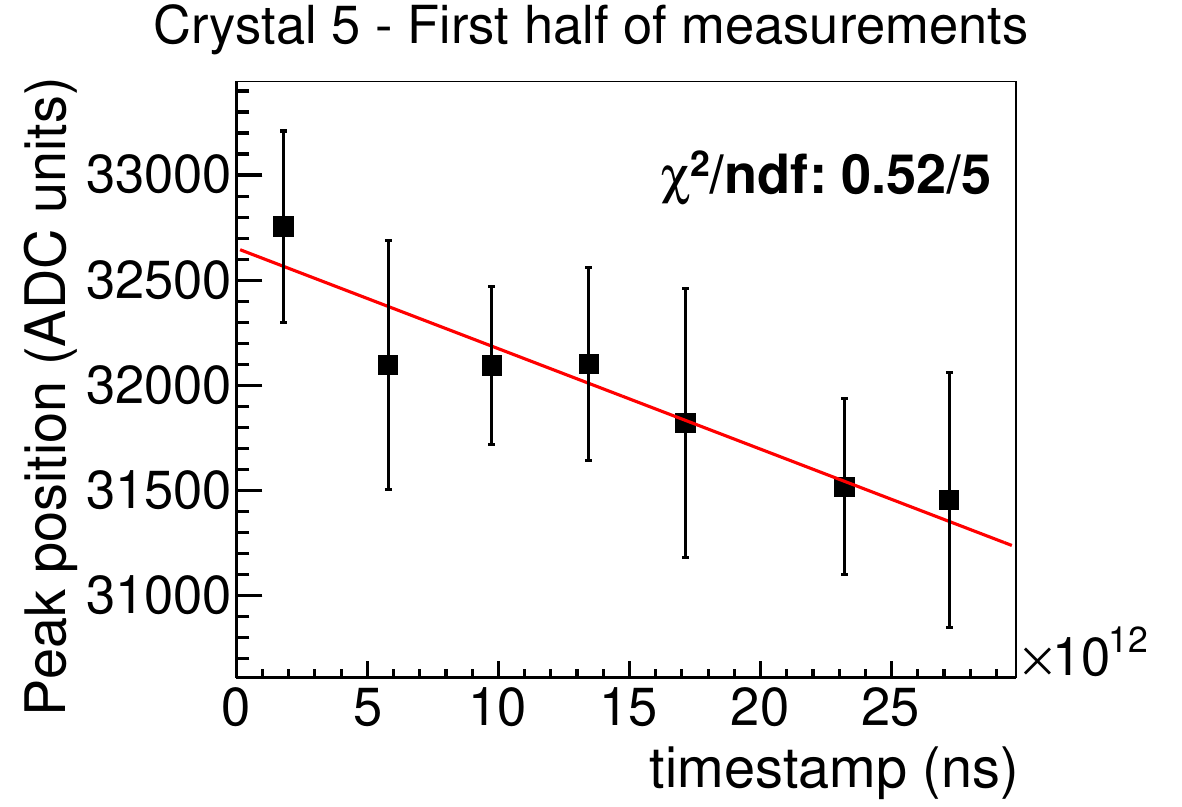}
\includegraphics[width=.3\textwidth]{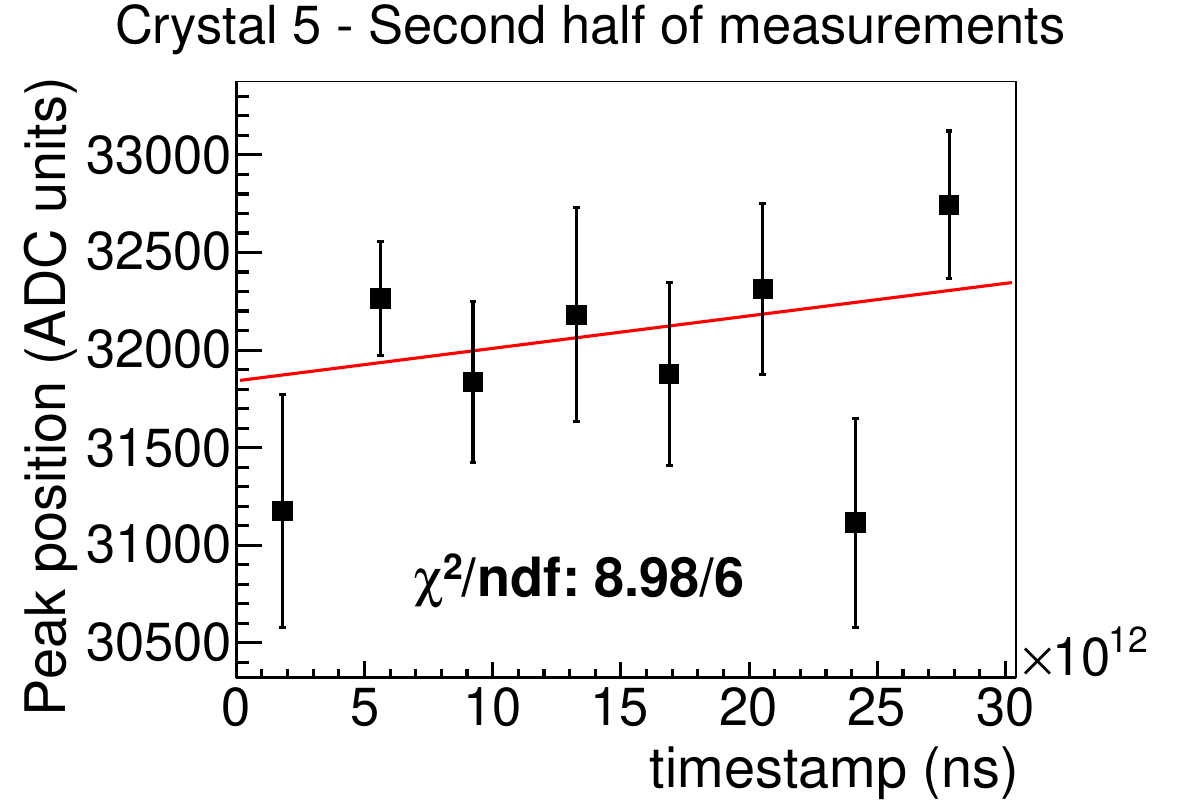} 
\caption{\label{fig:stability} Fits of the 57.6 keV peak mean value for the pulse area to a linear dependence with time. 
} 
\end{figure*}

For the gain stability control in this work, the 57.6~keV peak was analyzed every hour after applying the BDs neutron selection procedure explained in Section~\ref{sec:analisisTOF}. 
This peak was fitted to a gaussian function summed with a constant background, and the corresponding mean is shown as a function of the
time for all the crystals in Figure \ref{fig:stability}. A drift is clearly observed in crystals No.~1 and No.~4, while crystals No.~2, No.~3 and No.~5 show some variation in the positions of the mean, but without a distinct trend. The data of the crystal No.~5 was divided in two different periods, as it was observed a different behaviour after a calibration run, in the middle of the beam-on
measurements. For all the crystals, a linear dependence of the pulse integral with
time was used to model (and correct) this drift. This correction was applied to all the data of the beam-on measurements and the calibrations with \isotope{Ba}{133} by extrapolating the detector behavior at the time every dataset was acquired. The energy resolution of the 57.6 keV peak in crystal No.~4 improved from 14.0$\pm$0.2\% to 13.3$\pm$0.2\% after this correction was applied. For the other crystals, this correction only resulted in a slight improvement in resolution. The reason behind these gain instabilities was not identified, although they could be attributed to changes in the PMT-crystal coupling and/or PMT HV bias.
\par

In the calculation of the QF, the electron equivalent energy calibration of the energy depositions in the NaI(Tl) is one of the most relevant points. The light yield of the NaI(Tl) is non-proportional with the energy, with variations of a few percent up to about 20~keV~\cite{nonlinearity,nonlinearity2,nonlinarity3,nonlinearity4,nonlinearity5,nonlinearity6,nonlinearity7}. 
\par
Trying to evaluate the effect on the QF results of this non-proportional response, we have considered three different approaches for the calibration in electron equivalent energy
of the NaI(Tl) crystals, using the drift-corrected spectra both for \isotope{Ba}{133} calibrations and beam-on measurements:
\begin{enumerate}[(i)]
    \item Assuming a proportional response calibrating with the 57.6~keV inelastic peak from the $^{127}$I($n,n^{\prime}\gamma$) process. The proportionality constant relating the mean pulse integral of the peak and the energy (A/E) for the five crystals measured is shown in Table \ref{tab:c}. The very different conversion factors between pulse area and energy in the different crystals relates with the different light collection achieved, because operation conditions of the PMT and the electronic chain used in all the measurements were equivalent.

\begin{table}[htbp]
\centering 
\begin{tabular}{|c|c|}
\hline
  Crystal number & ADC units/keV \\

  \hline
1 & 1388 $\pm$ 3  \\
2 & 1421 $\pm$ 3  \\
3 & 1265 $\pm$ 3 \\
4 & 688 $\pm$ 3 \\
5 & 549 $\pm$ 6 \\
\hline
\end{tabular}
\caption{\label{tab:c} Proportionality parameter between pulse area and energy (A/E) for each crystal, determined with the 57.6~keV inelastic peak from the $^{127}$I($n,n^{\prime}\gamma$) process. } 
\end{table}
    
    \item  Applying a linear calibration in the ROI using the energy depositions produced by the interaction of the gamma and x-rays emitted by the external $^{133}$Ba source, associated to 6.6, 30.9 and 35.1~keV according to the GEANT4 simulation's results. As the GEANT4 simulation of the source did not  quantitatively reproduce the measured spectrum, 
    the calibration coefficients were determined by fitting the experimental spectrum to a model with three Gaussian peaks and only in a region close to their maximums.
    The fitting was done by building a PDF which included a flat background plus three Gaussians at the fixed energies previously commented and an energy resolution variable with energy, modelled with two free parameters:
    \begin{equation}
        \sigma=\sqrt{a+b E} 
    \end{equation}
    The conversion from pulse area into energy included two additional free parameters:
    \begin{equation}
        E=c_1 A +c_0
    \end{equation}
    Results of these fits are shown in Table \ref{tab:lincal} for all the crystals, and in Figure \ref{fig:Ba_fit} for crystal No.~1.

\begin{figure}[htbp]
\centering 
\includegraphics[width=.5\textwidth]{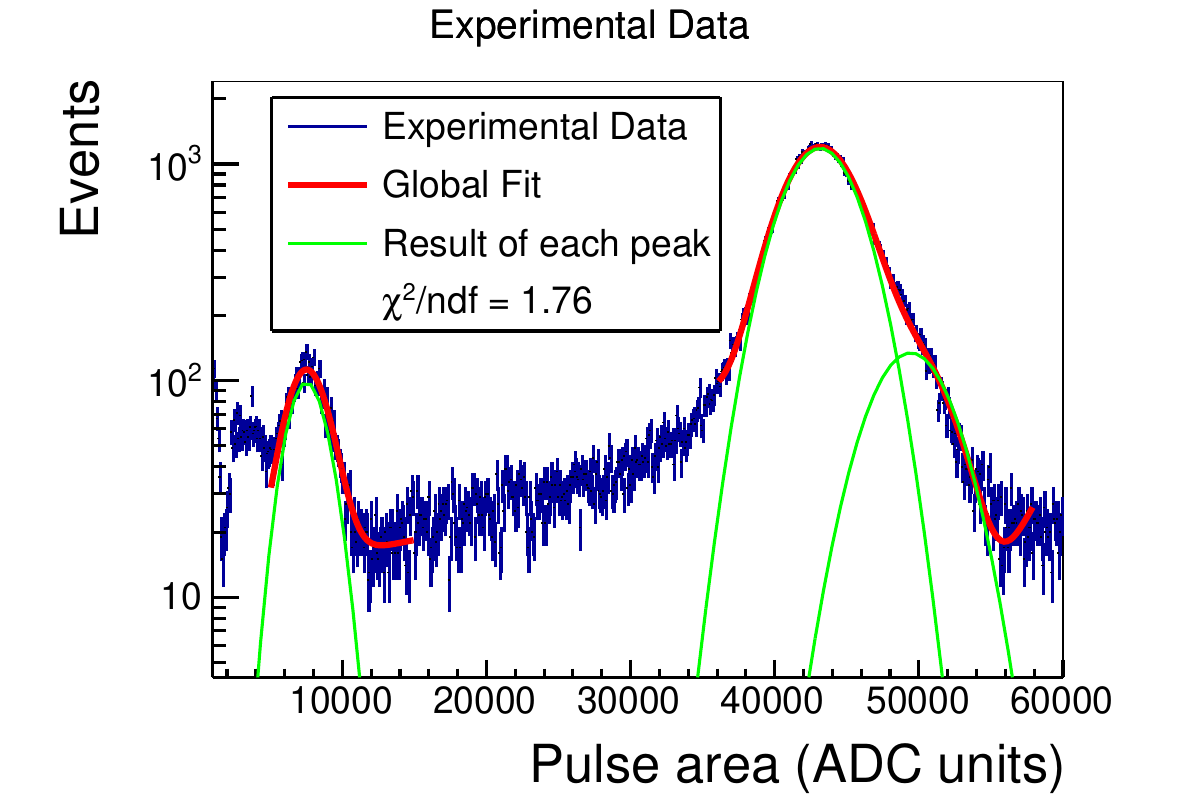}
\caption{\label{fig:Ba_fit} Fit of the \isotope{Ba}{133} spectrum for crystal No.~1 to a flat background plus three Gaussian peaks, with linear conversion between energy and pulse area and energy-dependent energy resolution.
} 
\end{figure}  

\begin{table*}[htbp]
\centering 
\begin{tabular}{|c|c|c|c|c|}
\hline
  Crystal number & a (keV)$^2$ & b (keV) & c$_0$ (keV) & c$_1$ (eV/ADC unit) \\

\hline
1 & 0.000 $\pm$ 0.018  & 0.103 $\pm$ 0.001 & 1.195 $\pm$ 0.032 & 0.688 $\pm$ 0.001\\
2 & 0.000 $\pm$ 0.064  & 0.084 $\pm$ 0.001 & 0.825 $\pm$ 0.031 & 0.673 $\pm$ 0.001\\
3 & 0.111 $\pm$ 0.084 & 0.095 $\pm$ 0.003 & 0.819 $\pm$ 0.034 & 0.754 $\pm$ 0.001\\
4 & 0.000 $\pm$ 0.634 & 0.176 $\pm$ 0.006 & 1.254 $\pm$ 0.082 & 1.366 $\pm$ 0.005\\
5 & 0.000 $\pm$ 0.501 & 0.275 $\pm$ 0.017 & 1.470 $\pm$ 0.080 & 1.674 $\pm$ 0.006\\
\hline
\end{tabular}
\caption{\label{tab:lincal} Energy-pulse area and energy resolution parameters derived from the fits to the \isotope{Ba}{133} PDF built as explained in the text for each crystal. } 
\end{table*}

    If we compare this calibration with the proportional described in (i), the difference at low energies is important. The peak corresponding nominally at 6.6~keV is found with this calibration approach at much lower energies with the proportional calibration, with a residual larger than 1~keV.    
    \item Combining both approaches aiming at better calibrating the ROI of our measurement by using the linear calibration above 6~keVee, while assuming a proportional response below this energy, using as reference the position of the 6.6~keV peak from the $^{133}$Ba source. 
    This approach relies on a typical solution to take into account non-linear behaviours, as the well-known non-proportionality in the light response of NaI(Tl), by combining different linear functions in smaller energy regions which overlap. Unfortunately, only a limited number of peaks were available, limiting the reach of the approach.
    
\end{enumerate} 

\subsection{Event Selection}
\label{sec:analisisTOF}
\label{sec:correlationBDNaI}
A robust protocol for identifying neutrons reaching the BDs after scattering off nuclei in the NaI(Tl) crystal compared to different backgrounds is essential in this measurement. Many of these backgrounds can be well identified and rejected. Pulse shape analysis allows the discrimination of neutron events from gamma events in liquid scintillators, such as those used as target in the BDs. A PSD variable is built, see Section~\ref{sec:analysischain}, for the BDs output signals in order to profit from the different scintillation times associated to neutron and gamma/electron events, as shown in Figure~\ref{fig:PSD}.
\par
\begin{figure}[htbp]
\centering 
\includegraphics[width=.5\textwidth]{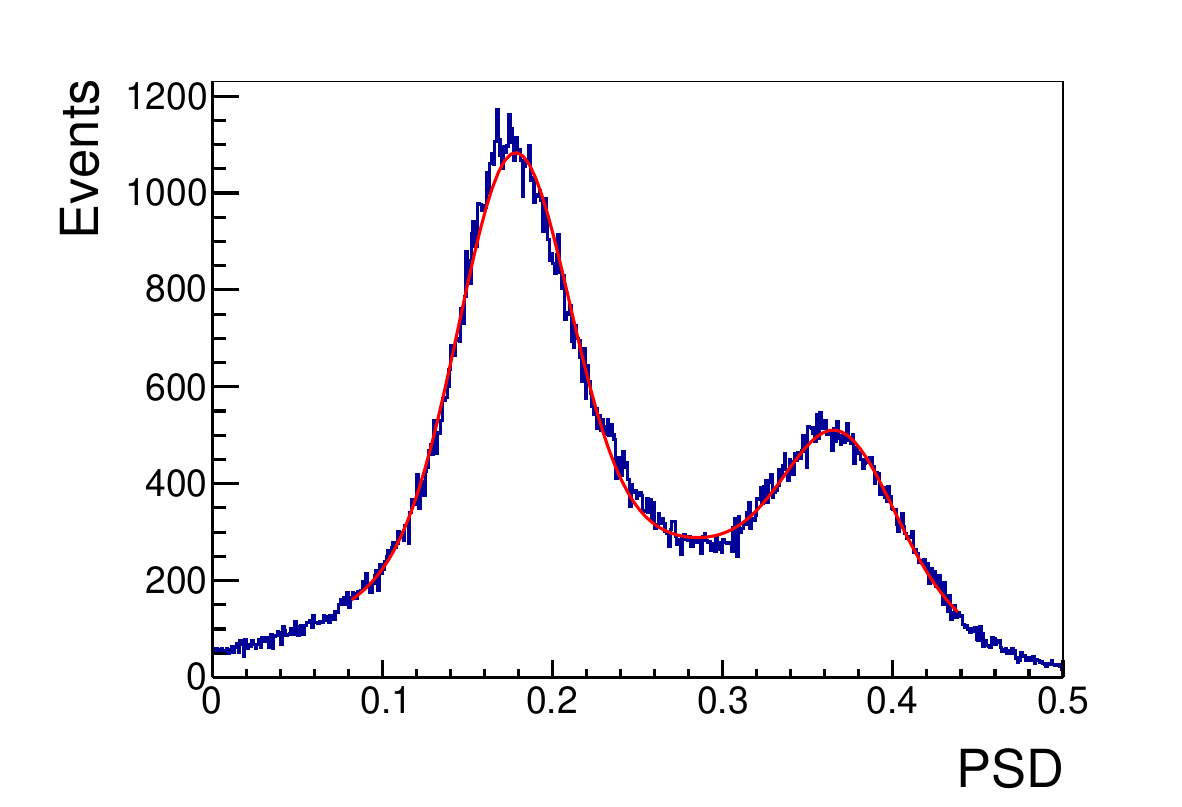}
\caption{\label{fig:PSDdist} PSD distribution for 10$^5$ events triggering any of the BDs. A clear discrimination between gammas (lower PSD value) and neutrons (higher PSD value) can be observed.  
} 
\end{figure}
\begin{figure}[htbp]
\centering 
\includegraphics[width=.5\textwidth]{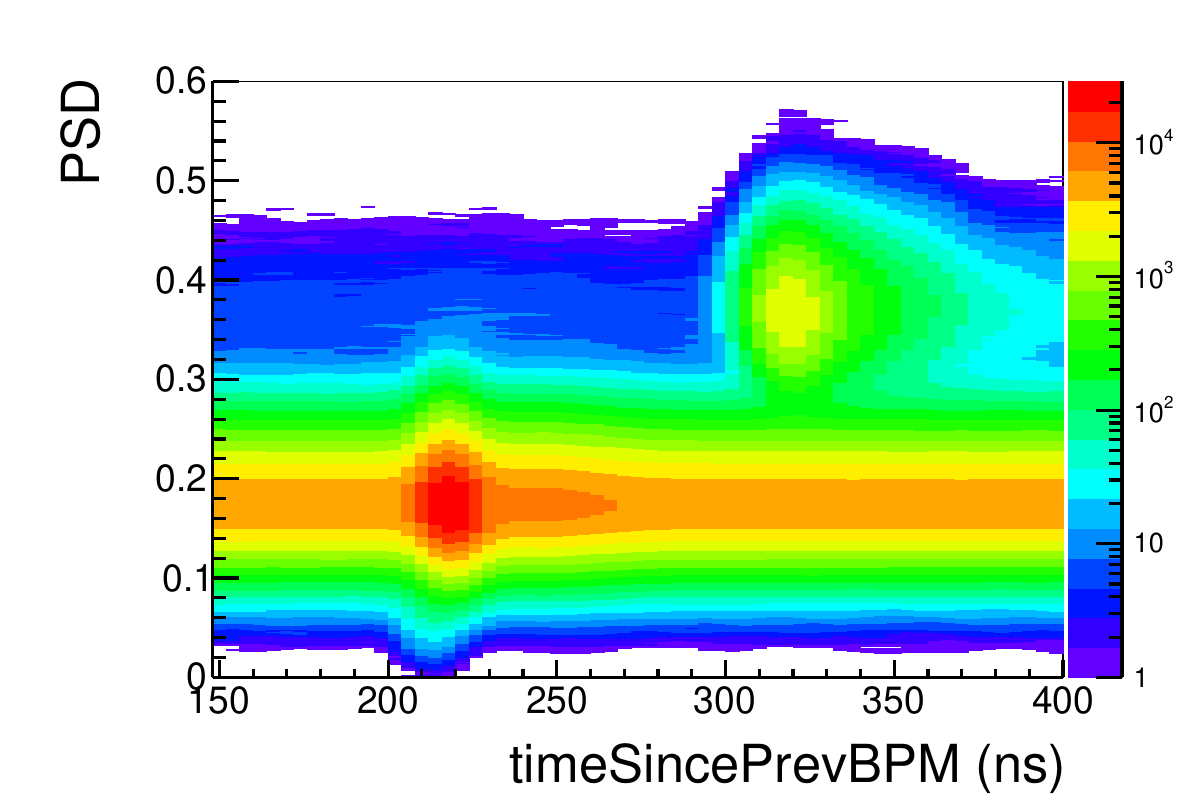}
\caption{\label{fig:PSD_TOF} PSD vs time after BPM signal corresponding to all the BD waveforms registered in the measurement of crystal No.~5. Neutrons correlated with the beam are found at high PSD and time after BPM signal above 300~ns.  
} 
\end{figure}
\begin{figure}[htbp]
\centering 
\includegraphics[width=.5\textwidth]{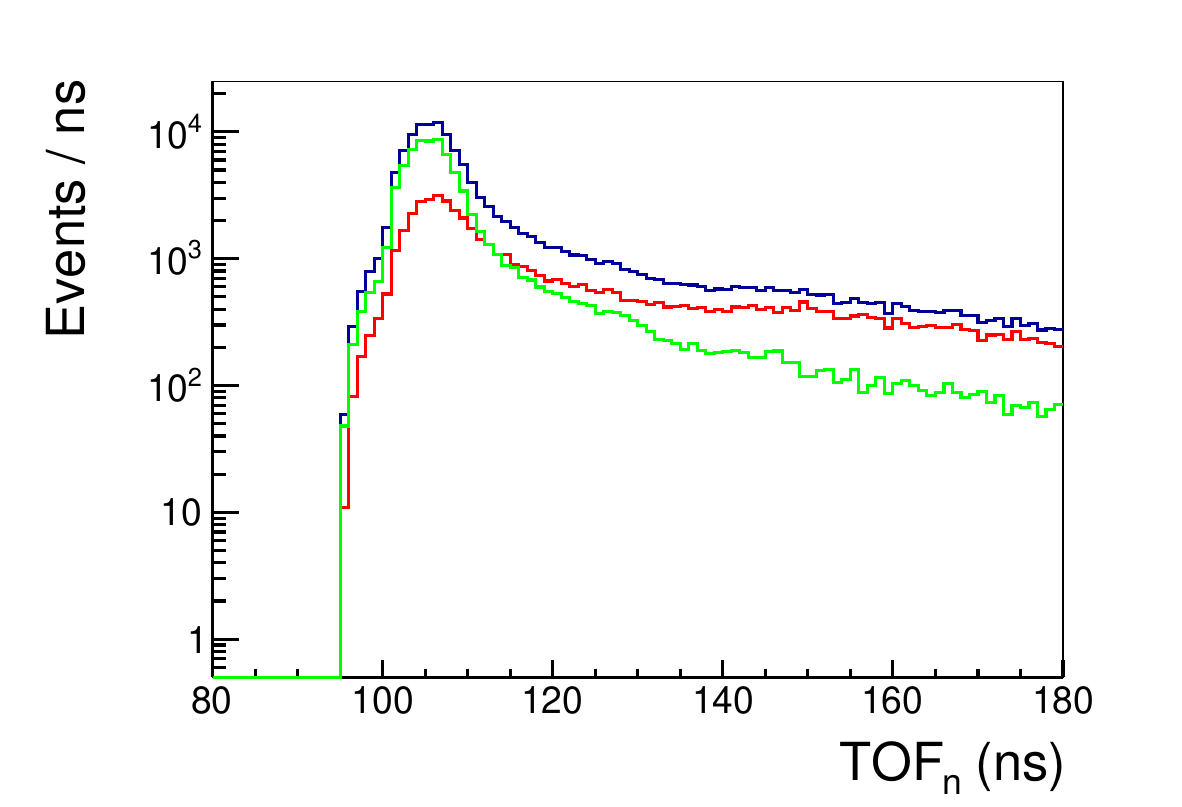}
\caption{\label{fig:t0} Time of arrival of the neutrons to the BDs for single scattered neutrons (green line), multiple scattered neutrons (red) and total (blue), according to the GEANT4 simulation of the setup.  
} 
\end{figure}

Figure~\ref{fig:PSDdist} shows the distribution of this variable. Higher PSD values correspond to neutrons, while gammas are found below 0.3. Better discrimination can be achieved by combining the information on PSD and TOF, as it can be observed in Figure~\ref{fig:PSD_TOF}. Gamma interactions are observed non correlated with the beam (flat distribution of the time after BPM signal with PSD $\approx$0.18), while two beam correlated populations can be identified: gammas produced in the LiF target (having PSD $\approx$0.18 and time after BPM signal $\approx$220~ns) and neutrons (having PSD $\approx$0.35 and time after BPM signal above 300~ns). The TOF distribution for neutrons hints at a contribution from neutron multiple scattering, which could reduce the energy of the neutron reaching the BD and consequently increase the TOF. This hypothesis was checked using simulation data. Figure~\ref{fig:t0} shows the distribution of the arrival time of neutrons to the BDs, using as time origin the neutron generation time. It can be observed that multiple scattering dominates for TOF beyond 25~ns from the most probable value. Moreover, we do not expect to detect neutrons in the BDs more than 10~ns before the most probable TOF. Then, we introduce a selection in the time after BPM signal between 304 and 340~ns. With the applied selection, the fraction of single scattered neutrons is increased from 68\% to 80\%. 

However, using the neutron TOF for event selection implies an indirect neutron energy selection. It was checked with the simulation that this event selection criterion did not imply any correlation between nuclear recoil energy transferred in the NaI(Tl) crystal and the time of the first energy deposit in the BDs. 
\par
\begin{figure}[htbp]
\centering 
\includegraphics[width=.5\textwidth]{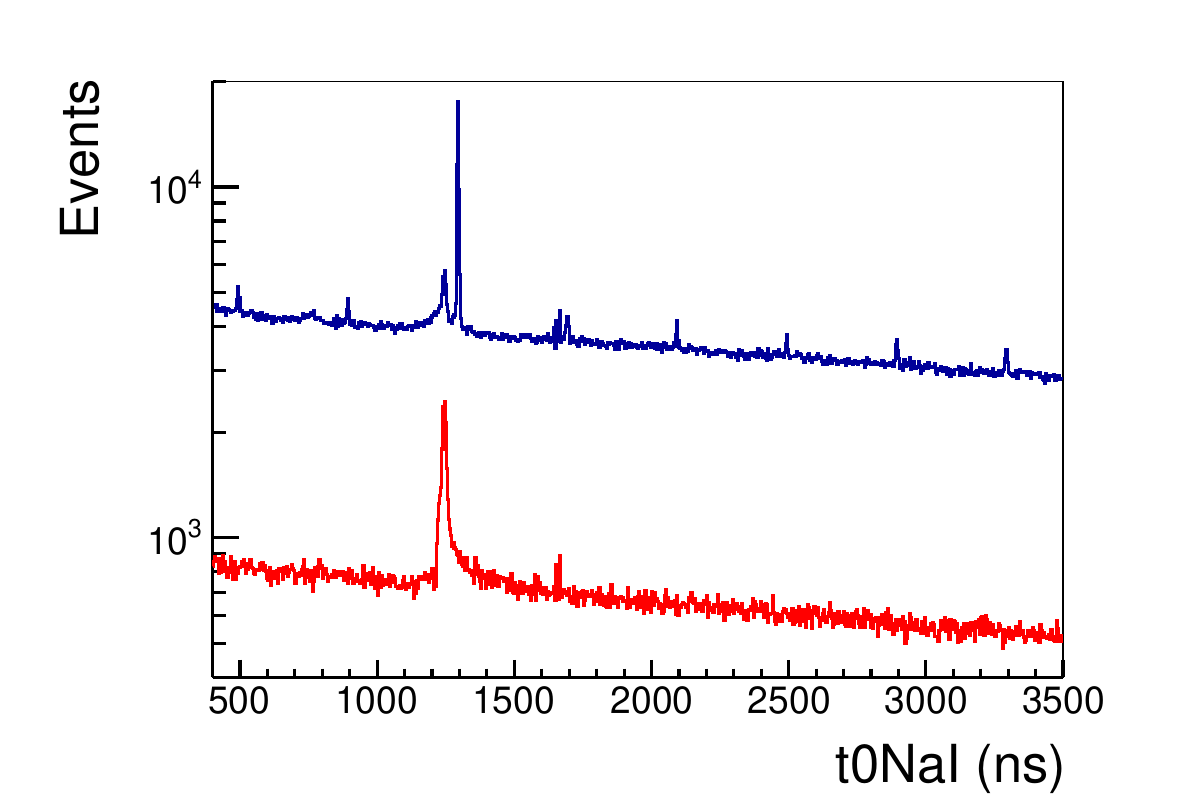}
\caption{\label{fig:t0NaI} Distribution of the $t_{0,\mbox{NaI}}$ variable before (blue line) and after (red line) selection of events compatible with neutrons in the BDs for crystal No. 5.  
} 
\end{figure}

Once the events induced by neutrons in the BDs are selected, the next step is to search for events correlated with them in the NaI(Tl) crystal. This is done with the $t_{0,\mbox{NaI}}$ variable previously defined, whose distribution is shown in Figure \ref{fig:t0NaI}, before and after the application of the neutron selection procedure. Rate is clearly dominated by gamma/electron events, but after removing them, only one peak in the distribution is observed that can be attributed to nuclear recoil energy deposited in the NaI(Tl) crystal by the neutron which later triggered the BD. This analysis shows that events with correlated neutron interactions in the BD and NaI(Tl) do not appear earlier than 1200~ns into the waveform. This allowed us to fix the integration time interval for signals in the NaI(Tl) from 1.2 to 3.2 $\mu$s. 

\subsection{Calculating the Quenching Factors in NaI(Tl)}
\label{sec:fit}
\subsubsection{Sodium Quenching Factor}
To obtain the sodium quenching factor, the gain-corrected and energy-calibrated experimental spectra associated with neutrons in each triggering backing detector (channel) and crystal were fit to the corresponding simulated nuclear recoil energy distributions shown in Figure \ref{fig:NR_Na}. Quenching factors, different for each channel, are floated along with a background component. In addition, modelling of the detector's energy resolution was required for the fitting procedure to succeed.
\par
The sources of background were difficult to identify. Background measurements were done in dedicated runs, but they had low statistics in the ROI and they did not include beam related events which are expected to be the most relevant background contribution. It was considered as a better option to use as background those events that do not fulfill the neutron selection criteria explained in Section \ref{sec:analisisTOF}. No clear differences were observed among different channels and crystals. These background spectra were gain corrected and converted into electron equivalent energy to build a PDF ($S_{bkg}$).
\par
In addition, the recoils of iodine nuclei contribute significantly to the data in some of the channels, and therefore were also included in the fit.  The corresponding recoil spectra were obtained from the simulation, and a constant QF for iodine was adopted in the fit, as the recoil peak could not be observed.  
\par
To account for the energy resolution, a gaussian function was used to convolve the signal of the sodium recoils for each channel, and two different modellings were applied for the standard deviation: energy independent and
proportional to the square root of the electron equivalent energy, as it would correspond to a poissonian resolution.
\par
\begin{figure}[htbp]
\centering 
\includegraphics[width=.5\textwidth]{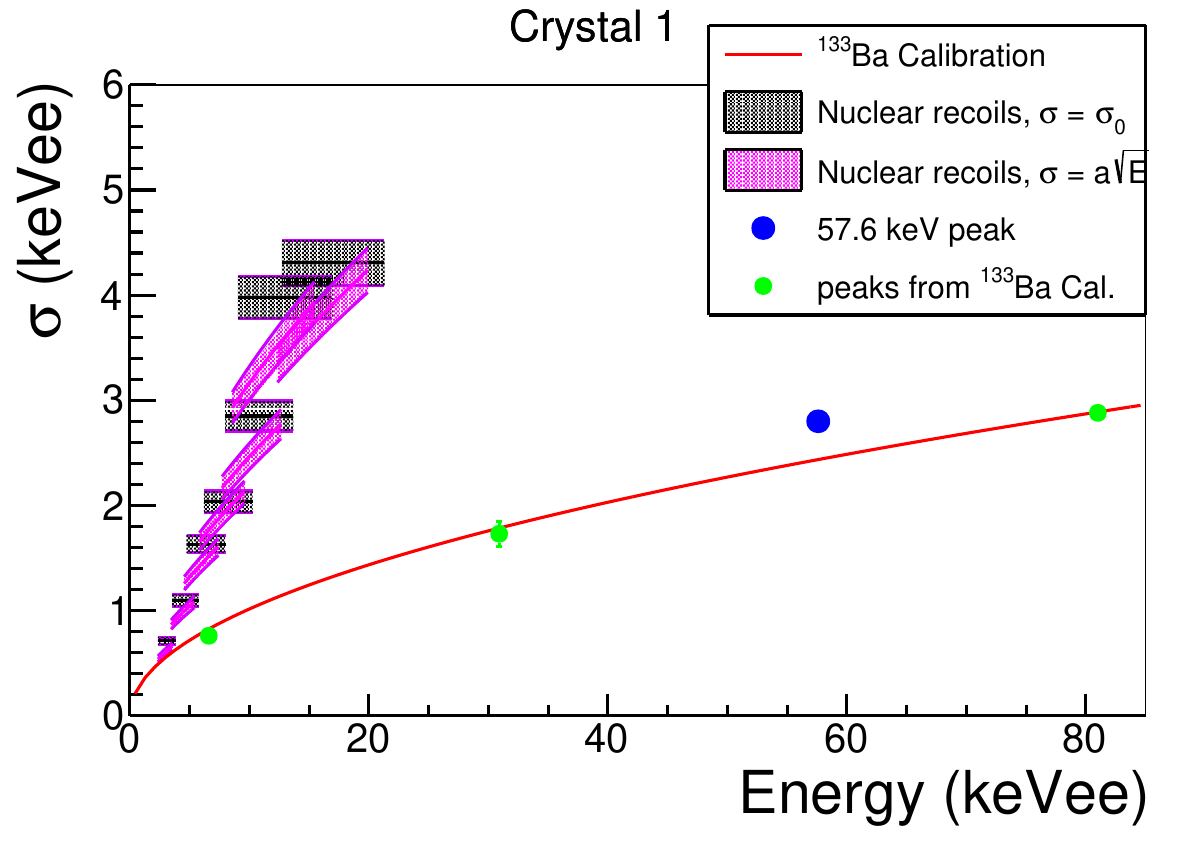}
\includegraphics[width=.5\textwidth]{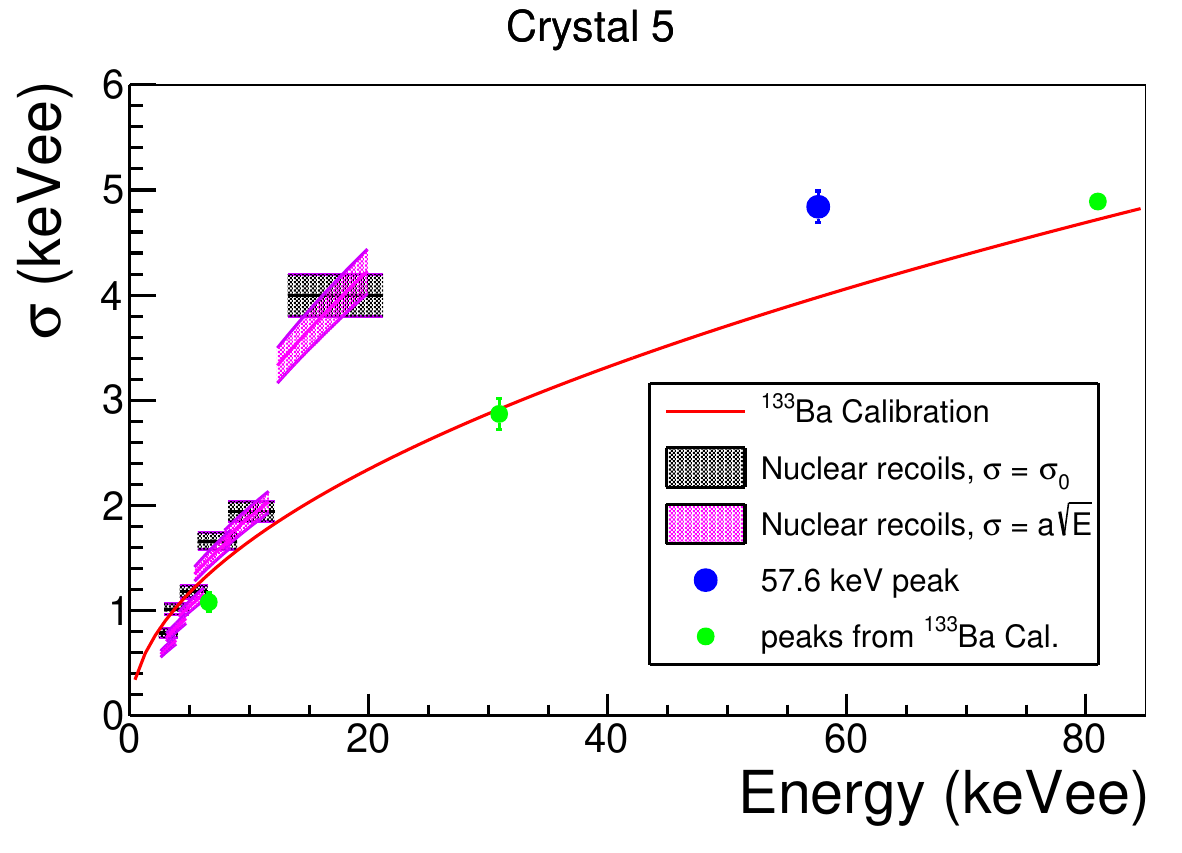}
\caption{\label{fig:resolution} Energy resolution as a function of the electron equivalent energy for crystal No.~1 and crystal No.~5 data fitted using the two different resolution modelling (shadowed regions). The solid red line represents the energy resolution derived from the \isotope{Ba}{133} calibration data. The energy resolution for the inelastic peak at 57.6~keV and for the 6.6 and 30.9 keV peaks from\isotope{Ba}{133} is also shown.} 

\end{figure}

The procedure for each fit is the following: First, the region above the sodium recoils peak for each channel (between 30 and 40 keV) is fitted to the background PDF ($S_{bkg}$) in order to determine a scaling factor.
Then, the simulated sodium recoil spectrum for that channel is converted into electron equivalent energy with a energy-independent free-floating QF$\mathrm{_{Na}}$ parameter, different for each channel, convolved with a gaussian with the standard deviation modelled as commented above. The resulting PDF is
called $S_{Na}$. The PDF for iodine recoils ($S_I$) is built following a similar procedure but with constant QF$\mathrm{_I}$ = 5\% and standard deviation of 1 keV. It was checked that the fitting procedure was not sensitive to slight variations of these values, and the systematic contribution of the change in these parameters to the final QF results was also analyzed, as it is explained next. Finally, the total PDF was constructed as
\begin{equation}
    N_{Na} S_{Na} + N_I S_I + N_{bkg} S_{bkg}
\label{eq:pdf}
\end{equation}
\noindent and the experimental spectrum for each channel was fitted using as free parameters $N_{Na}$ and $N_I$, the QF$\mathrm{_{Na}}$ and the parameter corresponding to the resolution model chosen. According to the p-values of the fits, no preference for any resolution modelling can be concluded. Figure~\ref{fig:resolution} shows the energy resolution derived for crystal No.~1 from the two different models in addition to the energy resolution obtained from the \isotope{Ba}{133} calibration. 
It was impossible to achieve a good fit by using the same resolution function for all the energy ranges associated to the different channels. 
It can be observed in Figure~\ref{fig:resolution} that the energy resolution obtained from the recoil data fitting is clearly worse than that obtained for electronic recoils using the peaks from the \isotope{Ba}{133} calibration. This result has still to be understood. The energy resolution obtained for the inelastic peak at 57.6~keV, also shown in Figure~\ref{fig:resolution}, is also worse than expected from the \isotope{Ba}{133} calibration data. The inelastic peak corresponds to energy depositions in the crystal bulk, while the interactions from the gamma and x-rays produced in the \isotope{Ba}{133} decay are more local. This could be considered as a hint of possible spatial dependences on the scintillation properties or light collection. On the other hand, the inelastic peak is dominated by the energy deposition of a gamma, with a mean free path in NaI(Tl) of 0.4~mm, while a recoiling Na nucleus with energy lower than 100~keV in NaI(Tl) has a range below 200~nm. This makes them sensitive to very different scales of possible spatial effects contributing to the light yield, for instance the distribution of the Tallium activator in NaI(Tl), that could be more homogeneous in scales of 0.1~mm than in the submicrometer range. Our result should be taken as a warning: it is necessary to better determine the response of NaI(Tl) detectors to nuclear recoils without assuming as valid the same parameters derived from conventional calibrations using electron recoils. 
The difference between the QF results obtained by fitting with the two resolution modellings were included in the presented results as a systematic contribution to the final uncertainty. 
\par

A procedure was followed systematically in all the channels to fix the range of energies considered in the fit. First, the experimental data was fitted to the PDF from Equation \ref{eq:pdf} from 2.5 to 40~keV applying the constant resolution modelling, thus obtaining preliminary QF ($QF_p$) and resolution values ($\sigma_p$). After this preliminary fit, and using the mean value of the sodium recoil energy distribution obtained from the simulation for each channel, corresponding to a mean Na recoil energy of $E_{nr}$, the upper electron equivalent energy considered in the fit will be $E_{nr} QF_p + 5 \sigma_p$. However, the fit results were much more dependent on the lower energy considered in the fit. An iterative procedure was designed for adjusting that minimum energy, which was increased step by step and fitted until the $\chi^2$ changed by less than 10\% after three iteration steps. 
\par
\begin{figure*}[htbp]
\centering 
\includegraphics[width=.3\textwidth]{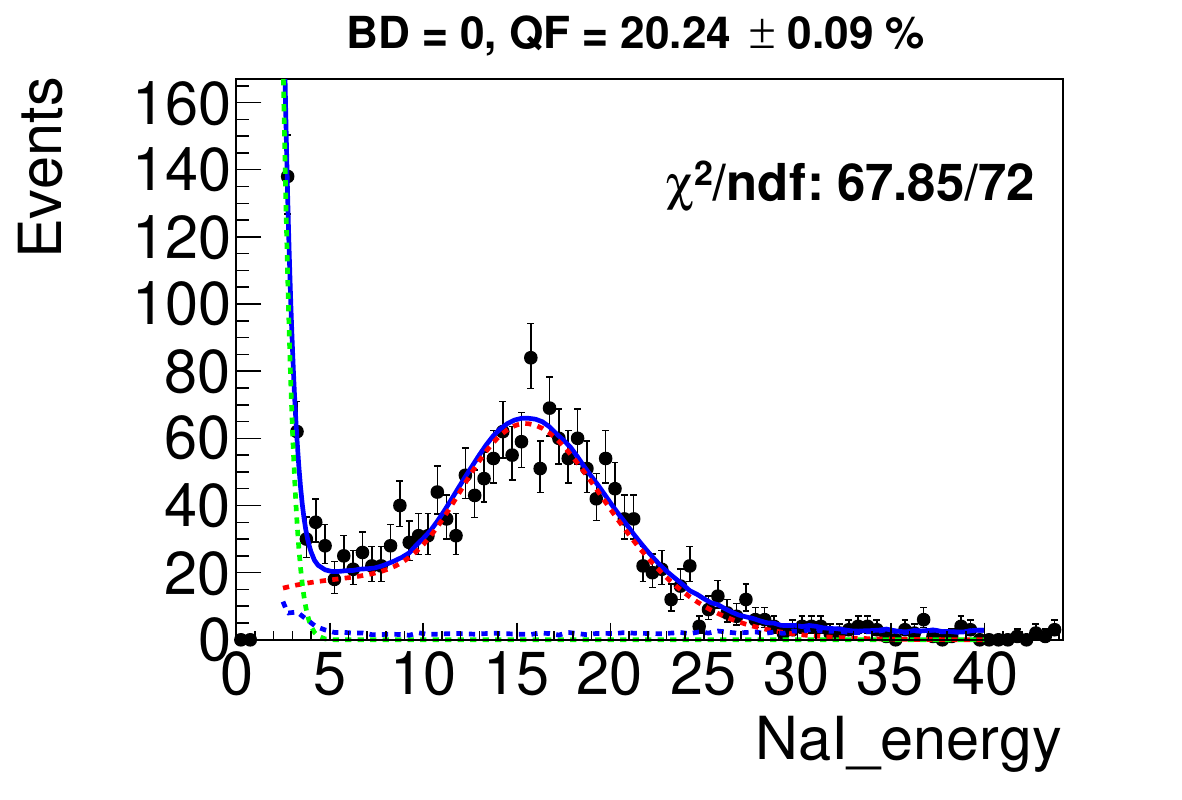}
\includegraphics[width=.3\textwidth]{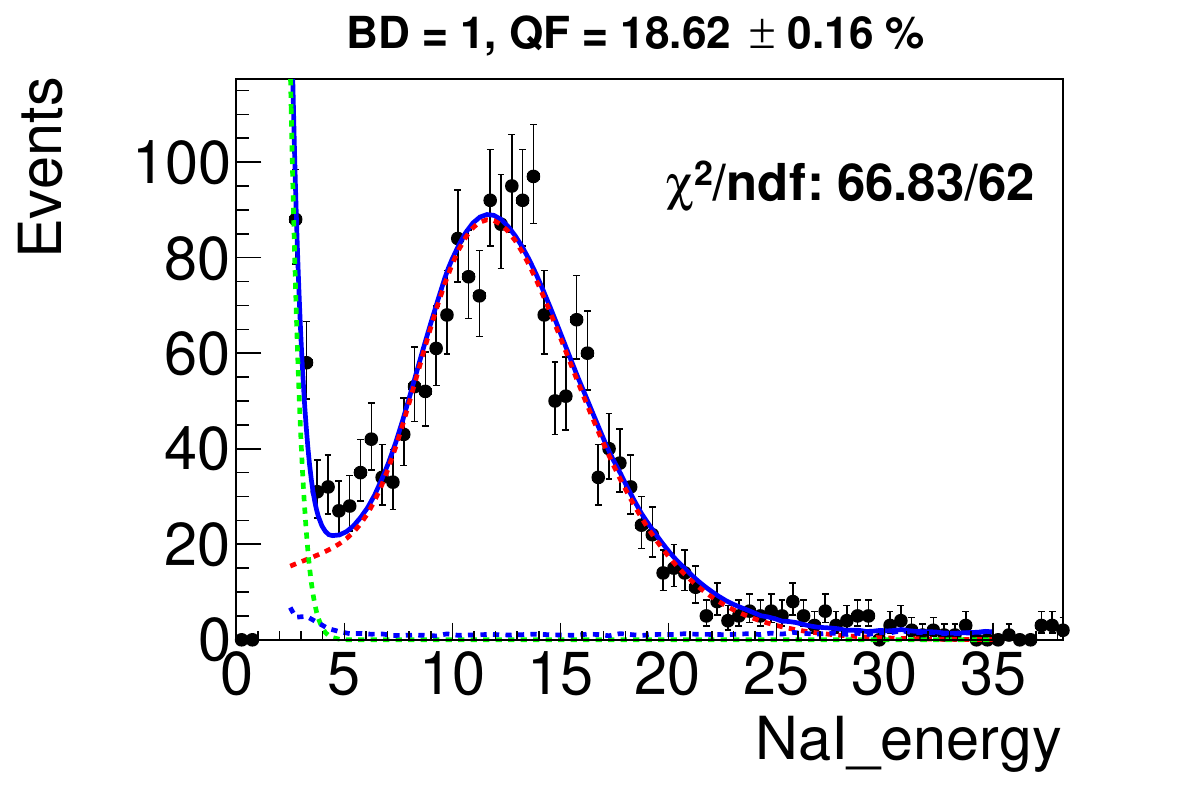}
\includegraphics[width=.3\textwidth]{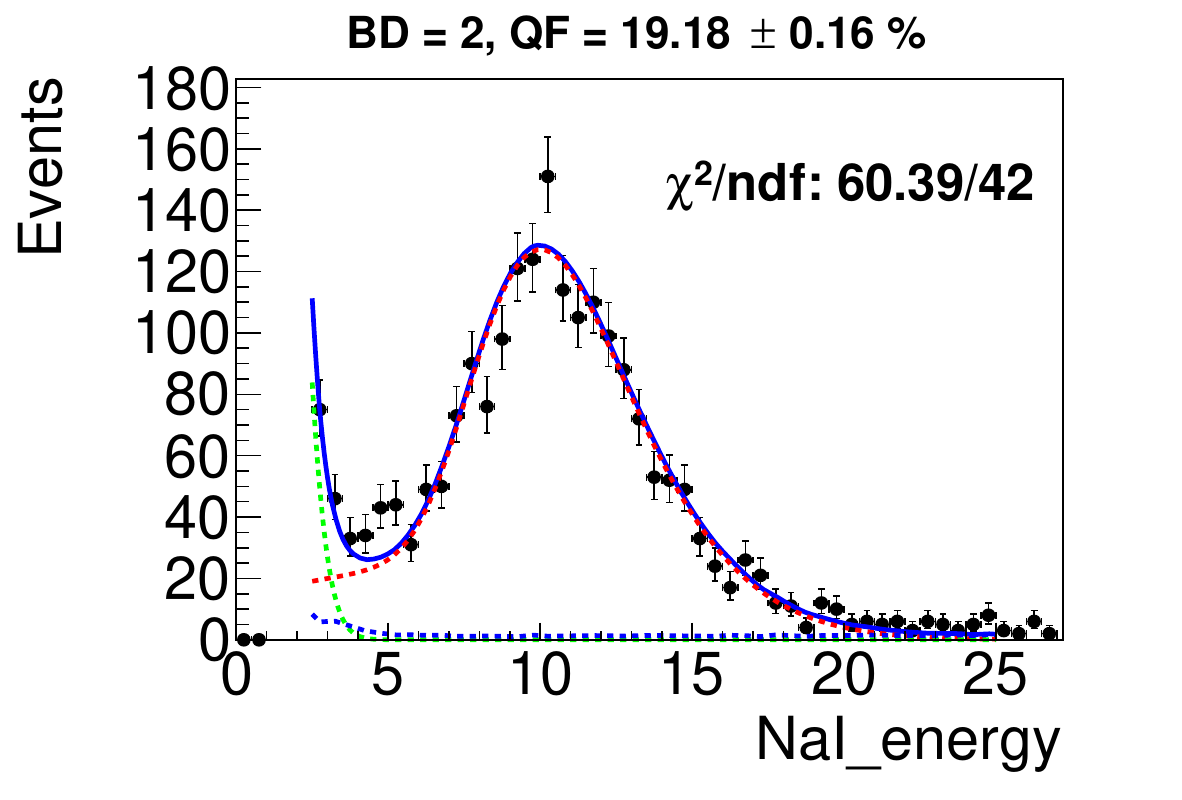}
\includegraphics[width=.3\textwidth]{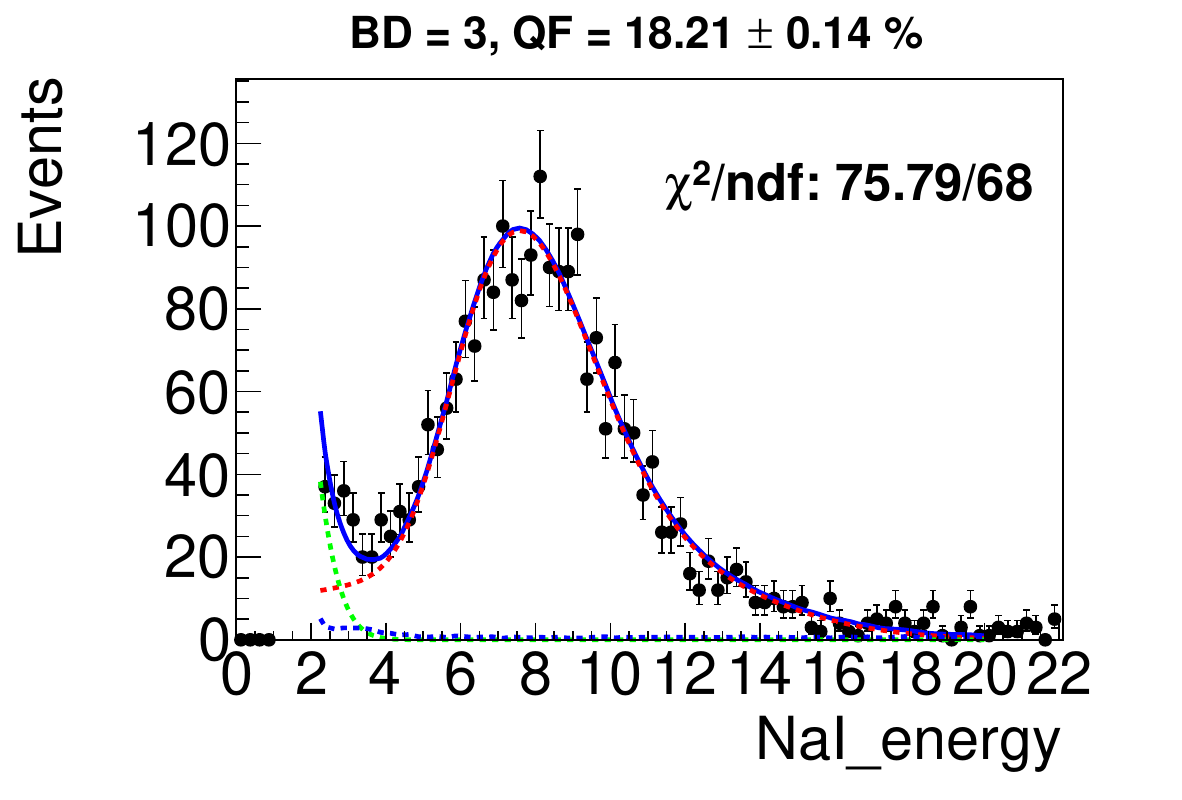}
\includegraphics[width=.3\textwidth]{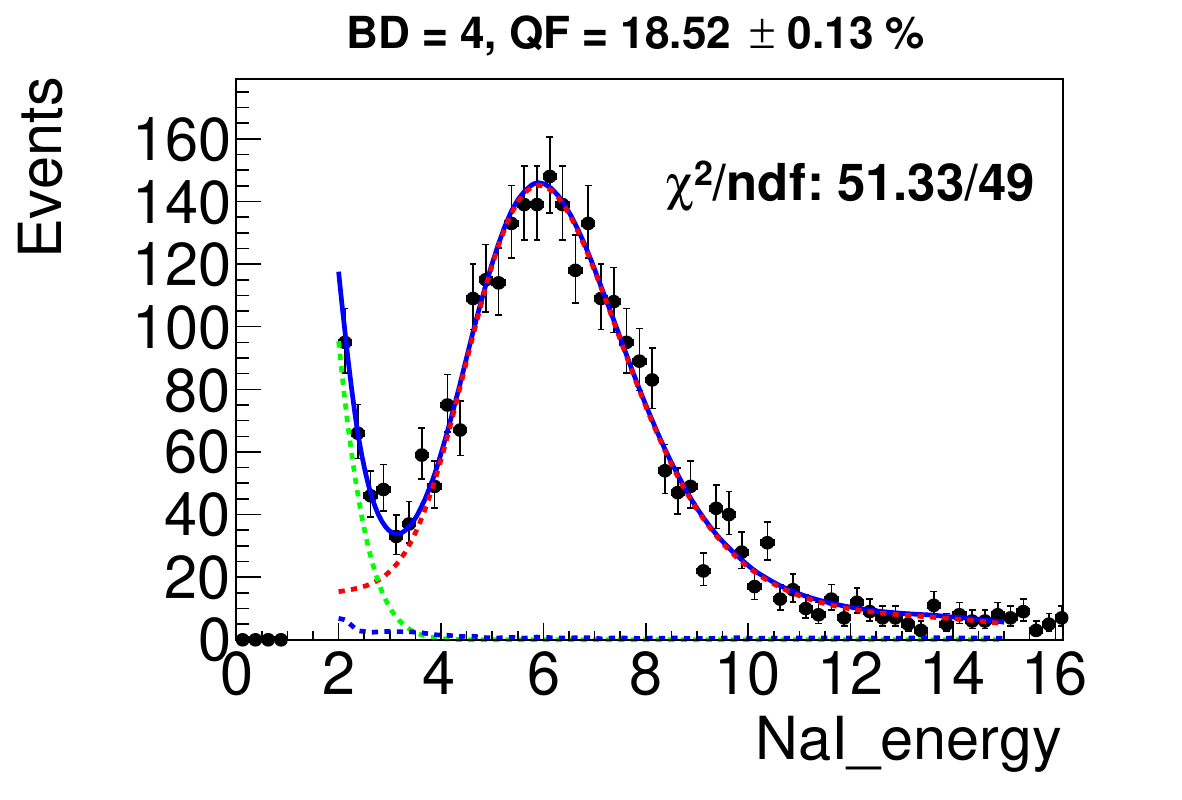}
\includegraphics[width=.3\textwidth]{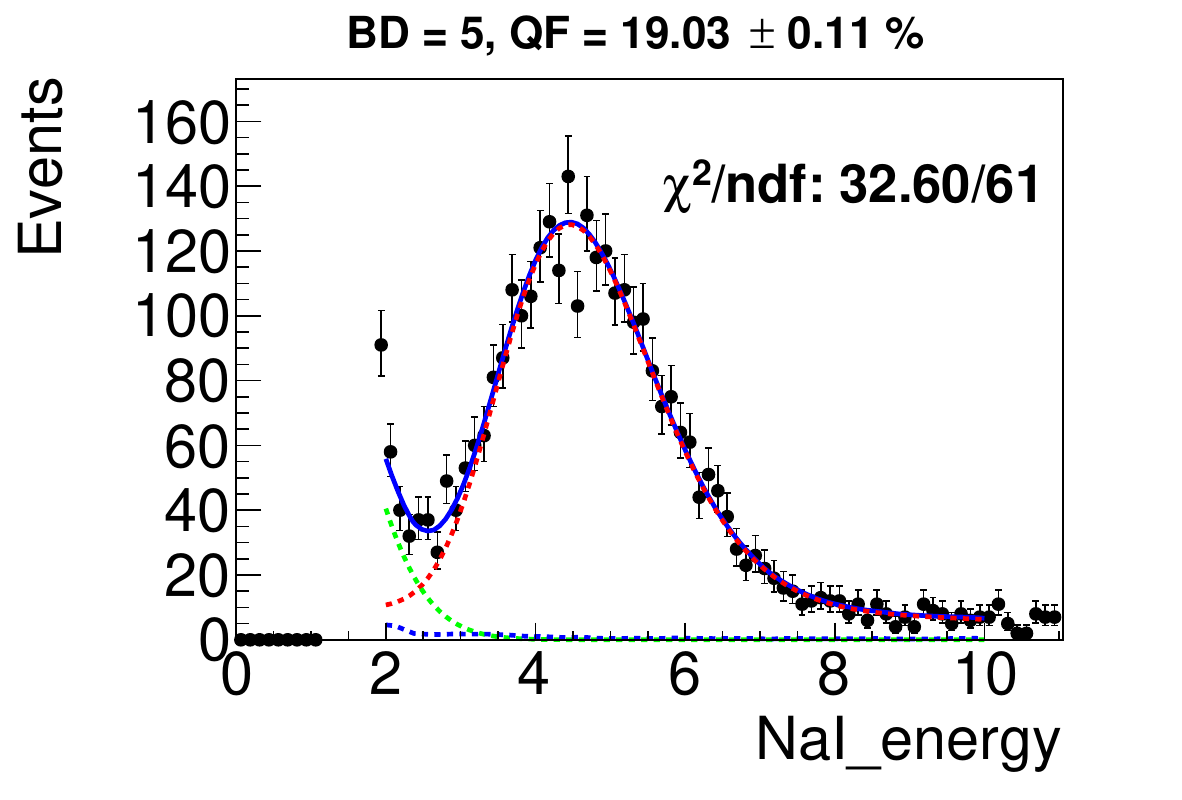}
\includegraphics[width=.3\textwidth]{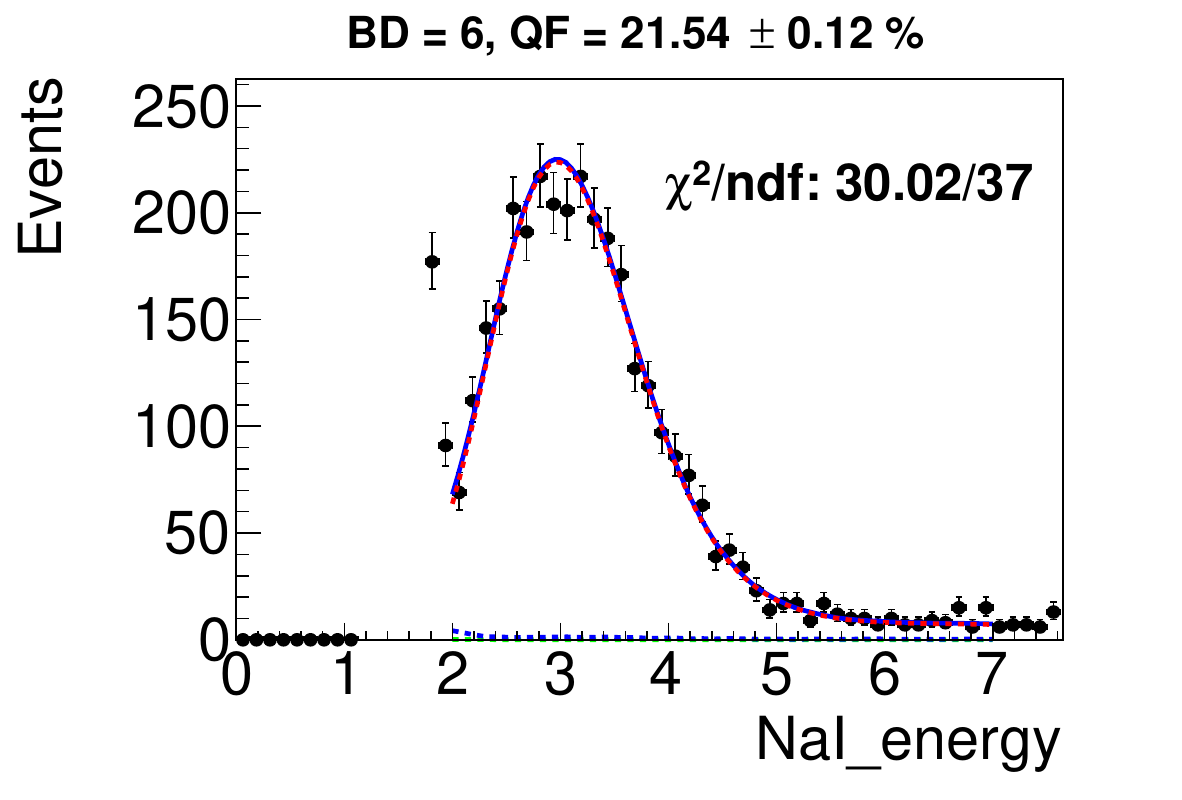}
\includegraphics[width=.3\textwidth]{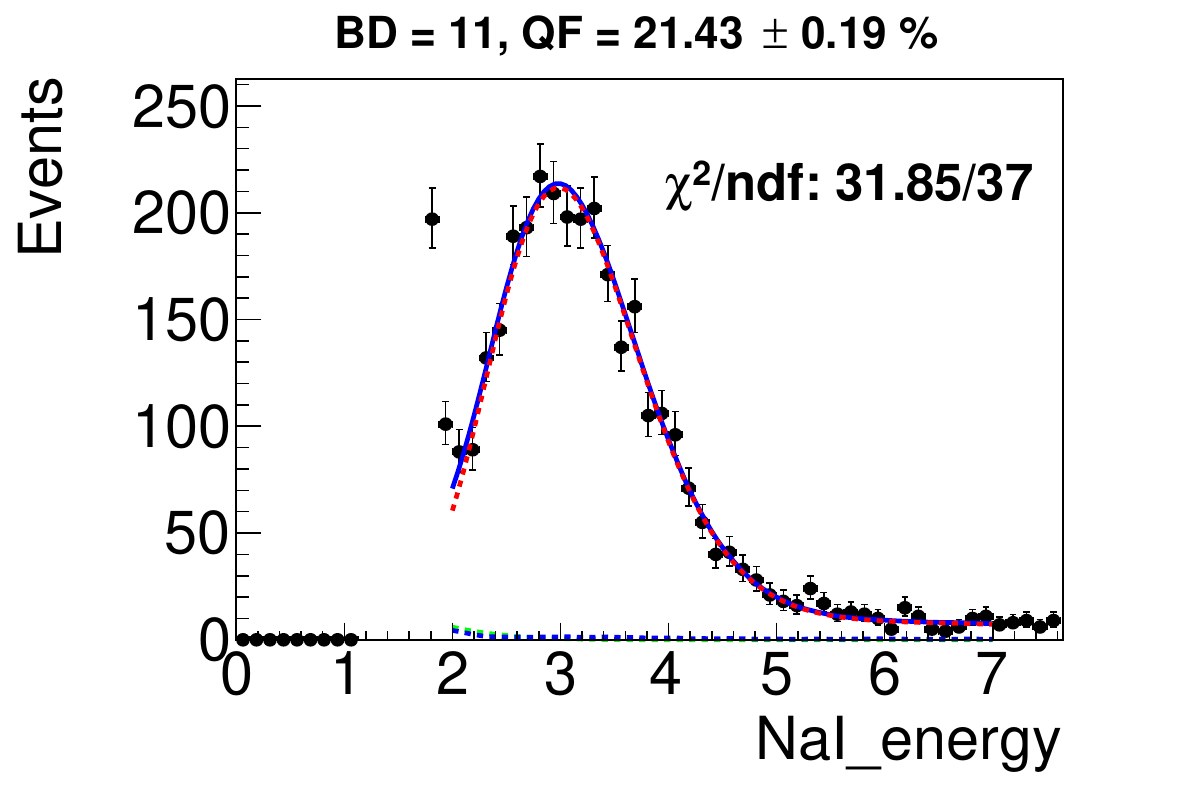}
\includegraphics[width=.3\textwidth]{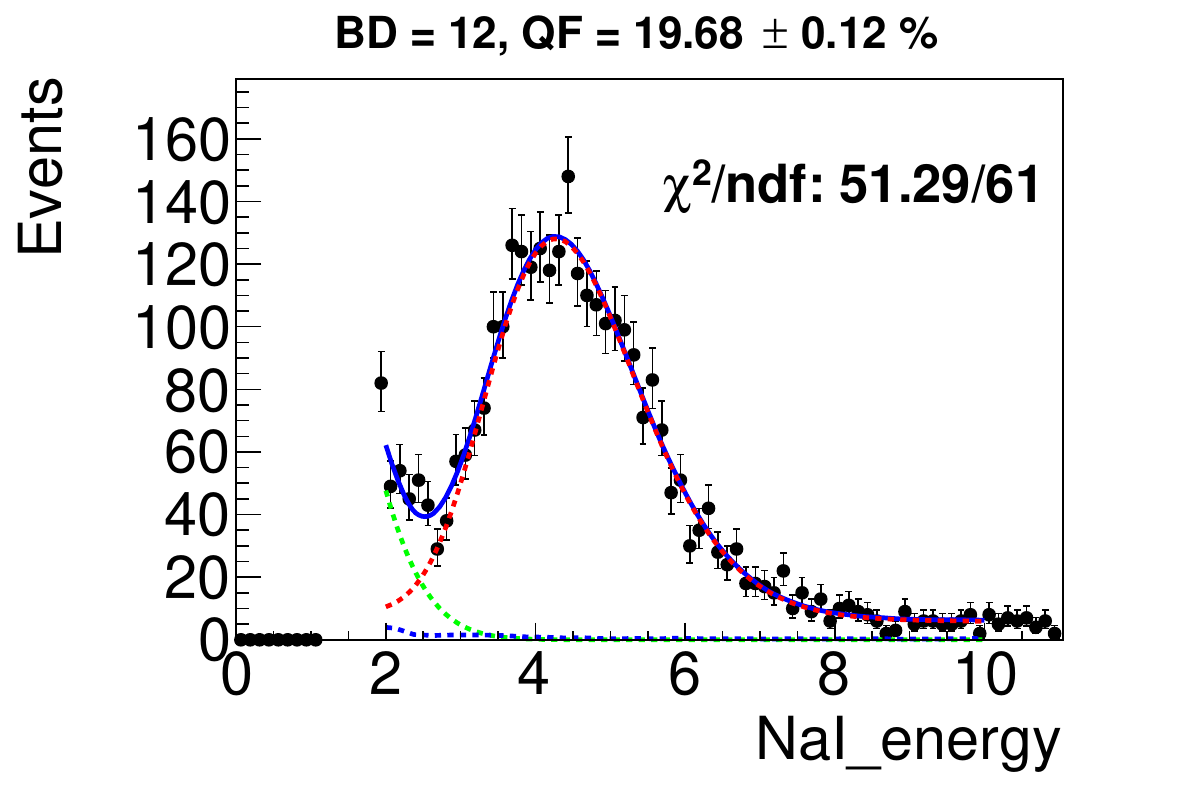}
\includegraphics[width=.3\textwidth]{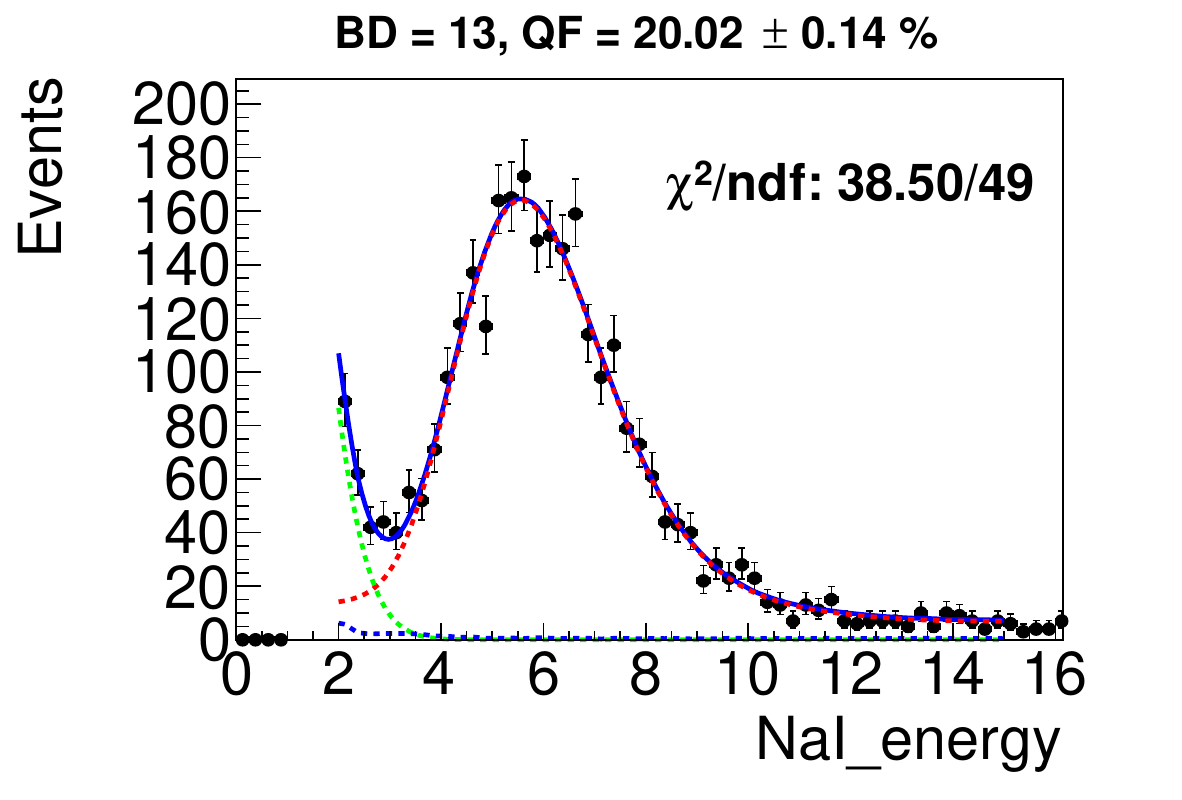}
\includegraphics[width=.3\textwidth]{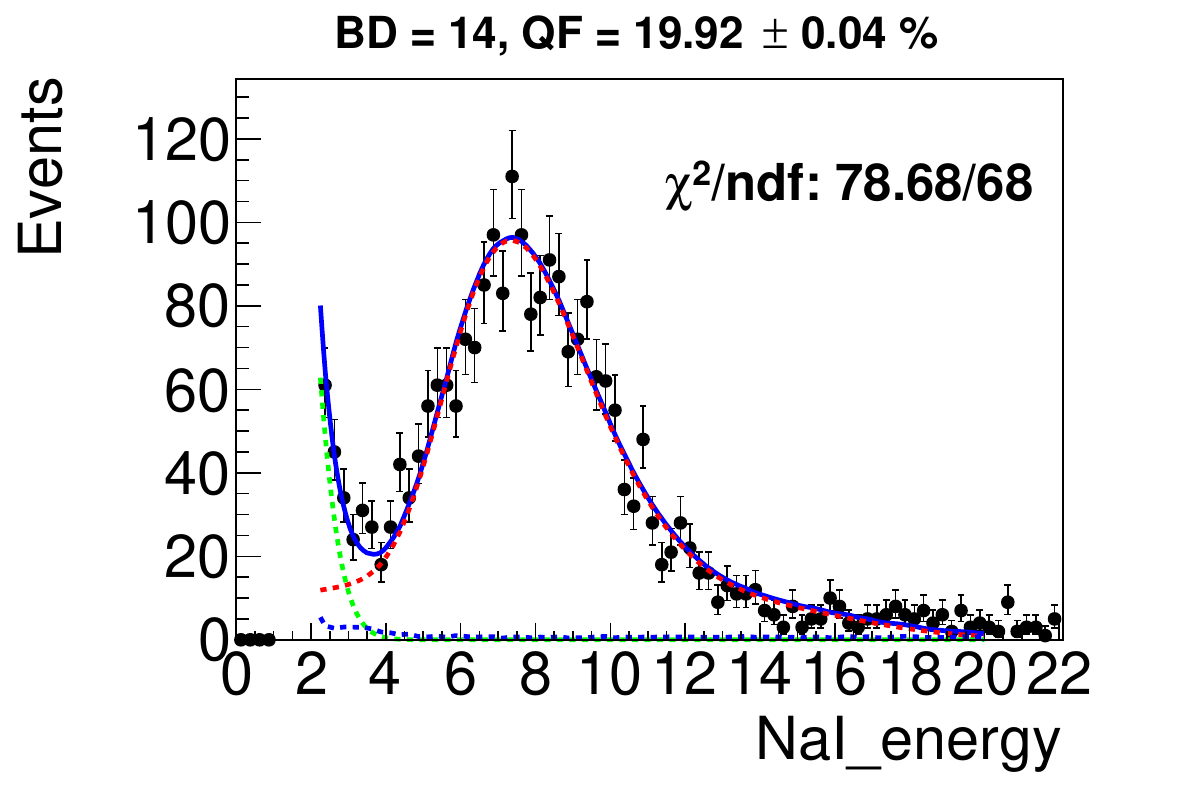}
\includegraphics[width=.3\textwidth]{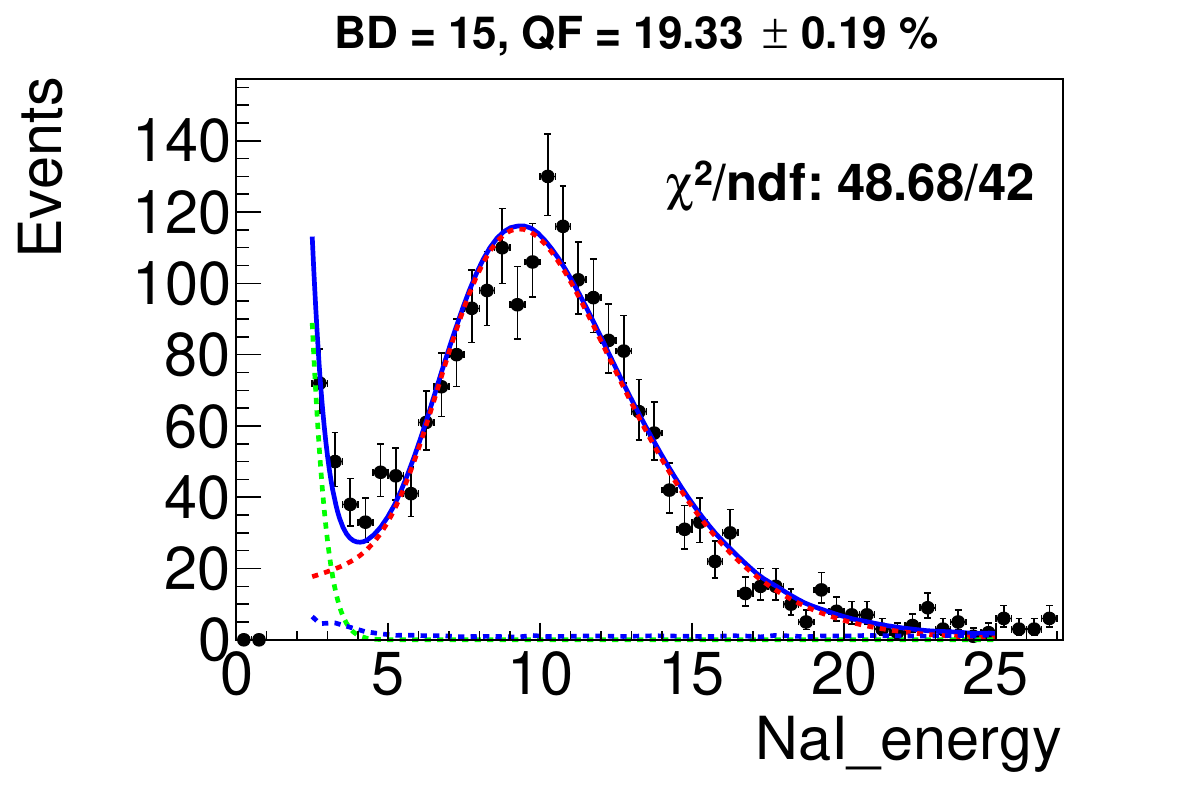}
\includegraphics[width=.3\textwidth]{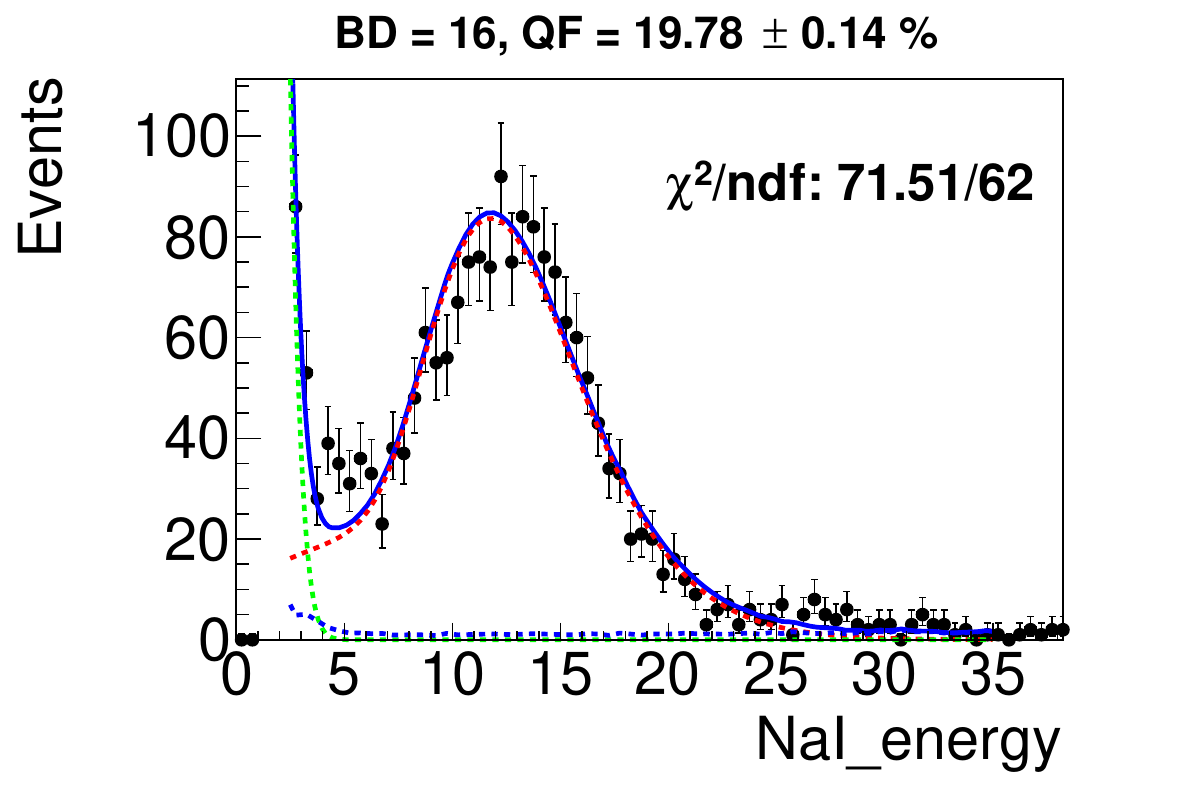}
\includegraphics[width=.3\textwidth]{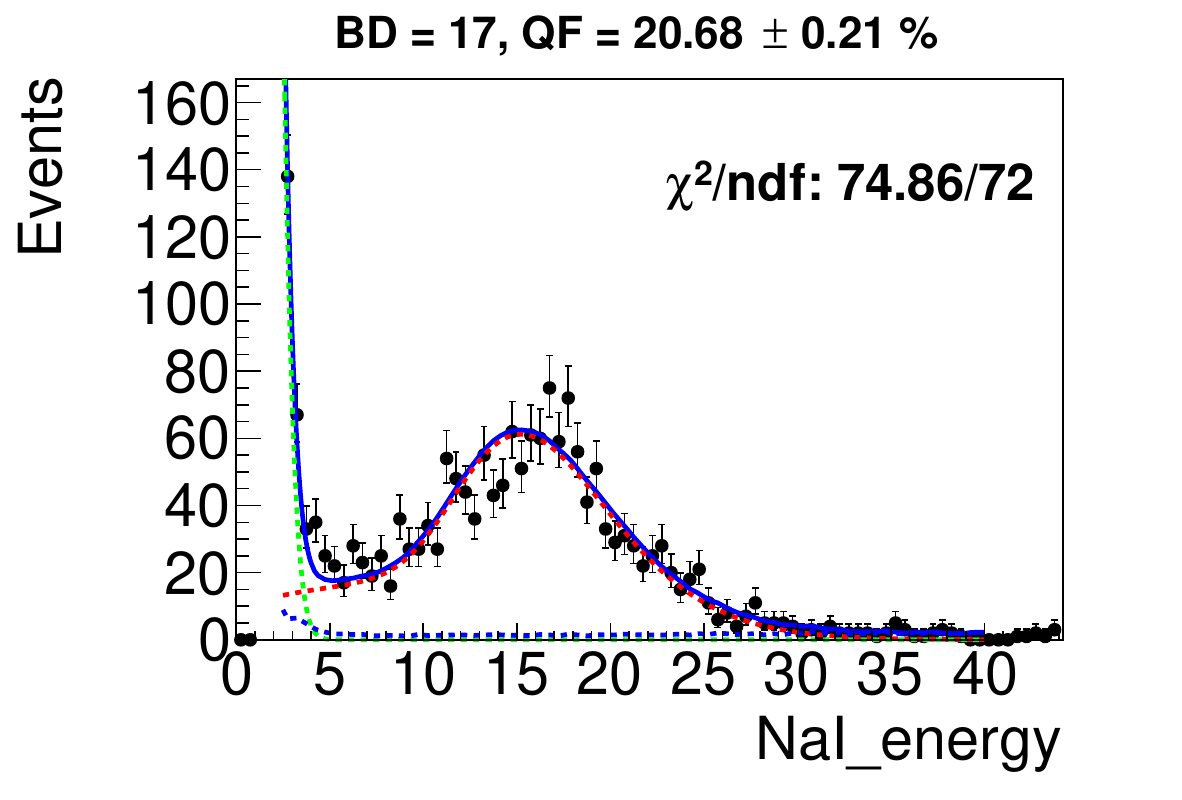}
\caption{\label{fig:Na_fit_crystal1} Results of the fits for crystal No.~1 using the PDF with the energy dependent resolution modelling and the non-proportional \isotope{Ba}{133} calibration (energy calibration method 2). 
} 
\end{figure*}
\begin{figure*}[htbp]
\centering 
\includegraphics[width=.3\textwidth]{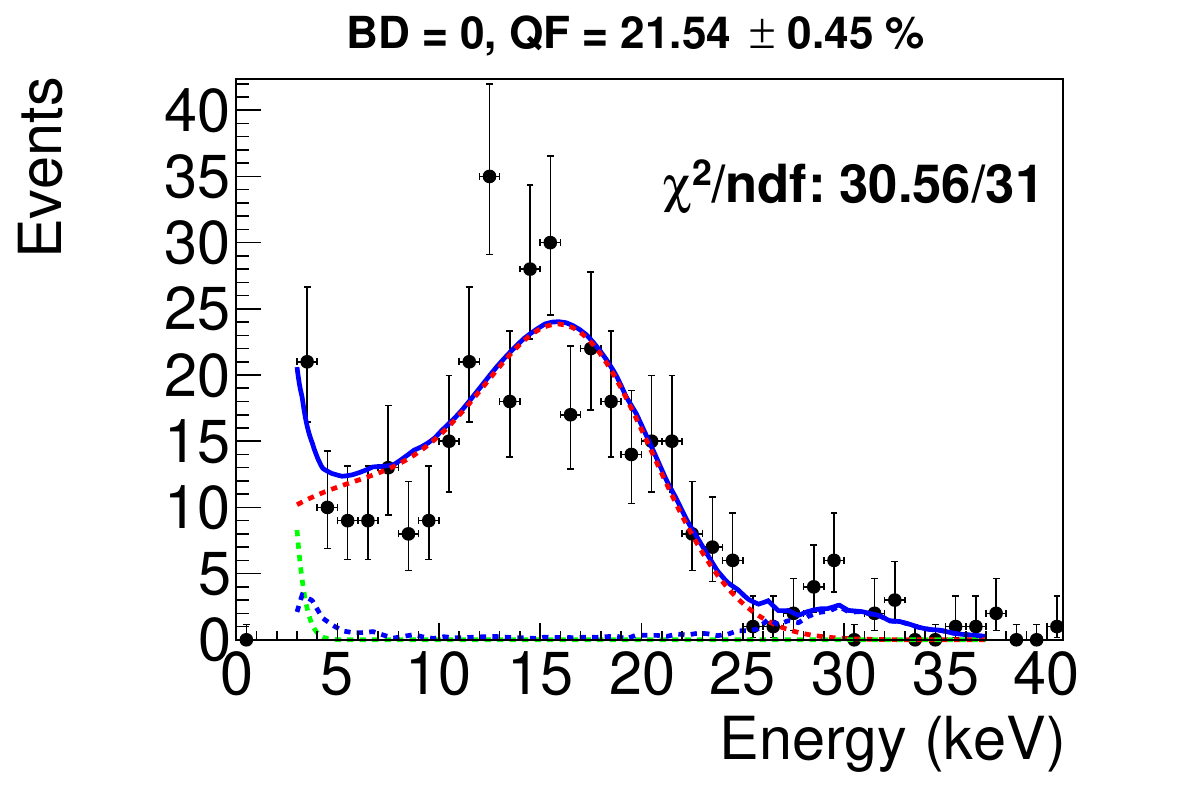}
\includegraphics[width=.3\textwidth]{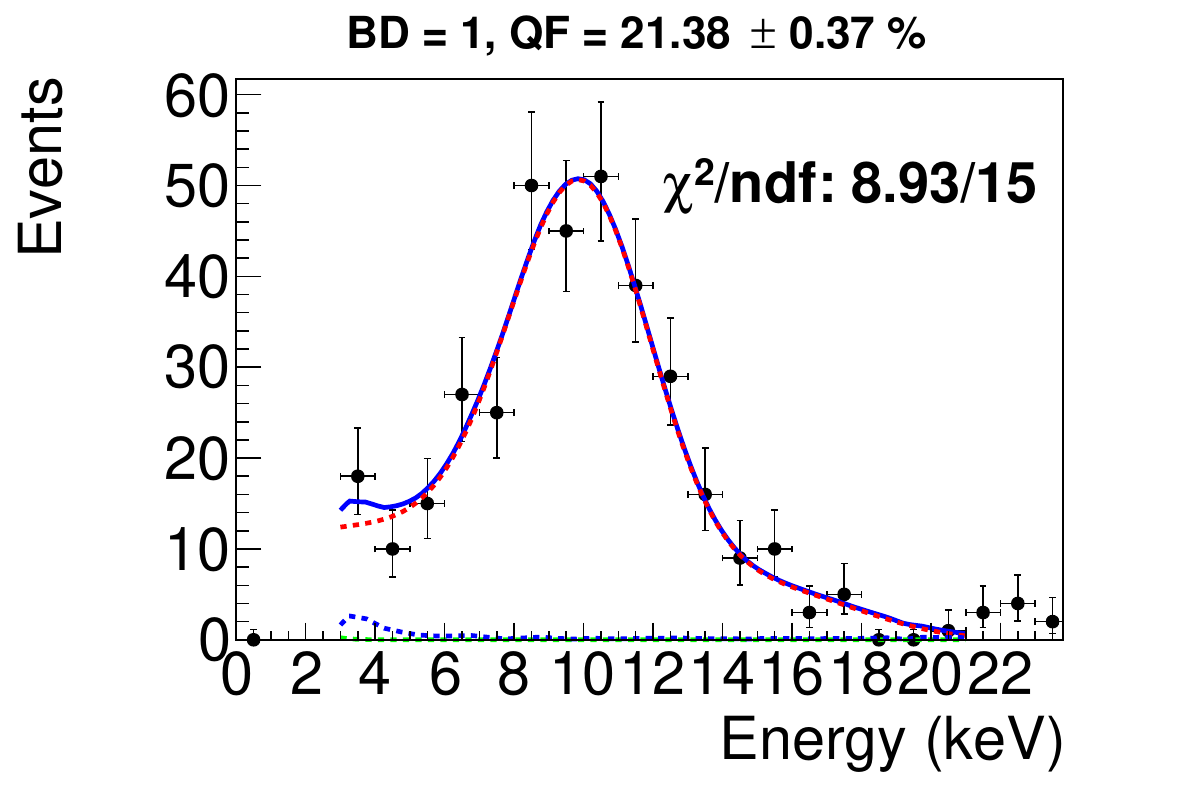}
\includegraphics[width=.3\textwidth]{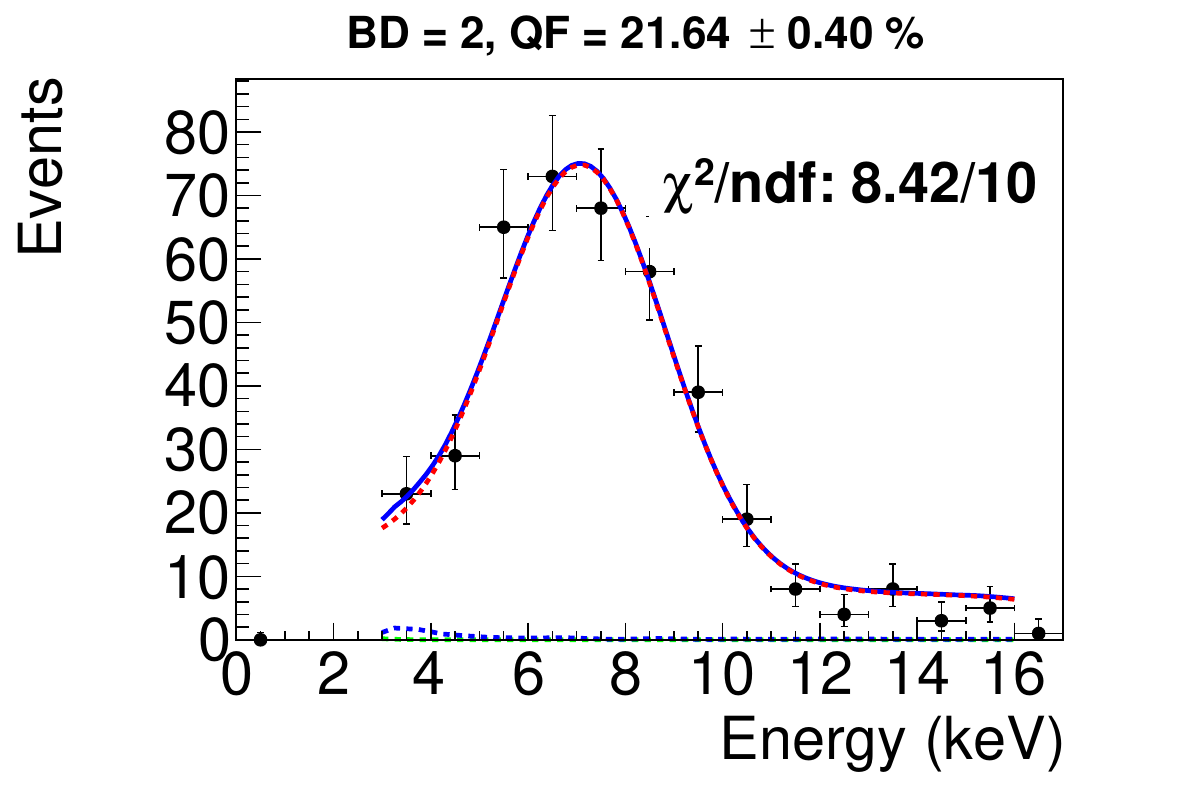}
\includegraphics[width=.3\textwidth]{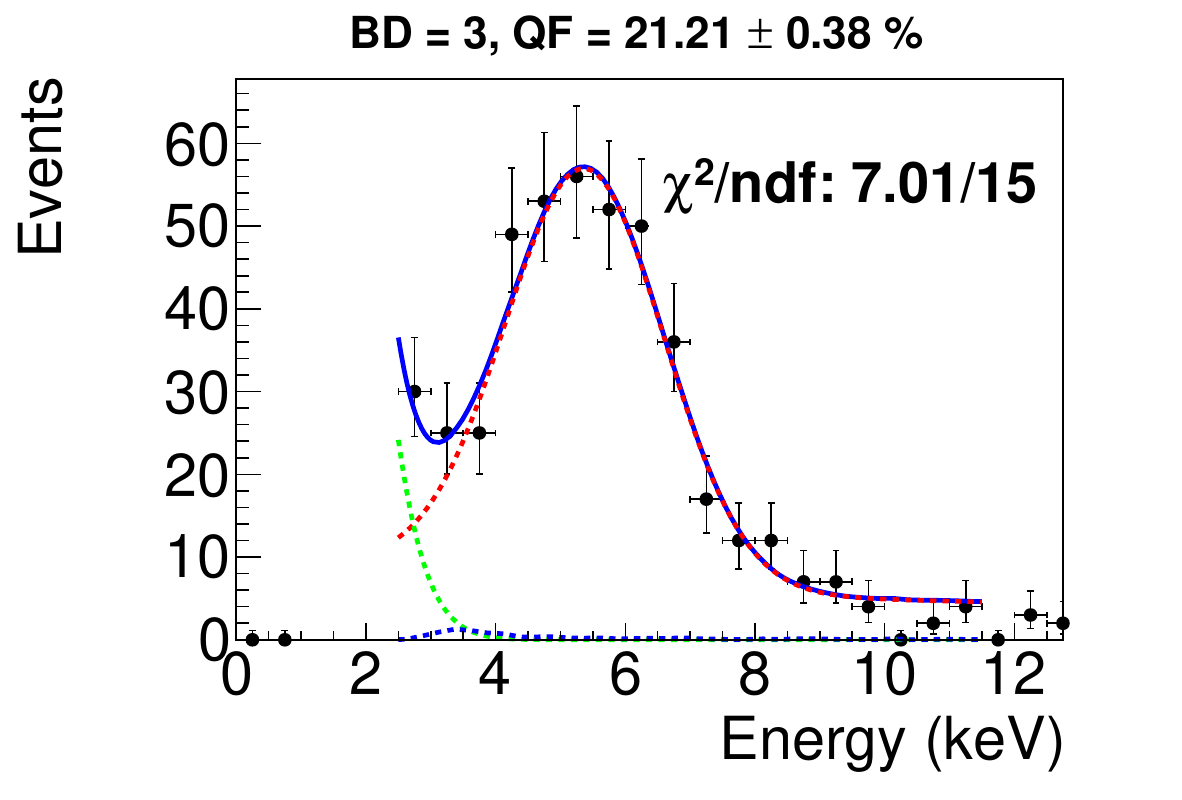}
\includegraphics[width=.3\textwidth]{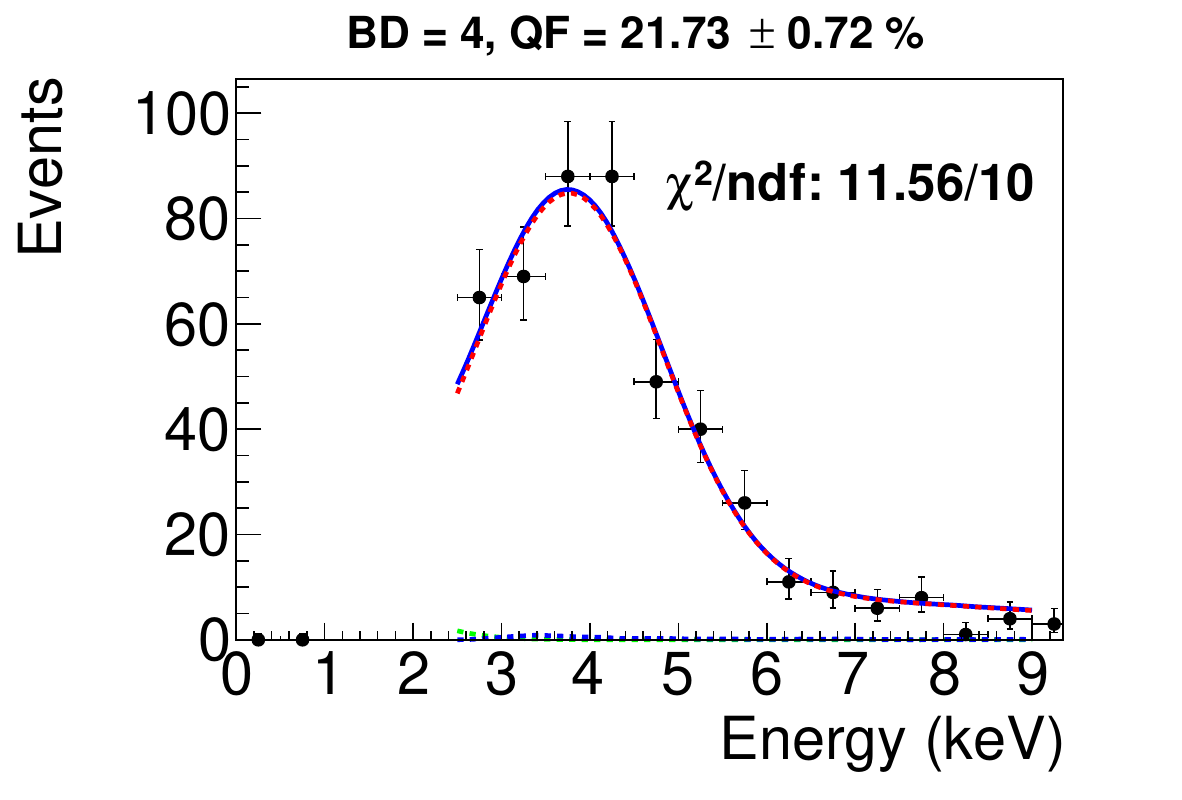}
\includegraphics[width=.3\textwidth]{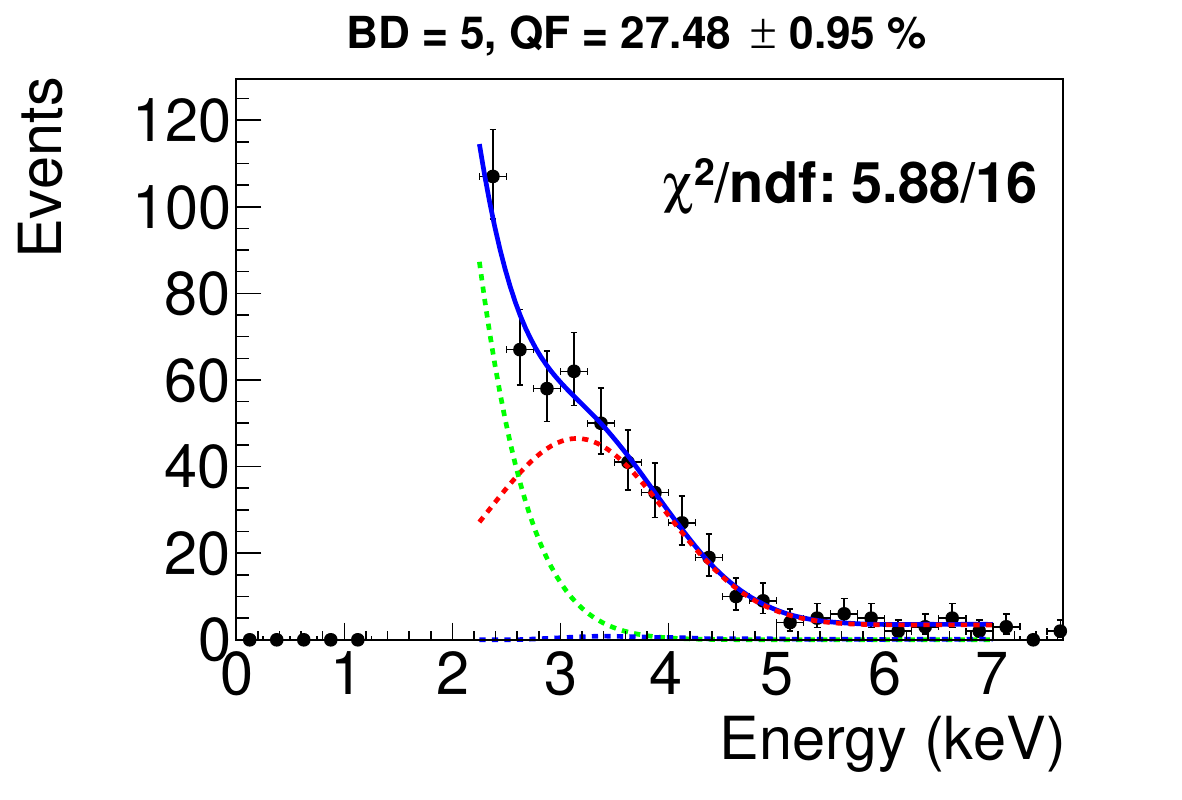}
\includegraphics[width=.3\textwidth]{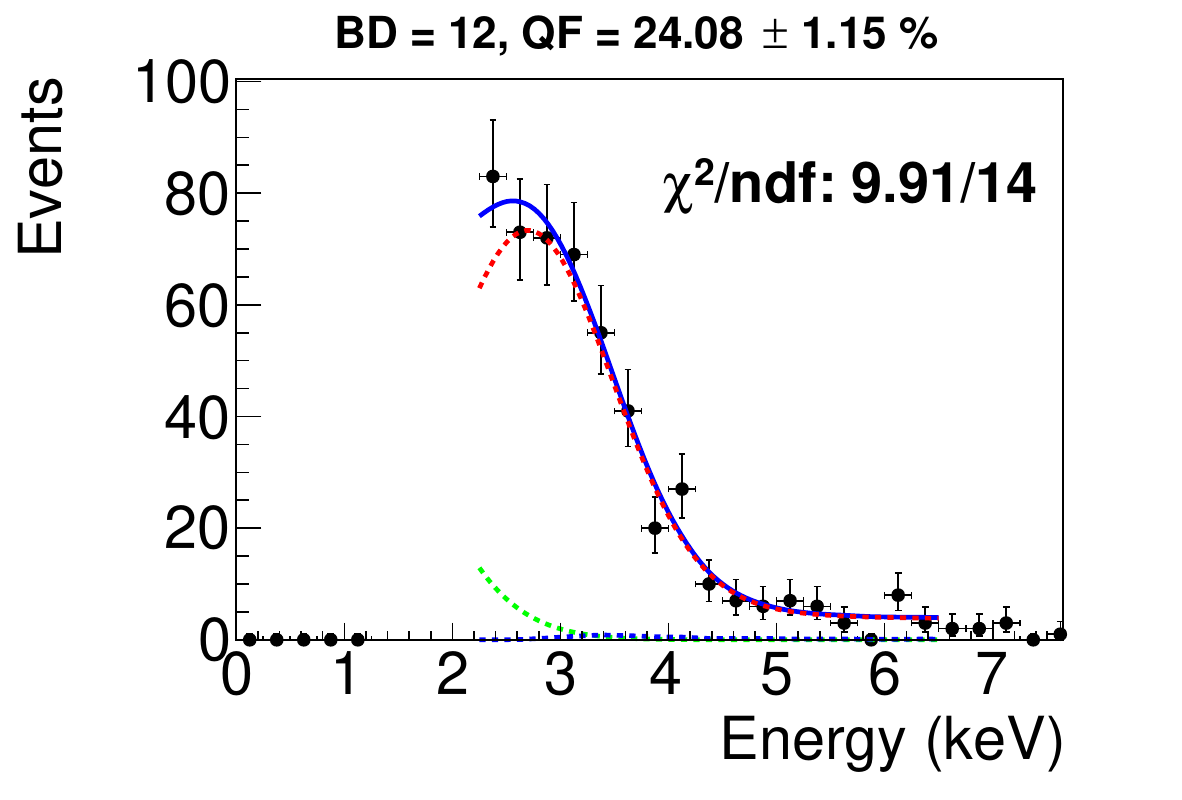}
\includegraphics[width=.3\textwidth]{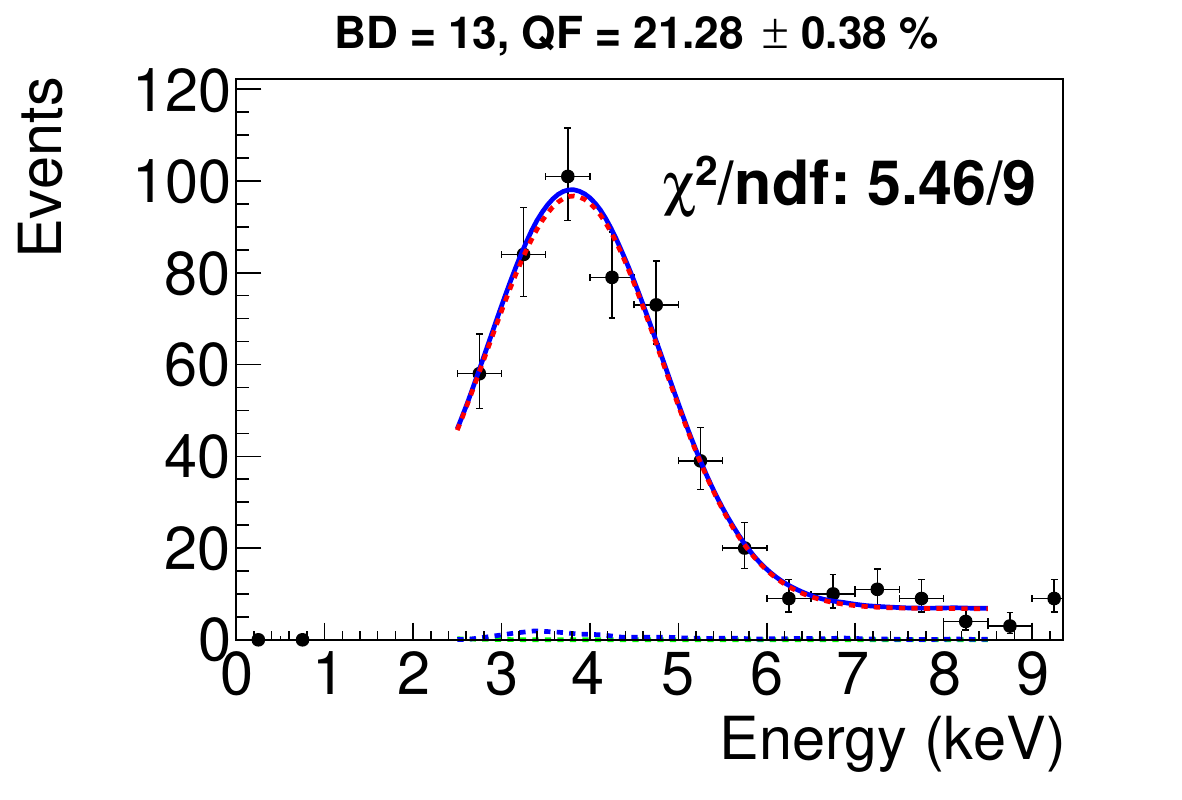}
\includegraphics[width=.3\textwidth]{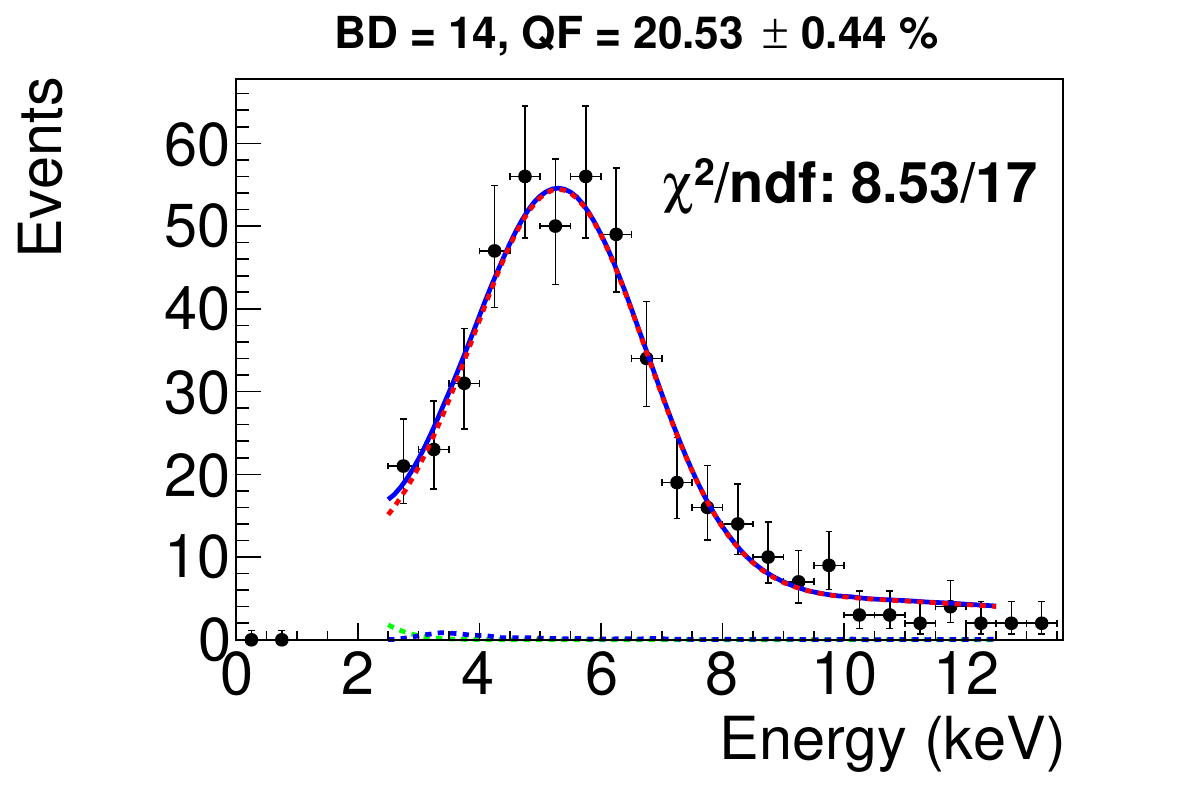}
\includegraphics[width=.3\textwidth]{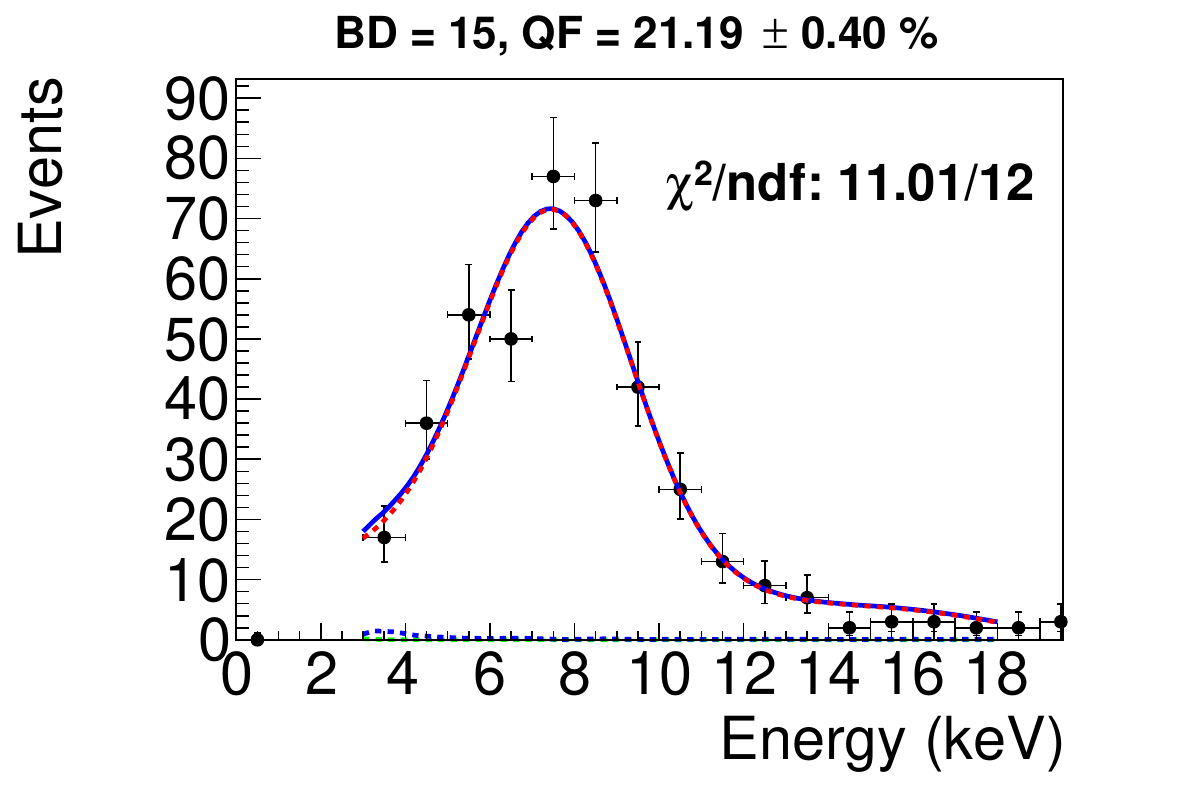}
\includegraphics[width=.3\textwidth]{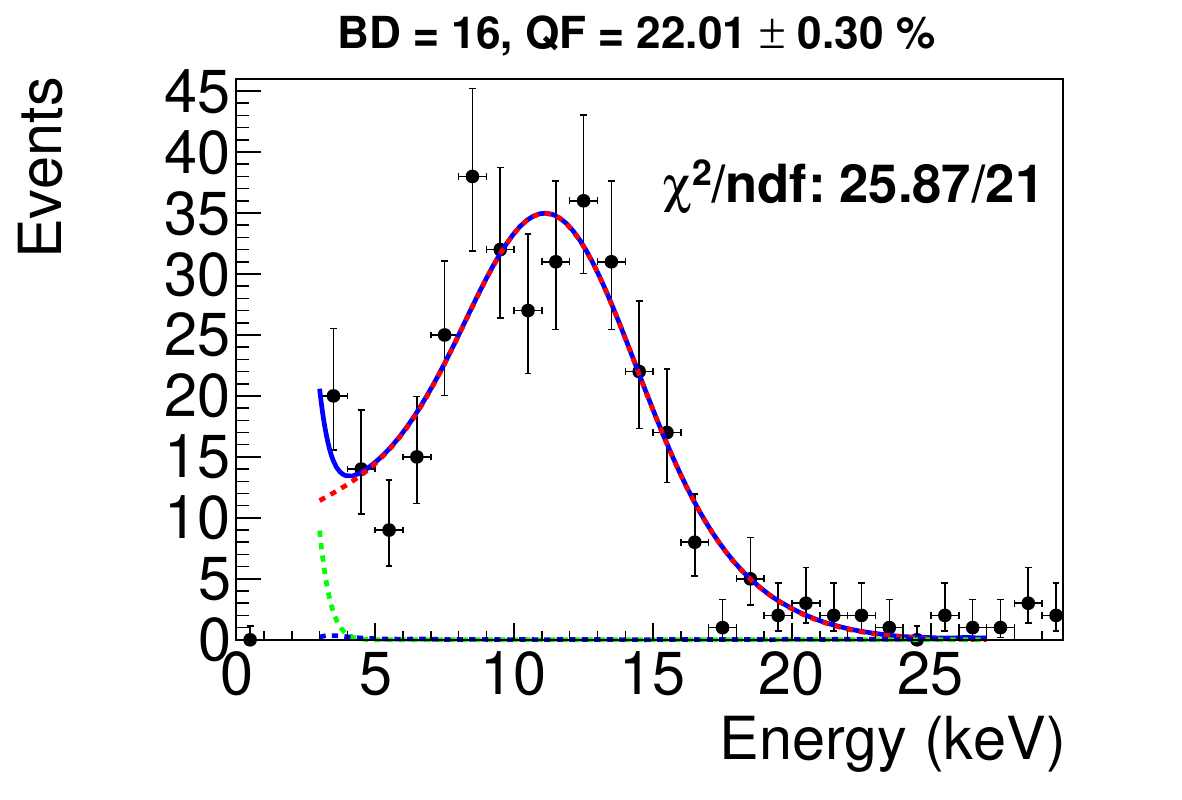}
\includegraphics[width=.3\textwidth]{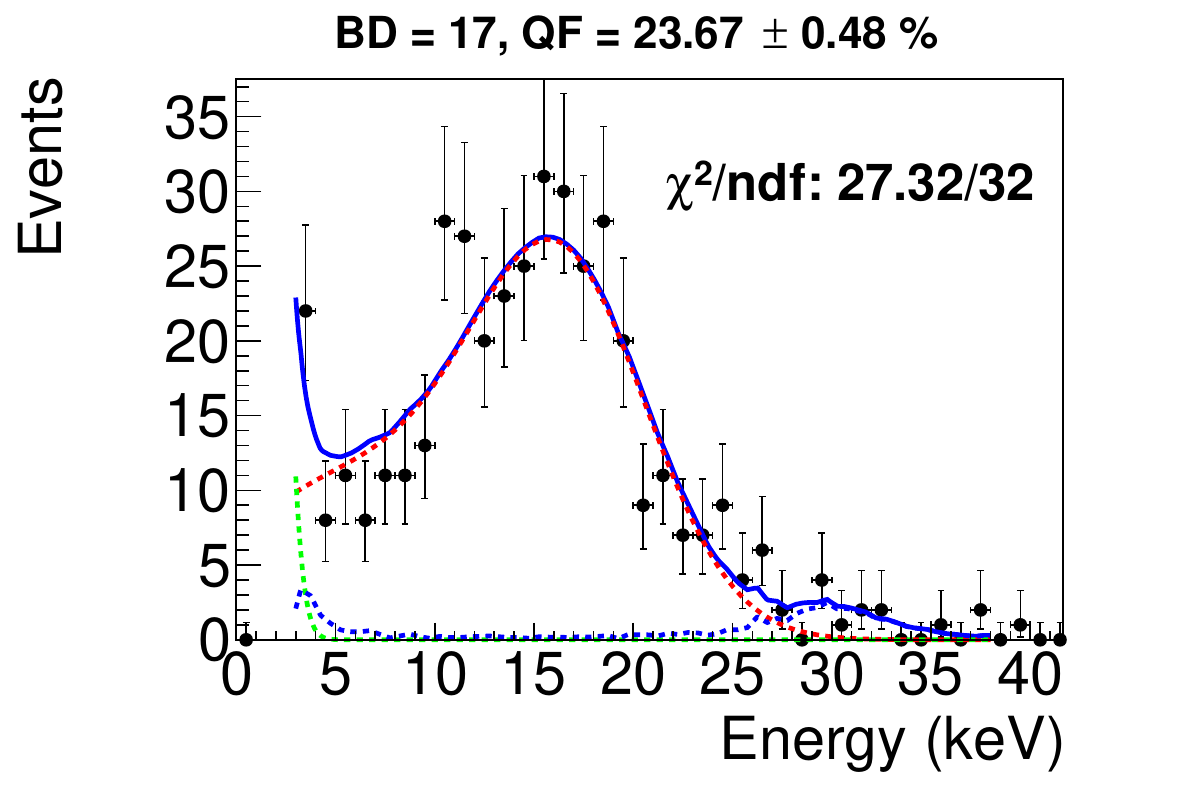}
\caption{\label{fig:Na_fit_crystal5} Results of the fits for crystal No.~5 using the PDF with the energy dependent resolution modelling and the non-proportional \isotope{Ba}{133} calibration (energy calibration method 2). 
} 
\end{figure*}

Figures \ref{fig:Na_fit_crystal1} and \ref{fig:Na_fit_crystal5} show the results of the fits for crystals No.~1 and No.~5, using the previously explained fitting protocol with the energy dependent resolution modelling and the non-proportional \isotope{Ba}{133} calibration (energy calibration method 2). It can be observed in Figure~\ref{fig:Na_fit_crystal5} that because of the reduced light collection in crystal No.~5 (and similarly in crystal No.~4) the Na-NR peak in BDs 5 and 12 could not be disentangled from the I-NR signal and noise peak. These BD channels will not be shown in the QF results presented below. 
\par

\begin{figure}[htbp]
\centering 
\includegraphics[width=.5\textwidth]{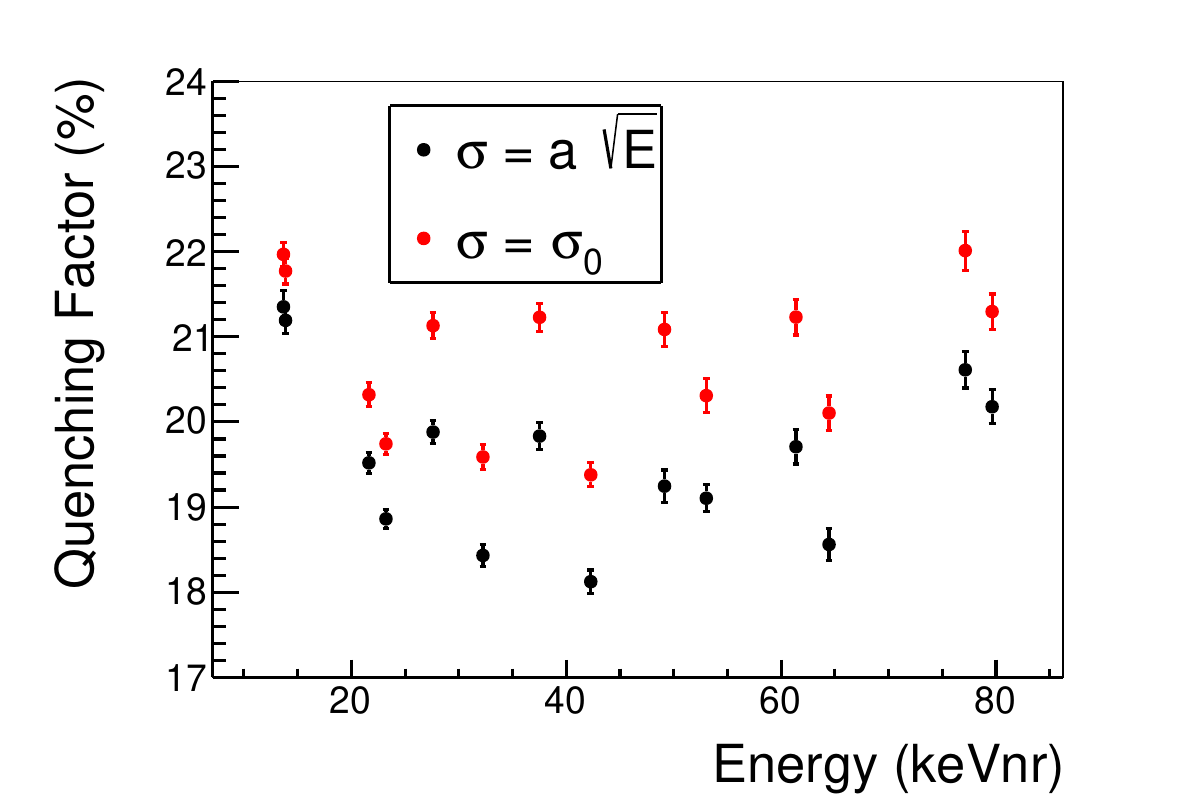}
\caption{\label{fig:QFresolution} Results obtained for the QF$\mathrm{_{Na}}$ using crystal No.~1 data for the two different energy resolution models considered. The QF$\mathrm{_{Na}}$ derived as result of this work is calculated as the average of both, and half the difference is taken as one of the contributions to the systematic uncertainty of the result. } 
\end{figure}

Because the values obtained using the two energy resolution models were not compatible with each other (as Figure \ref{fig:QFresolution} shows for crystal No.~1), and in fact, systematically lower QF were obtained for the constant energy resolution case, the QF derived from this work, and listed in Section \ref{sec:results}, have been calculated as the average of both, and half the difference taken as associated systematic uncertainty.
\par

Apart from this, other systematics contributing to the QF estimate have been analyzed: the uncertainty
in the positions of the components of the experiment (source, crystal and BDs), the uncertainty in the electron equivalent energy calibration and the value selected for the QF$\mathrm{_{I}}$ and the resolution considered for the iodine recoils. For the estimate of the first contribution, simulations with the neutron source, NaI(Tl) crystal and BDs displaced for their nominal positions within their corresponding uncertainties were carried out. The results of these simulations for each crystal and channel were used in a fit similar to the one previously explained, and the corresponding QF values were obtained. 
\par
For the second contribution, two fits were performed calibrating the spectra using energy calibration functions obtained by modifying the calibration parameters within one standard deviation. 
\par
For the third contribution, fits were done fixing the QF$\mathrm{_{I}}$ to 1\% and to 9\% and the resolution applied to the iodine recoil energy distribution to 0.8 keV and to 1.2~keV. 
\par
The corresponding systematic uncertainties of each contribution were calculated as the difference between the QF values obtained in these fits and those obtained in the original situation, for each
channel. All of these fits were done applying both resolution functions and fixing
the resolution parameters to that obtained in the original fits. The systematic uncertainties associated to the selection of the QF and the resolution of the iodine recoils were from one to two orders of magnitude lower than the statistical uncertainty, and therefore they were not considered in the error propagation. The uncertainties of the other two contributions were found to be compatible
for both resolution functions applied, and therefore the maximum of them was considered in the error calculation for each channel. They were also computed as symmetrical by considering the total uncertainties of the contribution as the maximum between the upper and lower errors. Finally, the three systematic uncertainties were combined with the statistical contribution to obtain the total uncertainty in each calculated QF$\mathrm{_{Na}}$. 
\par
It is worth to remind that all the analysis was carried out in parallel for the three calibration strategies followed to convert nuclear recoil energies into electron equivalent energies. In all the cases the contribution from the different uncertainties are similar: statistical contribution is at the level of 0.1\% while the total uncertainty is closed (but below) 1\%. 

\subsubsection{Iodine Quenching Factor}
The iodine recoils could not be disentangled from the background for any channel and crystal, so a different strategy for the estimate of the QF$\mathrm{_{I}}$ was followed. This was performed by studying the inelastic peak from \isotope{I}{127}, which corresponds to the sum of the light produced by the energy depositions of the gamma (57.6 keV) and the iodine recoil, the latter quenched by the corresponding QF$\mathrm{_{I}}$. In this case the ROI is centered around the 57.6~keV peak, therefore the proportional linear energy calibration (energy calibration method 1, see Section \ref{sec:Ecalibration}) was applied to the data. 
\par
From the simulation, channels 8 and 9 have iodine recoil energies below 0.2~keV, that result in a negligible shift in electron equivalent energy, and therefore the combination of both channels was used to build a reference for each crystal. On the other hand, channels 0 and 17, corresponding to the highest iodine recoil energies above 10~keV, were used for this analysis. After correcting the gain drift (as explained in Section~\ref{sec:Ecalibration}) and calibrating in energy as commented, the inelastic peak is fitted to a Gaussian to obtain the position of the peak and the corresponding statistical uncertainty. The difference between the mean energy obtained from the fit for each channel and that from the reference (referred to as $\Delta$) was calculated. The systematic uncertainty was estimated by changing the reference to channel 8 alone, and then calculating the difference between the corresponding $\Delta$ values as uncertainty (systematic error 1). Moreover, channels 0 and 17 have a similar recoil energy in some of the crystals. In those cases, an additional systematic uncertainty was also estimated from the difference between the $\Delta$ values derived for both channels (systematic error 2). 
\par
\begin{figure}[htbp]
\centering 
\includegraphics[width=.5\textwidth]{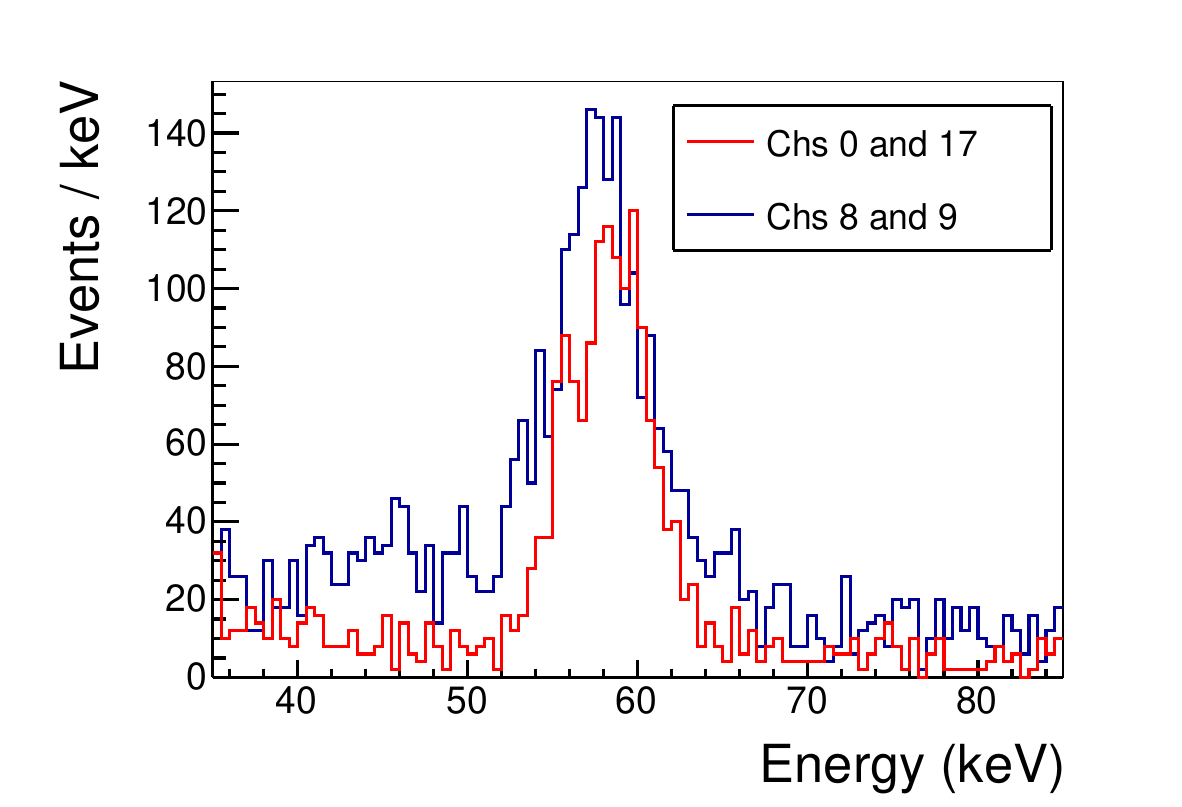}
\caption{\label{fig:DeltaE} Comparison between the \isotope{I}{127} inelastic peak for the reference channels (with negligible nuclear recoil energy, 8 and 9), blue line, and the analysed channels, red line (with the maximum nuclear recoil energy available added, 0 and 17) for crystal No.~3. 
} 
\end{figure}

Results for the peak shift $\Delta$ including statistical and systematic uncertainties at 1$\sigma$ are presented in Table~\ref{tab:Delta_I} for crystals No.~2 and No.~3. The comparison between the \isotope{I}{127} inelastic peak for the reference channels (8 or 9) and the analysed channels (0 or 17) for the measurements with the crystal No.~3 is shown in Figure \ref{fig:DeltaE}. A difference in the mean energy of the distributions (lower than 1 keV) is observed, which allows to determine the corresponding QF$\mathrm{_{I}}$ for the recoil energy of those channels, which is about 14~keV.

\begin{table}[htbp]
\centering 
\begin{tabular}{|c|c|c|c|c|c|}
\hline
  Crystal & $\Delta$  & \multicolumn{4}{c|}{Uncertainties (keV)} \\
  number  & (keV) & Stat.& Sys. 1 & Sys. 2 & Total \\
\hline
2 & 0.73  & 0.24 & 0.29 & 0.01 & 0.38\\
3 & 0.93  & 0.26 & 0.01 & 002  & 0.26\\
\hline
\end{tabular}
\caption{\label{tab:Delta_I} \isotope{I}{127} peak shift values ($\Delta$) between reference and analysed channels obtained for crystals No.~2 and No.~3, together with their statistical and systematic uncertainties. } 
\end{table}

\section{Results \& Discussion}
\label{sec:results}

\subsection{Sodium Quenching Factor Results and Comparison to Prior Measurements}
\begin{figure}[htbp]
\centering 
\includegraphics[width=.5\textwidth]{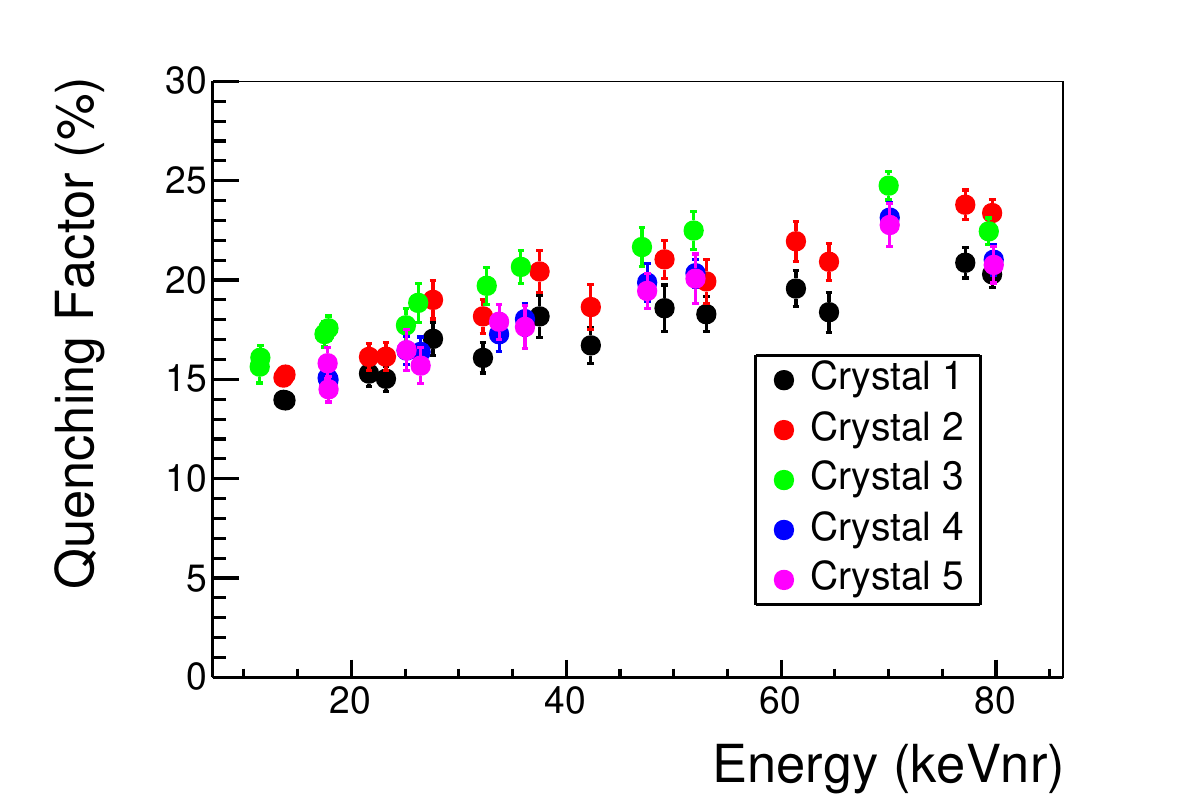}
\caption{\label{fig:Na_QF_results1} Sodium QF results for the five crystals measured using the first calibration method (proportional using as reference the 57.6 keV line). Uncertainties shown include statistical and systematic contributions. 
} 
\end{figure}

\begin{figure}[htbp]
\centering 
\includegraphics[width=.5\textwidth]{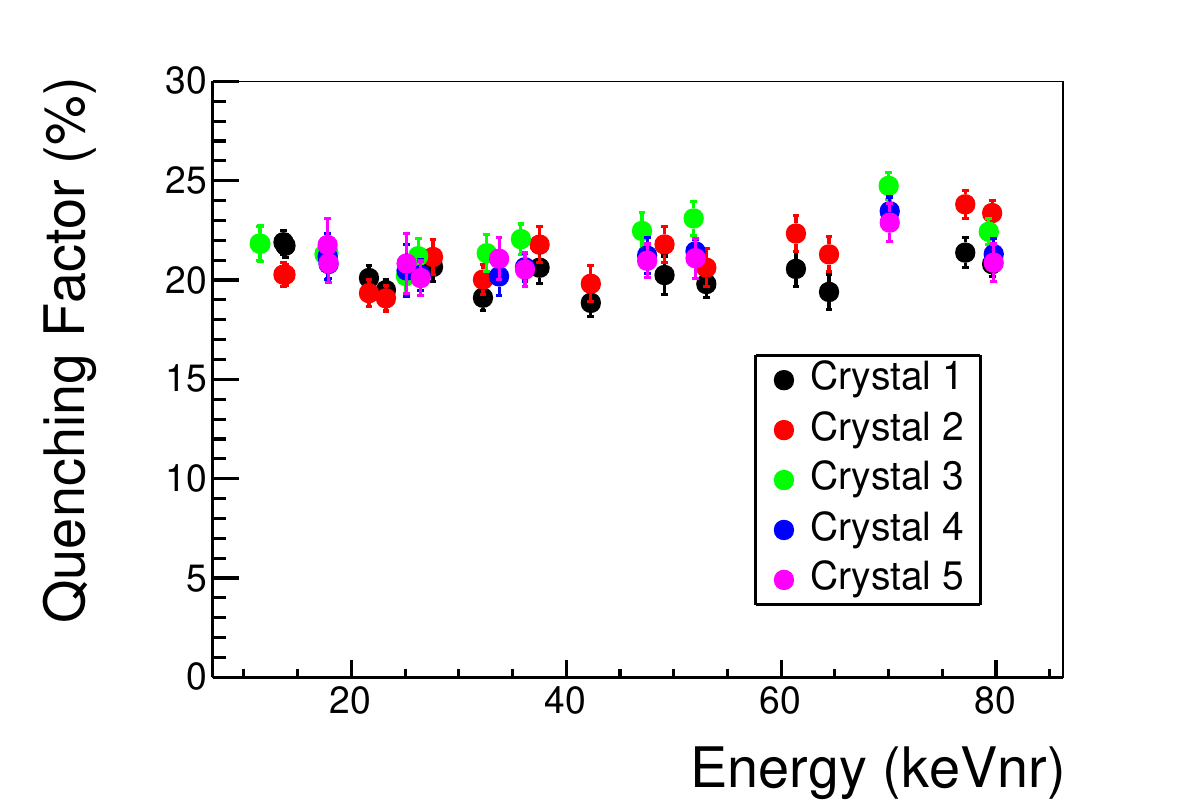}
\caption{\label{fig:Na_QF_results2} Sodium QF results for the five crystals measured using the second calibration method (linear using the \isotope{Ba}{133} lines). Uncertainties shown include statistical and systematic contributions. 
} 
\end{figure}

\begin{figure}[htbp]
\centering 
\includegraphics[width=.5\textwidth]{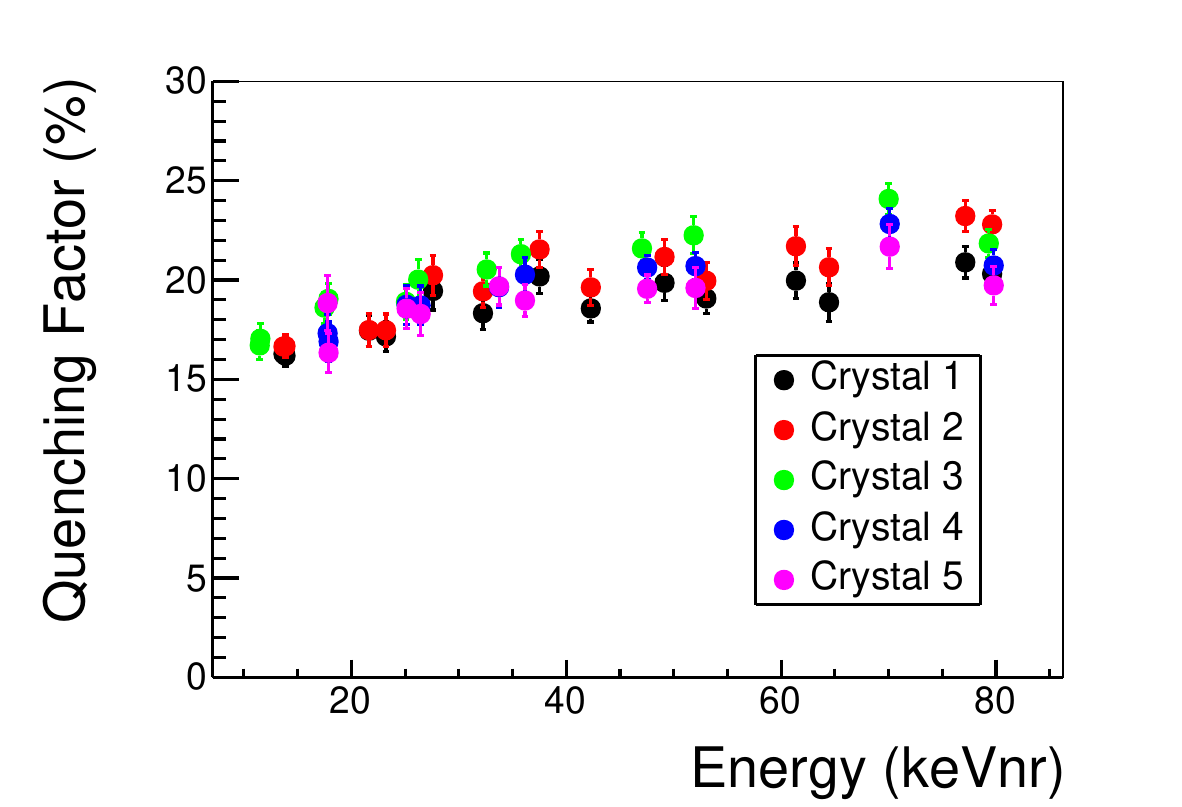}
\caption{\label{fig:Na_QF_results3} Sodium QF results for the five crystals measured using the third calibration method (linear above 6~keV following the second calibration method, but proportional below 6~keV using as reference the 6.6~keV line measured in the \isotope{Ba}{133}). Uncertainties shown include statistical and systematic contributions.  
} 
\end{figure}

QF$\mathrm{_{Na}}$ results for the five measured crystals are shown in Figures \ref{fig:Na_QF_results1}, \ref{fig:Na_QF_results2} and \ref{fig:Na_QF_results3}, using the three calibration methods explained in Section \ref{sec:Ecalibration}, respectively. 
\par

\begin{figure*}[htbp]
\centering 
\includegraphics[width=.7\textwidth]{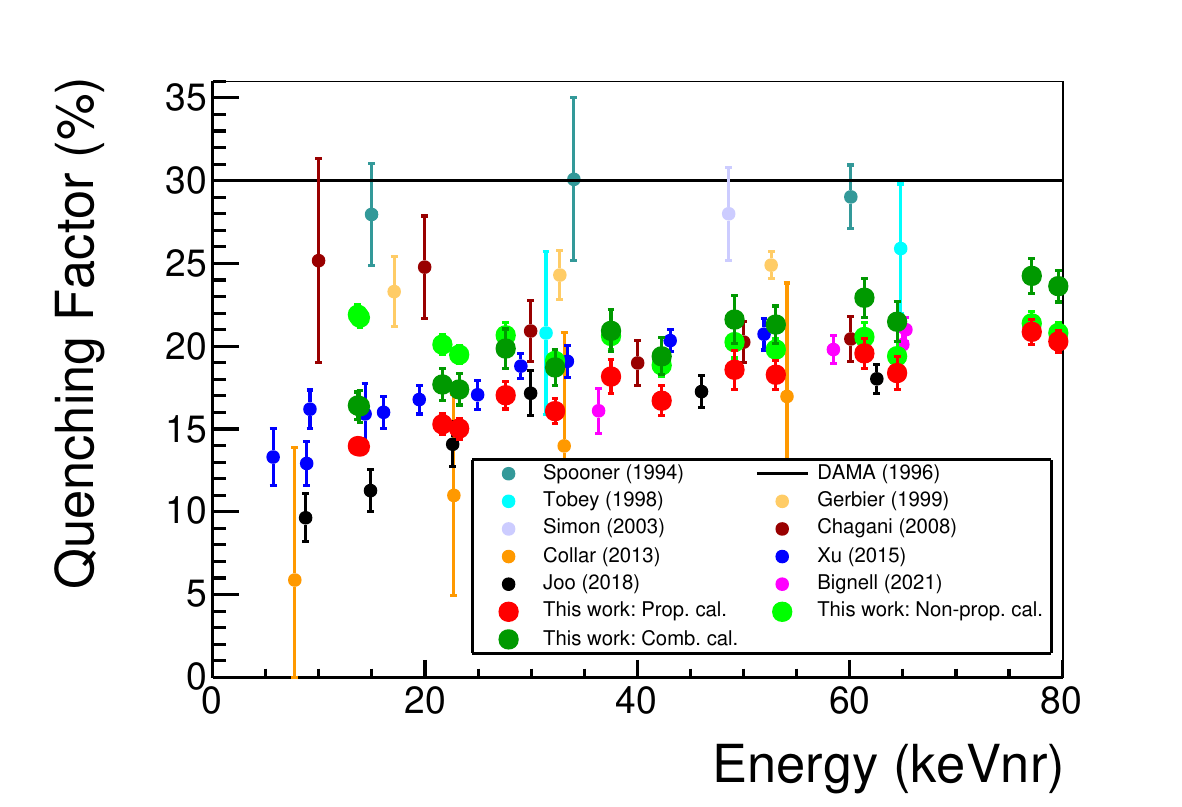}
\caption{\label{fig:Na_results} QF$\mathrm{_{Na}}$ results for crystal No.~1 for the three calibration approaches. Previous measurements are also shown \cite{xu-quench,joo-quench,dama-quench-hypo,gerbier-quench,chagani-quench,bignell-quench,collar-quench,spooner-quench}. 
} 
\end{figure*}

The results of all the five crystals are consistent with each other despite their light collection and energy resolution vary significantly. A comparison between the results of this analysis for the crystal No.~1 and the results from previous measurements is shown in Figure~\ref{fig:Na_results}. We observe a decrease in QF$\mathrm{_{Na}}$ at lower energies as reported by~\cite{xu-quench,joo-quench,gerbier-quench,bignell-quench,collar-quench,lee-quench} when using the first and third calibration methods. In the case of some of the previously cited results~\cite{collar-quench,lee-quench}, non-linearity in the response of the NaI(Tl) to electron recoils have been accounted for and still a monotonically decreasing QF$\mathrm{_{Na}}$ with decreasing energy is found~\footnote{Results presented in \cite{lee-quench} appeared when this article was already in press, and they have not been included in figure~\ref{fig:Na_results}.}.

On the other hand, in our analysis there is no clear dependency with energy of the QF$\mathrm{_{Na}}$ when the non-proportional but linear energy calibration using lines from 6.6 to 35~keV is applied (second calibration method), similar to the quenching factors reported by the earlier measurements by Spooner et al., DAMA, and Chagani~\cite{spooner-quench, dama-quench-original, chagani-quench}. 
For all the assumptions and modelling considered in our analysis, a value for the QF$\mathrm{_{Na}}$ clearly lower than the used by DAMA/LIBRA is obtained.

Considering the independence of the QF$\mathrm{_{Na}}$ on the nuclear recoil energy (obtained when the second calibration method is applied) in the range of energies accessed in these measurements (from 10~keV$_{nr}$ to 80~keV$_{nr}$), the weighted mean values of the QF$\mathrm{_{Na}}$ for each crystal have been obtained. They are shown in Table \ref{tab:QF}, being the mean for all the crystals 21.0 $\pm$ 0.3\%.

\begin{table}[htbp]
\centering 
\begin{tabular}{|c|c|}
\hline
  Crystal  & QF$\mathrm{_{Na}}$\\
  number & (\%) \\
  \hline
1 & 20.04 $\pm$ 0.7  \\
2 & 21.0 $\pm$ 0.8  \\
3 & 22.1 $\pm$ 0.8 \\
4 & 21.1 $\pm$ 0.8\\
5 & 21.1 $\pm$ 0.6 \\
\hline
\end{tabular}
\caption{\label{tab:QF} Mean sodium QF values obtained for the five crystals measured using the second calibration method. } 
\end{table}

\subsection{Iodine Quenching Factor Results and Comparison to Prior Measurements} 
\begin{figure}[htbp]
\centering 
\includegraphics[width=.5\textwidth]{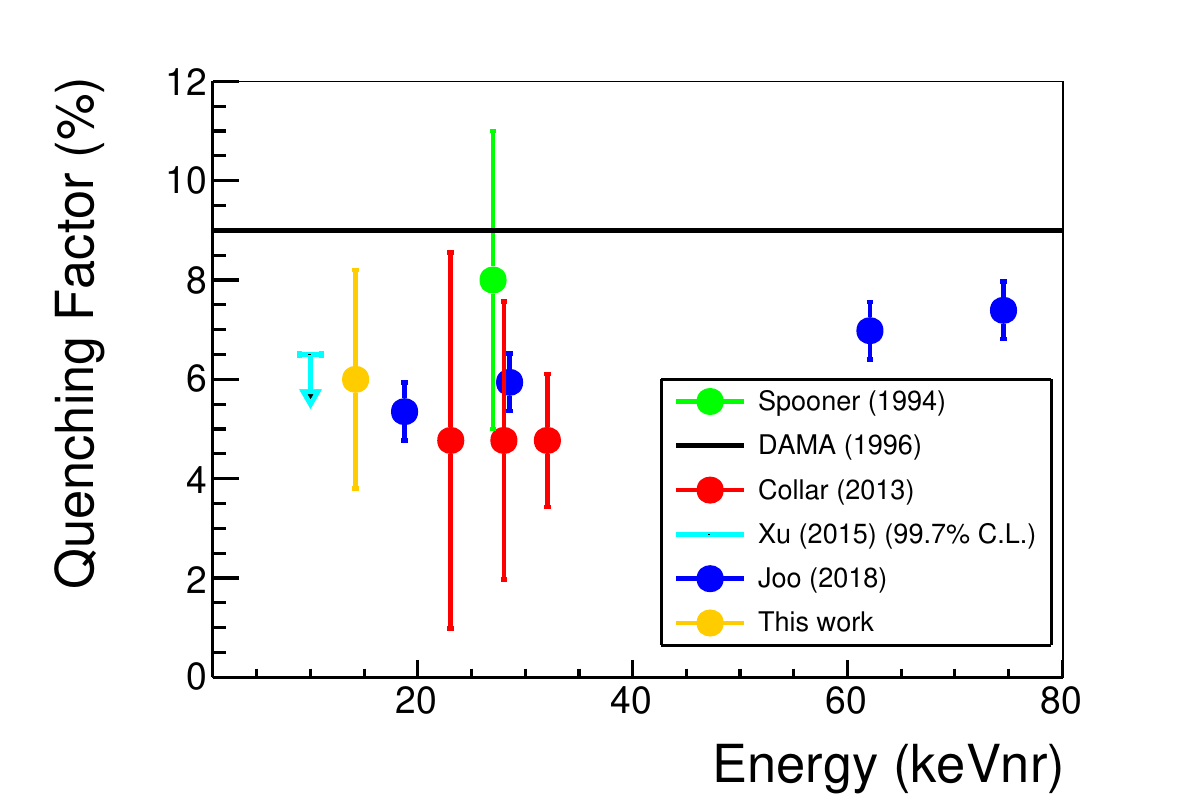}
\caption{\label{fig:I_results} QF$\mathrm{_I}$ results for the combination of the values obtained for crystals No.~2 and No.~3 and for previous measurements \cite{dama-quench-original,spooner-quench,collar-quench,xu-quench,joo-quench}.  }

\end{figure}

We estimate the QF$\mathrm{_I}$ for crystals No.~2 and No.~3, resulting values of (5.1 $\pm$ 2.7)\% and (6.5 $\pm$ 1.8)\%, respectively. As both measurements correspond to the same recoil energy (14.2 keV), they were combined together, obtaining a weighted mean of (6.0 $\pm$2.2)\%. Figure \ref{fig:I_results} shows this value together
with those obtained in previous measurements \cite{dama-quench-original,spooner-quench,collar-quench,xu-quench,joo-quench}, showing a good agreement with the most recent ones, and a value clearly lower than the used by DAMA/LIBRA.

\section{Conclusions}
\label{sec:conclusions}

We have carried out measurements of the sodium and iodine quenching factors for five small NaI(Tl) crystals, all of them performed in the same experimental setup to control systematic effects and using the same analysis protocols. Special care has been devoted to minimize the contribution from systematics in the final results. The sodium quenching factor results are compatible between crystals and the most relevant systematic effect identified is related with the energy calibration. This systematic effect may also be present in most of the previous measurements, and it is related with the well-known non-proportional behaviour of the NaI(Tl) light yield. The iodine quenching factor has been only determined with data from two of the five tested crystals and no information on the possible dependency with energy can be derived from these measurements. 
\par
\begin{figure}[t!]
\centering 
\includegraphics[width=.5\textwidth]{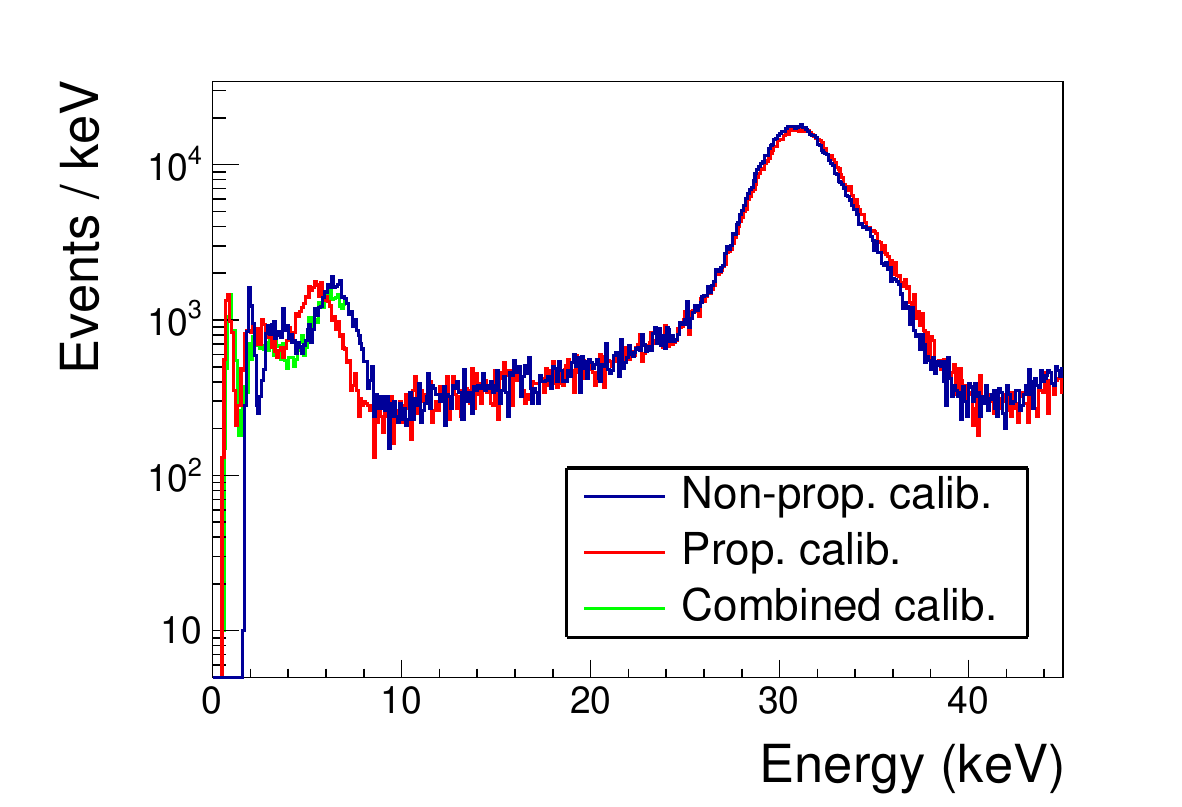}
\caption{\label{fig:prop} {\isotope{Ba}{133}} calibration spectrum for crystal No.~1 with calibration methods 1 (red line), 2 (blue line) and 3 (orange line). See text for further discussion.}
\end{figure}

This work shows the relevance of taking into account non-linearity in NaI(Tl) in the estimate of the QFs in the energy range of interest for dark matter and \cevns. Using the same datasets and different calibration methods, this work has derived QFs affected by a large dispersion, similar to that observed when comparing the previous available measurements (see Figure~\ref{fig:Na_results}). The effect is systematically observed in the five crystals measured. Figure~\ref{fig:prop} shows the non-proportionality in the response of NaI(Tl) in the range below 40 keV for crystal No.~1. The blue line corresponds to the \isotope{Ba}{133} spectrum calibrated with the lines from 6.6, 30.9 and 35.1~keV (calibration  method 2), while the red line corresponds to a proportional calibration using the 57.6 keV line as reference (calibration method 1) and the orange to the combined calibration (method 3). The line at 6.6~keV is found with calibration method 1 more than 1~keV away from the nominal energy. However, to reach the lowest nuclear recoil energies observed, we have to extrapolate the calibration below 5 keV (which corresponds to about 25~keV$_{nr}$). This implies, that the range where the behaviour of the QF$\mathrm{_{Na}}$ is most interesting is not properly calibrated with calibration method 2. By comparing the spectra in Figure~\ref{fig:prop} with the expectations from the simulation (see Figure~\ref{fig:Ba_esp_sim}), we observe that below 5 keV, calibration method 2 is not longer valid. Calibration method 3, on the other hand, linearizes in two steps the range of interest, accommodating the data from \isotope{Ba}{133} calibration but assuming proportionality in a smaller energy range in terms of electron equivalent energy. Because of this, we consider the most sound results for QF$\mathrm{_{Na}}$ those obtained with calibration method 3. However, these results highlight the relevance of stablishing sound calibration protocols at very low energies and better understanding the light production mechanisms in NaI(Tl), in particular, for the energy deposited by nuclear recoils. The results obtained using method 3, for instance, are compatible with previous measurements which conveniently corrected their data by the non-linearity in the response of NaI(Tl)~\cite{collar-quench,lee-quench}.
\par

Although further work is required to improve our understanding of scintillation quenching factors for nuclear recoils in NaI(Tl), other works complementary to the presented in this article are ongoing, for instance calibrations onsite of the ANAIS-112 detectors using \isotope{Cf}{252} sources, being preliminary results presented recently \cite{pardo2023neutron}. This work supports that energy-dependent quenching factor for sodium provides a better description of all the measurements, and it is aligned with most of the previous quenching factor estimates for sodium nuclear recoils. It is also worth to highlight that both, sodium and iodine quenching factors in our five tested crystals are smaller than those reported by DAMA/LIBRA for all the considered assumptions and modellings in the analysis. 
\par

\section*{Acknowledgments}
The work from DC, MM and MLS has been financially supported by MCIN/AEI/10.13039/501100011033/FEDER, UE under grants PID2022-138357NB-C21, PID2019-104374GB-I00 and FPA2017-83133-P, by the Gobierno de Arag\'on and the European Social Fund (Group in Nuclear and Astroparticle Physics) and funds from the European Union NextGenerationEU/PRTR (Planes complementarios, Programa de Astrof\'{\i}sica y F\'{\i}sica de Altas Energ\'{\i}as).
WGT, EBS, JHJ, and RHM acknowledge support from the National Science Foundation No. PHY-1913742 and DGE-1122492. This research was supported by the U.S. Department of Department of Energy, Office of Nuclear Physics,
under grant number DE-FG02-97ER41033 (TUNL).

\end{document}